%% file: ATHOS.tex
\documentclass{aa}  

\usepackage{graphicx}
\usepackage{txfonts}
\usepackage{natbib}
\usepackage{arydshln}
\usepackage{color, colortbl}
\usepackage[colorlinks, citecolor=blue, linkcolor=blue, urlcolor=blue]{hyperref}
\usepackage{blindtext}

\newcommand{\numberofTeffrel}{nine~}
\newcommand{\numberofmetrel}{31~}
\newcommand{\Teffsyserr}{97}
\newcommand{\metsyserr}{0.16}
\newcommand{\loggsyserr}{0.26}
\newcommand{\nameoftool}{ATHOS}

\begin{document}

   \title{ATHOS: On-the-fly stellar parameter determination of FGK stars based on flux ratios from optical spectra\thanks{\url{https://github.com/mihanke/athos}} 
         }

   \author{Michael Hanke\inst{1}
          \and
          Camilla Juul Hansen\inst{2,3}
          \and
          Andreas Koch\inst{1}
          \and
          Eva K. Grebel\inst{1}
          }

   \institute{Astronomisches Rechen-Institut, Zentrum für Astronomie der Universität Heidelberg, Mönchhofstr. 12-14, D-69120  Heidelberg, Germany\\
              \email{mhanke@ari.uni-heidelberg.de}
              \and
              Max Planck Institute for Astronomy, Königstuhl 17, D-69117 Heidelberg, Germany
              \and
              Dark Cosmology Centre, The Niels Bohr Institute, Juliane Maries Vej 30, 2100 Copenhagen, Denmark
              }

   \date{\today}
   
\abstract{The rapidly increasing number of stellar spectra obtained by existing and future large-scale spectroscopic surveys feeds a demand for fast and efficient tools for the spectroscopic determination of fundamental stellar parameters. Such tools should not only comprise customized solutions for one particular survey or instrument, but, in order to enable cross-survey comparability, they should also be capable of dealing with spectra from a variety of spectrographs, resolutions, and wavelength coverages. To meet these ambitious specifications, we developed ATHOS (A Tool for HOmogenizing Stellar parameters), a fundamentally new analysis tool that adopts easy-to-use, computationally inexpensive analytical relations tying flux ratios (FRs) of designated wavelength regions in optical spectra to the stellar parameters effective temperature ($T_\mathrm{eff}$), iron abundance ([Fe/H]), and surface gravity ($\log{g}$). Our $T_\mathrm{eff}$ estimator is based on FRs from nine pairs of wavelength ranges around the Balmer lines H$\beta$ and H$\alpha$, while for [Fe/H] and $\log{g}$ we provide \numberofmetrel and 11 FRs, respectively, which are spread between $\sim4800$~{\AA} and $\sim6500$~{\AA}; a region covered by most optical surveys. The analytical relations employing these FRs were trained on $N=124$ real spectra of a stellar benchmark sample that covers a large parameter space of $T_\mathrm{eff}\approx4000$ to $6500$~K (spectral types F to K), [Fe/H$]\approx-4.5$ to 0.3~dex, and $\log{g}\approx1$ to 5~dex, which at the same time reflects ATHOS' range of applicability. We find accuracies of \Teffsyserr~K for $T_\mathrm{eff}$, \metsyserr~dex for [Fe/H], and \loggsyserr~dex for $\log{g}$, which are merely bounded by finite uncertainties in the training sample parameters. ATHOS' internal precisions can be better by up to 70\%. We tested ATHOS on six independent large surveys spanning a wide range of resolutions ($R=\lambda/\Delta\lambda\approx2000$ to 52000), amongst which are the Gaia-ESO and the SDSS/SEGUE surveys. The exceptionally low execution time ($<30$~ms per spectrum per CPU core) together with a comparison to the literature parameters showed that ATHOS can successfully achieve its main objectives, in other words fast stellar parametrization with cross-survey validity, high accuracy, and high precision. These are key to homogenize the output from future surveys, such as 4MOST or WEAVE.}
\keywords{methods: data analysis -- surveys -- techniques: spectroscopic -- stars: abundances -- stars: fundamental parameters} 
\titlerunning{Constraining stellar parameters from spectral flux ratios}
\maketitle

\section{Introduction}
While spectroscopic campaigns aimed at dissecting the formation history of the Milky Way have a long-standing history \citep[e.g.,][]{Beers1985,Christlieb2001,Yanny2009}, the astronomical landscape of the next decades will be governed by ever-larger spectroscopic surveys that aim at painting a complete chemo-dynamic map of our Galaxy. Amongst these are the surveys RAVE \citep{Steinmetz06}, SEGUE \citep{Yanny2009}, APOGEE \citep{Majewski17}, GALAH \citep{DeSilva15}, Gaia-RVS \citep{Cropper18}, LAMOST \citep{Zhao12}, Gaia-ESO \citep{Gilmore2012}, 4MOST \citep{deJong2012}, and WEAVE \citep{Dalton2012}. All these build on the multiplexing capacities of present and future spectrographs and have the goal of expanding the six-dimensional phase space into a multidimensional information space by adding chemical abundance measurements of a large number of tracers of chemical evolution for several hundred thousands of stars. Inevitably, this requires high spectral resolution ($R = \lambda / \mathrm{FWHM} \gtrsim$20000; \citealt{Caffau2013}), but also a precise and accurate knowledge of the stellar parameters\footnote{Here taken as effective temperature, T$_{\rm eff}$, surface gravity $\log{g}$, microturbulence, $v_\mathrm{t}$, and the overall metallicity [M/H], which we will use synonymously with [Fe/H] in the following, though we are aware that the latter nomenclature is at odds with the formally correct definition. We chose, however, to follow the common usage in the literature. Higher order parameters such as stellar rotation will only be briefly discussed.} of the target stars. 

Various methods for parameter determination are in use, ranging from photometric calibrations of a temperature scale \citep[e.g.,][]{Alonso96, Alonso99a}, excitation equilibrium using large numbers of Fe lines, Balmer-line scrutiny, to least-squares fitting of spectral templates or line indices over a broad parameter grid \citep{Lee2008}. Systematic effects can, to first order, be decreased by using analysis techniques differentially to a standard star of known parameters \citep{Fulbright2006,Koch08}. To ensure success, all these methods, in turn, require accurate atomic data and stellar model atmospheres \citep{Barklem02}, and yet, degeneracies and covariances, in particular between $T_\mathrm{eff}$ and $\log{g}$, are often inevitable \citep{McWilliam1995,Hansen2011}. Further problems arise with large data sets, where the homogenization of parameter scales \citep{Venn2004,Smiljanic2014} and the sheer computational time for spectral analysis become an issue.

Here, we introduce a new, fast, and efficient algorithm for stellar parameter determination, named ATHOS (A Tool for HOmogenizing Stellar parameters). ATHOS relies on the measurement of flux ratios (FRs) between well-tested spectral regions that are sensitive to specific parameter combinations and that we optimize to reproduce a compilation of training spectra and their parameters, amongst which are the accurate and precise parameters of the Gaia-ESO benchmark stars \citep{Jofre14}. 

Line-depth ratios (LDRs) of metal lines have long been used as $T_\mathrm{eff}$ indicators, taking advantage of  variations of the lines’ temperature sensitivity over broad ranges, while at the same time being largely pressure-independent as long as the lines are not saturated \citep{Gray1991,Kovtyukh03}. An extension of these methods rather employs the ratios of flux points that do not necessarily coincide with the line cores, but with other parts of the lines that show empirical, strong sensitivities to the parameter of choice. \citet{Kovtyukh03} provided a very precise calibration of LDRs to the temperatures of F to K dwarfs. However, their stellar sample restricted its applicability to a narrow metallicity window of $-0.5<$[Fe/H]$<+0.5$ dex and the way of measuring line depths of the lines -- that is profile fits to the line -- made this method prone to the uncertainties of continuum normalization, which is circumvented by using FRs with rather narrow wavelength spacing. By not relying on pairs of low- and high-excitation lines, ATHOS further allows for measurements of parameters down to much lower metallicities.

Notable features of ATHOS are its fast performance ($\sim25$~ms/ <10~ms for a high-/ low-resolution spectrum), applicability over a wide range of resolutions ($R \gtrsim 10000$ for all parameters; $R \gtrsim 2000$ for the $T_\mathrm{eff}$ scale), and validity over a broad range of stellar parameters ($T_\mathrm{eff}\approx 4000$ to $6500$~K, [Fe/H$]\approx-4.5$ to  0.3~dex, $\log{g}\approx1$ to 5~dex). This makes it an ideal tool to provide precise and accurate stellar parameters for large samples within seconds -- an important asset in the era of future spectroscopic missions.

This tool is meant to work for all optical spectra, not just for stars originating from one survey as most tailored pipelines do, but it offers a way to homogenize large samples from different surveys. It is by no means an attempt to supersede various survey pipelines, but a simple way to put the millions of stars to be observed on the same scale, so that these are homogeneously treated and not biased by the choices or methods adopted within individual surveys.

This paper is organized as follows: In Sect. \ref{Sec: Training set} we present the set of training spectra used in this study together with a brief discussion of the stellar parameter derivation. Furthermore, the homogenization procedure and treatment of telluric contamination is described. Sect. \ref{Sec: Method} presents the concept of FRs and how we used correlation coefficients to find the FRs carrying information about stellar parameters, while in Sect. \ref{Sec: PAWN} the computational implementation of the deduced analytical relations is outlined. ATHOS' stability against resolution and signal-to-noise ratio (S/N) is discussed in Sect. \ref{Sec: Discussion}, where we also compare our parameter scales to various spectroscopic surveys. Finally, we conclude with a summary of our results in Sect. \ref{Sec: Summary}.

\section{Training set}\label{Sec: Training set}
This project was originally meant to be model-driven. Therefore, we initially synthesized spectra of a homogeneous and dense coverage of the parameter space and conducted the analysis outlined in Sect. \ref{Sec: Method}. Unfortunately, it turned out that the optimal FRs deduced from theory alone cannot be reproduced in real spectra, and vice versa. A possible explanation is the oversimplification made throughout the modeling by preferring the assumptions of local thermodynamic equilibrium (LTE) and plane-parallel, static atmospheres over a fully three-dimensional, hydrodynamic treatment under non-LTE conditions. This, in turn, would come with enormous computational costs. Further, we identify inaccurate line data as an additional caveat of the theoretical treatment. Due to these limitations, the decision was made to base this work entirely on observed spectra with accurately determined stellar parameters.

To this end, we have compiled a grid of in total 195 high-resolution, high S/N spectra of 124 stars covering the visual wavelength range including the regions around the strong features of the H$\beta$ profile at $4861.3$~{\AA}, the \ion{Mg}{i}~b triplet at $\sim5175$~{\AA}, the Na-D doublet at $\sim5890$~{\AA}, and H$\alpha$ at $6562.8$~{\AA}. A valuable part of the sample is the Gaia FGK benchmark star library of \citet{Blanco-Cuaresma14} (GBS) with attributed stellar parameters from \citet{Jofre14} and \citet{Heiter15} for the more metal-rich stars, and \citet{Hawkins16} for the metal-poor targets. Their data were obtained with four different spectrometers and resolutions. These are the HARPS spectrograph at the ESO, La Silla 3.6~m telescope; UVES on the VLT/UT2 at Cerro Paranal, Chile; ESPaDOnS on the Canada-France-Hawaii Telescope at the Mauna Kea observatory, Hawaii, and its twin NARVAL mounted on the 2~m Telescope Bernard Lyot on Pic du Midi, France. Most of the stars in the library have data available from at least two of these instruments and the S/N is mostly well above 100 pixel$^{-1}$. Moreover, the GBS library includes high-quality spectra of the Sun and Arcturus observed by \citet{Hinkle00} with the Echelle spectrograph at the Coud\'e Feed telescope on Kitt Peak, Arizona. We added UVES dwarf and giant spectra from \citet{Hansen12} (hereafter CJH12) to the grid. The metal-poor (-1~dex to -4.5~dex) end was additionally populated by stars characterized in \citet{Roederer14} (henceforth R14). Unfortunately, their spectra are not publicly available. Consequently, we used the ESO Advanced Data Products (ADP) query to cross-check for publicly available data, yielding matches for 48 stars with mainly UVES and some HARPS observations. In order to fill the otherwise sparsely populated horizontal branch (HB), we once again employed the ADP to retrieve spectra for the cooler targets in the spectroscopic HB studies of \citet{For10} and \citet{Afsar12} (red HB). A list of all the data including the respective stellar parameters and sources thereof can be found in Table \ref{Table:TRAINING_SET}.

\subsection{Stellar parameters}\label{Subsec: stellar parameters}
For the GBS sample, \citet{Heiter15} and \citet{Hawkins16} inferred $T_\mathrm{eff}$ and $\log{g}$ from angular diameters and bolometric fluxes, as well as from fitting stellar evolutionary tracks. The deduced errors for these procedures range from $\sim20$~K to 100~K and 0.01~dex to 0.15~dex for stars other than the Sun, which exhibits comparatively negligible errors. In light of the new interferometric temperature measurements for HD~140283, HD~122563, and HD~103095 by \citet{Karovicova18}, who found earlier measurements to have suffered from systematic effects much larger than the provided errors, we decided to use their more recent values and errors for these stars. For the metallicity, \citet{Jofre14} averaged line-by-line abundances that stem from up to seven different analysis codes. These have been each corrected for departures from LTE. Uncertainties achieved in this way span 0.03~dex up to 0.40~dex.

CJH12 derived effective temperatures for their sample using photometric color-$T_\mathrm{eff}$ relations. The same study provides $\log{g}$ based on either parallaxes in conjunction with stellar structure equations, or by enforcing ionization balance, that is requiring deduced abundances from \ion{Fe}{i} and \ion{Fe}{ii} to agree with each other. The provided [Fe/H] values originate from an LTE analysis of \ion{Fe}{i} lines. 

R14 employed a strictly spectroscopic determination of $T_\mathrm{eff}$ by balancing abundances of \ion{Fe}{i} transitions at low and high excitation potential. For stars cooler than the main sequence turnoff (MSTO), $\log{g}$ was based on fits to theoretical isochrones, while for the hotter stars it was inferred from ionization balance. [Fe/H] values for the R14 targets were calculated using LTE \ion{Fe}{ii} abundances, which ought to experience smaller corrections in an NLTE treatment \citep[e.g.,][]{Bergemann12, Lind12}. 

\begin{table}
\caption{Line list used in Sect. \ref{Subsec: stellar parameters}}
\label{Table: line list}
\centering
\resizebox{\columnwidth}{!}{%
\begin{tabular}{@{\extracolsep{6pt}}ccr ccr}
\hline\hline
$\lambda$ & $\chi_\mathrm{ex}$ & $\log{gf}$ & $\lambda$ & $\chi_\mathrm{ex}$ & $\log{gf}$\\
\phantom{.}[{\AA}] & [eV] & [dex] & [{\AA}] & [eV] & [dex]\\
\hline
\multicolumn{3}{c}{\ion{Fe}{i}} & \multicolumn{3}{c}{\ion{Fe}{ii}}\\
\cline{1-3}\cline{4-6}
3536.556 & 2.880 &  0.115 & 3406.757 & 3.944 & -2.747\\  
3640.389 & 2.730 & -0.107 & 3436.107 & 3.967 & -2.216\\  
3917.181 & 0.990 & -2.155 & 3535.619 & 3.892 & -2.968\\  
4021.867 & 2.760 & -0.729 & 4178.862 & 2.583 & -2.535\\  
4072.510 & 3.430 & -1.440 & 4233.172 & 2.583 & -1.947\\  
4076.640 & 3.210 & -0.530 & 4416.830 & 2.778 & -2.602\\
  \hline
  \end{tabular}}
  \tablefoot{The full table is available through the CDS.}
  \end{table}
Due to the former two studies pursuing different approaches to obtain stellar parameters -- notably $T_\mathrm{eff}$ from photometry or from excitation balance and [Fe/H] from \ion{Fe}{i} or from \ion{Fe}{ii} abundances -- we decided to re-analyze both samples in a homogeneous investigation\footnote{For further information on the discrepancy of photometric and spectroscopic parameter scales, see Sect. \ref{Subsec: comparison S4N}.}. This was done by employing a common, carefully selected and inspected Fe line list, in conjunction with equivalent widths (EWs). The Fe line list was compiled for CJH12 and used therein. The lines were chosen such that there was no trend with wavelength, and so that excitation and ionization trends were tight (little scatter). The \ion{Fe}{i} lines are from the Vienna Atomic Line Database \citep[VALD,][2008 version]{Piskunov95}, the Oxford group \citep{Blackwell79a,Blackwell79b,Blackwell79c,Blackwell82b,Blackwell82c,Blackwell82a}, \citet{O'Brian91}, and \citet{Nissen07}. For \ion{Fe}{ii} the list was based on VALD, \citet{Blackwell80}, and \citet{Nissen07}. The line list is presented in Table \ref{Table: line list}. 

\begin{figure}
    \centering
    \resizebox{\hsize}{!}{\includegraphics{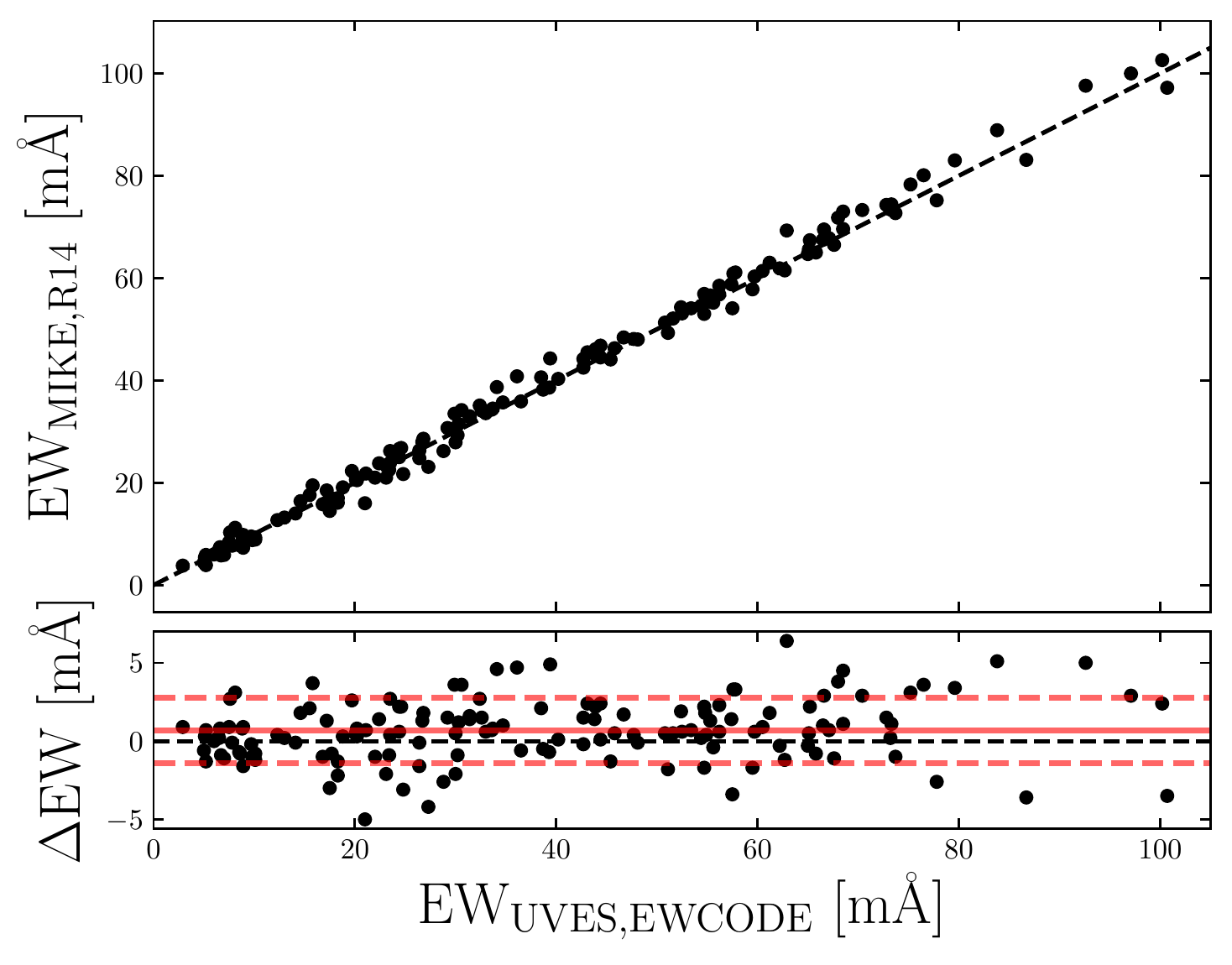}}
      \caption{Upper panel: Comparison of our EWCODE runs on UVES spectra with the literature values determined from MIKE spectra by R14. The one-to-one trend is shown by the black dashed line. Lower panel: Residual distribution. The solid and dashed red lines indicate the mean deviation of 0.67~m{\AA} and the rms scatter of 2.07~m{\AA}, respectively. 
              }
      \label{Fig:EWs_R14_vs_EWCODE}
\end{figure}

In case of the CJH12 spectra, EWs were computed from our library spectra (see next Sect.) using our own, semi-automated EW tool EWCODE \citep{Hanke17}. For spectra in the R14 sample, we relied on published EWs after cross-matching our line list with \citet{Roederer14}. Systematic differences in the methods to determine EWs and/or the spectrographs (UVES compared to MIKE) could be excluded by checking the EW results of Fe lines for the five stars in common between R14 and CJH12. At a mean deviation of $0.67\pm2.07$~m{\AA} (root mean square deviation, rms) no significant discrepancy was found (see Fig. \ref{Fig:EWs_R14_vs_EWCODE}). We note that not all of these EWs entered our subsequent analysis, because the stars in question were present in the GBS sample, which supersedes our parameters.
\begin{figure*}
    \centering
    \resizebox{\hsize}{!}{\includegraphics{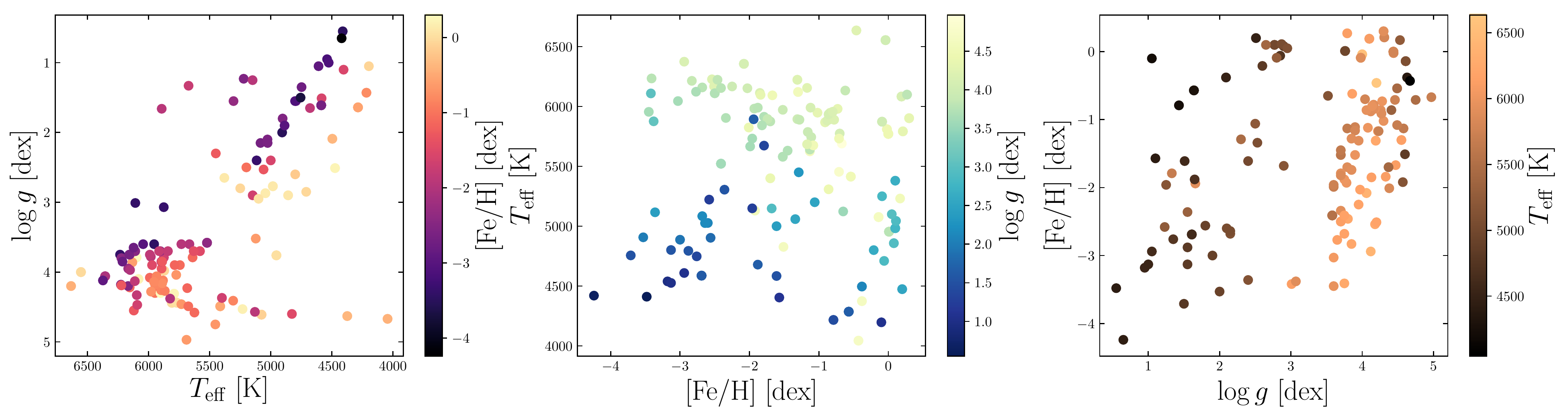}}
      \caption{Distribution of the training set in the parameters $T_\mathrm{eff}$, [Fe/H], and $log{g}$. The color coding in each plot indicates the respective missing dimension.
              }
      \label{Fig:grid_parameters}
\end{figure*}
We used EWs to constrain $T_\mathrm{eff}$ by enforcing excitation equilibrium of \ion{Fe}{i} lines adopting plane-parallel ATLAS9 model atmospheres interpolated from the grid by \citet{Castelli04}. Line-by-line abundances were computed using the LTE analysis code MOOG \citep[][July 2014 release]{Sneden73}. Typical temperature uncertainties are of order 100~K. In parallel to excitation equilibrium, the empirical atmospheric microturbulence parameter, $v_\mathrm{t}$, was tuned to satisfy agreement between weak and strong lines. For the majority of stars, where CJH12 and R14 used parallaxes and photometry to deduce $\log{g}$, we did not allow the model gravity to vary and used the literature values instead. For the other stars, ionization equilibrium was required to derive $\log{g}$ (labeled in Table \ref{Table:TRAINING_SET}). Finally, our estimate for [Fe/H] is based on \ion{Fe}{ii} abundances derived from the optimal set of model atmosphere parameters. Following, for example, \citet{Roederer14}, here we preferred the ionized species over the neutral one at the expense of number statistics, because it is less prone to NLTE effects. For the iron abundance, we adopted errors of 0.10~dex. 

For the red HB sample, \citet{For10} and \citet{Afsar12} used only spectroscopic indicators, that is excitation equilibrium for $T_\mathrm{eff}$ and LTE ionization balance for $\log{g}$ and consequently [Fe/H] from Fe lines. The provided uncertainties are 150~K, 0.16~dex, and $\sim0.1$~dex, respectively. Here, we adopted the literature parameters.

Since a few stars have been covered by more than one of the individual subsets discussed above, we had to homogenize the deduced stellar parameters from the different studies. For those stars that occur in the GBS, we chose the GBS parameters as reference. If this was not the case we averaged over the parameters we have derived from different EW sources and used the deviations and respective uncertainties for a new uncertainty estimate. In the present study we relied on the solar chemical composition by \citet{Asplund09} stating $\log{\epsilon(\mathrm{Fe})}_\odot=7.50$~dex. Hence, GBS metallicities, which are based on $\log{\epsilon(\mathrm{Fe})}_\odot=7.45$~dex by \citet{Grevesse07}, had to be adjusted accordingly. The final training parameters of the spectra entering our analysis can be found in Table \ref{Table:TRAINING_SET}.  

The selected spectra cover stars in the most relevant parts of the Hertzsprung-Russel diagram (left panel of Fig. \ref{Fig:grid_parameters}), viz. on the main sequence (MS), the MSTO, the subgiant branch (SGB), the red-giant branch (RGB), and the HB. In terms of stellar parameters, the training set spans a parameter space from $T_\mathrm{eff}\approx4000$ to $6500$~K, $\log{g}\approx1$ to 5~dex, and [Fe/H$]\approx-4.5$ to 0.30~dex. Fig. \ref{Fig:grid_parameters} illustrates the distribution of our training sample in stellar parameter space ($T_\mathrm{eff}$, $\log{g}$, [Fe/H]). 

\begin{table*}
\caption{Training set information}
\label{Table:TRAINING_SET}
\centering
\begin{tabular}{lcrccrcll}
\hline\hline
name        & $T_\mathrm{eff}$ & $\sigma_{T_\mathrm{eff}}$ & $\log{g}$ & $\sigma_{\log{g}}$  & [Fe/H] & $\sigma_\mathrm{[Fe/H]}$ & spectrograph & source \\
& [K] & [K] & [dex] & [dex] & [dex] & [dex]  & & \\
\hline
$\alpha$~Boo         & 4286\phantom{$^{(a)}$} &  35 & 1.64\phantom{$^{(b)}$} & 0.09 & -0.57\phantom{$^{(a)}$} & 0.08 & Coudé    & GBS   \\
$\omega^2$~Sco       & 5380\phantom{$^{(a)}$} & 150 & 2.65\phantom{$^{(b)}$} & 0.16 &  0.10\phantom{$^{(a)}$} & 0.10 & HARPS    & red HB\\
BD+20~571            & 5935$^{(a)}$           & 100 & 4.06$^{(b)}$           & 0.20 & -0.86$^{(a)}$           & 0.10 & UVES     & CJH12 \\
BD+24~1676           & 6110$^{(a)}$           &  71 & 3.70\phantom{$^{(b)}$} & 0.05 & -2.50$^{(a)}$           & 0.07 & UVES     & R14   \\
\hline
\end{tabular}
\tablefoot{
\tablefoottext{a}{Parameters redetermined from excitation balance ($T_\mathrm{eff}$) and \ion{Fe}{ii} abundances ([Fe/H]) using new line data (see Sect. \ref{Sec: Training set}).}
\tablefoottext{b}{$\log{g}$ inferred from ionization equilibrium of \ion{Fe}{i} and \ion{Fe}{ii}. Otherwise, the literature parameter was used. The full table is made available through the CDS.}
}
\end{table*}

\subsection{Grid homogenization}\label{Sec:Grid homogenization}
For the purpose of spectral homogeneity and computationally efficient access, spectra from different sources and spectrographs were first shifted to rest wavelengths. This was achieved by using radial velocities determined from a cross-correlation, either with a template spectrum of the Sun or Arcturus ($\alpha$~Boo), depending on which of those is closer to the target in stellar parameter space. Imprecisions introduced by using these metal-rich templates for metal-poor targets are unproblematic for this investigation (effect of less than 1~km~s$^{-1}$), which we validated by cross-correlating some of the RV-shifted, metal-poor targets against each other. Next, the spectra were degraded to match a resolution of $R=45000$ by convolution with a Gaussian kernel of appropriate width. Some of the spectra in the R14 sample are originally at resolutions slightly below the desired one, but still well above $40000$. We kept those at their original value and point out that this has negligible effects on this study (see Sect. \ref{Subsec:Resolution}). Finally, the data were re-binned to a common, linear wavelength scale with equidistant spacing of $\delta\lambda=0.017$~{\AA}~pixel$^{-1}$, this configuration being representative for a typical UVES580 setup \citep{Pasquini00}. We did not normalize the training spectra since the method introduced here (Sect. \ref{Sec: Method}) considers relative fluxes and is consequently scale-free.

\subsection{Telluric contamination}\label{Subsec: Tellurics}
\begin{figure*}
    \centering
    \resizebox{\hsize}{!}{\includegraphics{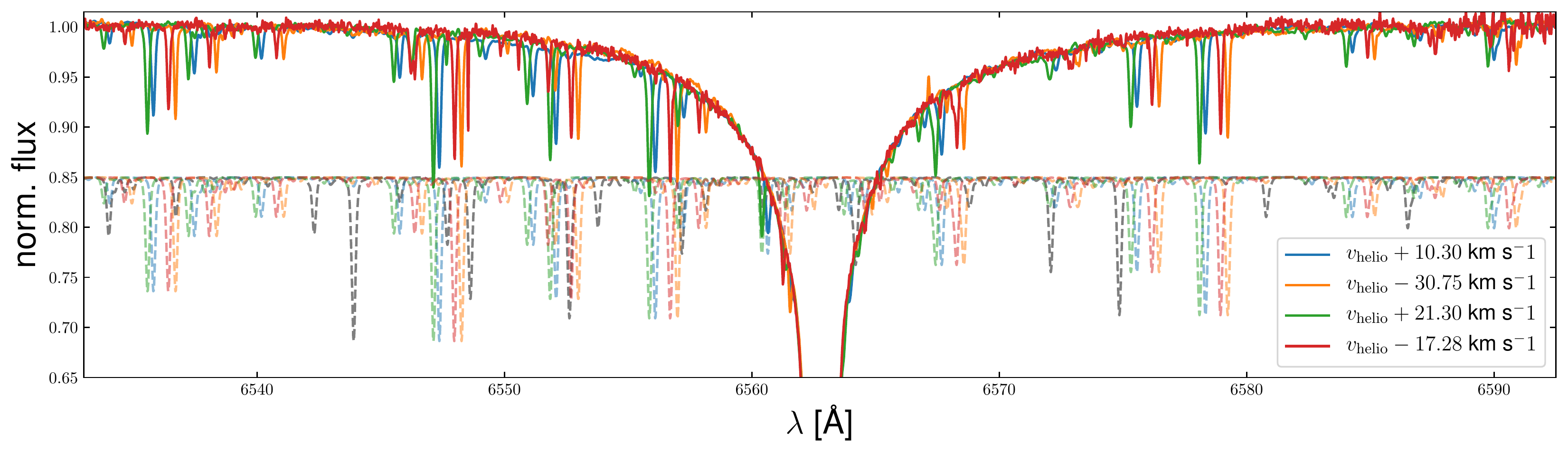}}
      \caption{Telluric contamination due to H$_2$O vapor in Earth's lower atmosphere. Shown as solid lines are rest-frame spectra of HD~140283 taken at different epochs and line-of-sight radial velocities. The apparent motion with respect to the heliocentric velocity $v_\mathrm{helio}=-169.00$ km~s$^{-1}$ is indicated in the legend. All dashed lines represent the same topocentric telluric absorption model (gray, see text for details) but being shifted in wavelength space to match the radial velocity of their observed counterpart (light-colored). For better visibility the models are shifted in flux direction as well.
              }
      \label{Fig:tellurics}
\end{figure*}
In the H$\alpha$ region, telluric absorption plays a non-negligible role in the line shape of this feature. As briefly noted in, for example, \citet{Eaton95} and \citet{Cayrel11}, there is a wealth of absorption features caused by H$_2$O vapor in the Earth's lower atmosphere falling right in the spectral region around H$\alpha$. None of the archival spectra in our set has been corrected for telluric contamination. We address this issue in Fig. \ref{Fig:tellurics}, where we plotted a portion around H$\alpha$ of four out of the six available spectra of the metal-poor SGB star HD~140283 ($T_\mathrm{eff}=5522$ K, $\log{g}=3.58$ dex, [Fe/H$] = -2.41$ dex; \citet{Jofre14, Heiter15}) acquired at different epochs. Being fairly metal-poor and hot, HD~140283 is not expected to show substantial stellar absorption in the presented region except for H$\alpha$ itself. However, there is a clear indication of contamination from lines moving with the topocentric -- that is the telescope's -- rest frame. We retrieved a telluric absorption model using SkyCalc \citep{Noll12,Jones13}, a tool dedicated to compute sky models for the VLT observatory on Cerro Paranal at $2640$~m above sea level. We did not attempt to match the ambient conditions of the observations of HD~140283, but used a global model for a zenith pointing and a seasonal averaged precipitable water vapor of 2.5 mm. A comparison of the models to the observed spectra reveals that the vast majority of the small-scale features originate from telluric absorbers. We note the variations of the depths of the real tellurics between observations and attribute them to varying observing conditions such as airmass and/or water vapor content in the lower atmosphere. 

As we are aiming for stellar parameter determinations irrespective of the targets' relative motions, we have to take into account the fact that -- depending on radial velocity -- contamination can in principle prevail at any wavelength in the vicinity of H$\alpha$. In the training set with precisely known velocities, we achieve this by masking out tellurics in the individual spectra based on their velocity shifts. This was done by looking for flux minima with line depths above 3\% in the telluric model described above and masking the neighboring wavelength ranges of one $\mathrm{FHWM}$ on either side. We favor this masking procedure over a detailed modeling and removal of telluric features because of the lack of knowledge of ambient observing conditions for all the spectra and the possible caveats coming from model uncertainties. Moreover, since many of the stars in the grid are represented by more than one spectrum with varying topocentric velocity, a range masked in the one may well be accessible in the other (as can be seen in Fig. \ref{Fig:tellurics}). Keeping the tellurics in would increase the rms scatter in our H$\alpha$-based temperature scales to 280~K compared to the 122~K we find below.

\section{Method}\label{Sec: Method}
For this study, we investigated how FRs are affected by stellar parameters and, vice versa, can be used to constrain them. We define a specific FR as the ratio of the two mean flux levels of a spectrum $F$ in the open intervals of width $w$ around the central wavelengths $\lambda_1$ and $\lambda_2$, that is,
\begin{equation}
\label{Eq:FR}
 \mathrm{FR}_{\lambda_1,\lambda_2} = \frac{\langle{F}_{\lambda_1}\rangle}{\langle{F}_{\lambda_2}\rangle}
\end{equation}
with
\begin{equation}
\label{Eq:medflux}
\langle{F}_{\lambda_i} \rangle = \frac{1}{n}\sum_{j=1}^n F(\lambda_j), \quad \lambda_j \in (\lambda - w/2, \lambda +  w/2).
\end{equation}
Using FRs bears the main benefit that they are scale-free, meaning they circumvent the caveats of normalization procedures. These are heavily affected by, among others, S/N and resolution of the spectrum, as well as by intrinsic physical quantities such as metallicity and temperature. Provided that the two dispersion points from which an FR is computed are closely spaced in wavelength, the local continuum can be approximated to be constant. This holds true even for merged echelle spectra with clearly extrinsic large-scale continuum variation. Another advantage of measuring FRs over employing iterative minimization approaches, such as profile fits or full spectrum fits -- on which to our knowledge any other approach of determining stellar parameters relies -- is the comparatively reduced computation time. Therefore, per spectrum, the demand for computational resources can be tremendously lowered.

In order to quantify the information content of an FR of a set of two wavelength regions in the grid with respect to a stellar parameter $x$, we chose the Pearson product-moment correlation coefficient
\begin{equation}\label{Eq:corr coeff}
 r_{\mathrm{FR},x} = \frac{\sum_{i=1}^{N}{(\mathrm{FR}_i - \langle\mathrm{FR}\rangle)(x_i - \langle x\rangle)}}{\sqrt{\sum_{i=1}^{N}{(\mathrm{FR}_i - \langle\mathrm{FR}\rangle)^2}}\sqrt{\sum_{i=1}^{N}{(x_i - \langle x\rangle)^2\phantom{\overline{\mathrm{I}}}}}},
\end{equation}
where $N$ corresponds to the number of grid points that do not contain pixels masked as tellurics in either of the two wavelength bins, or which are not accessible for other reasons. Here, $\langle\mathrm{FR}\rangle$ represents the mean of the measured FR$_i$ and $\langle x\rangle$ the mean of the investigated parameter values $x_i$. According to Eq. \ref{Eq:corr coeff}, both strong anticorrelations and correlations, that is absolute values $\left|r_{\mathrm{FR},x}\right|$ close to unity, indicate a tight linear relation between the tested FR and the quantity $x$. The demand for monotonic and linear -- i.e. with constant sensitivity -- analytical functions describing the behavior of a parameter with an FR justifies the use of Pearson's correlation coefficient as test measure.

Looking for the strongest correlations for each of the parameters $T_\mathrm{eff}$, [Fe/H], and $\log{g}$, we tested all possible wavelength combinations in a window of width 17~{\AA} -- or 1000 pixels in the grid -- around each dispersion point in the training set and ranked them by decreasing absolute value of the correlation coefficient. The width of 1000 pixels was chosen so that the maximum spacing between two ranges was 8.5~{\AA}, thus ensuring the aforementioned condition of a close-to-constant continuum. In a subsequent step we excluded those FRs containing a wavelength interval that is overlapping with one of the other FRs of higher $\left|r_{\mathrm{FR},x}\right|$. In doing so we made sure that individually measured ratios are independent of each other, thus enabling linear combinations of observables. In the following, we comment on the individual stellar parameters and the relations derived from them.

\subsection{Effective temperature}\label{Subsec:Effective temperature}
  \begin{figure*}
    \centering
    \resizebox{\hsize}{!}{\includegraphics{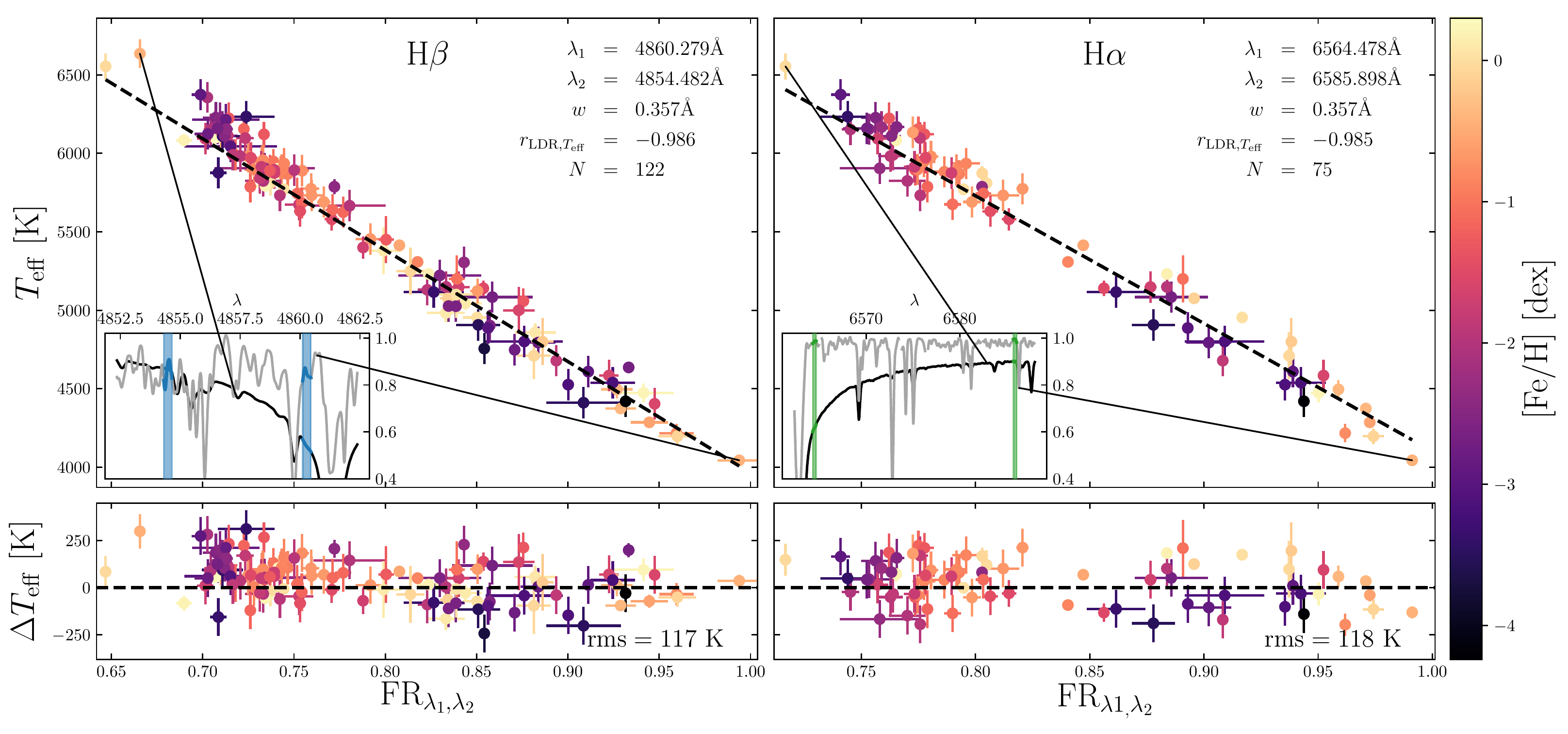}}
      \caption{Upper panels: Exemplary scatter plots for the FR-$T_\mathrm{eff}$ relations of the two strongest correlations $\left|r_{\mathrm{FR},T_\mathrm{eff}}\right|$ in our test grid around H$\beta$ (left) and H$\alpha$ (right). The color coding is the same as in Fig. \ref{Fig:grid_parameters}. Black-dashed lines resemble the linear ODR fits to the data. The small inlays at the bottom show cut-out spectra normalized by their 99th percentile for the highest (black) and lowest (gray) corresponding $T_\mathrm{eff}$, the former being offset by $-0.1$ in flux direction. Gaps in the spectra mark the expected positions of strong telluric contamination for the observed radial velocity (Sect. \ref{Subsec: Tellurics}). The blue and green shaded regions in the inlays indicate the ranges from which the respective FRs were computed using Eqs. \ref{Eq:FR} and \ref{Eq:medflux} with the values provided in the upper corner of each panel. Lower panels: Residual distribution.
              }
      \label{Fig:ldr_teff_demo}
  \end{figure*}
  According to the method outlined above, the best spectral regions to derive $T_\mathrm{eff}$ irrespective of the other stellar classifiers appear to be the Balmer lines of neutral hydrogen, H$\alpha$ and H$\beta$. This does not come by surprise, as in FGK stars -- that is at temperatures below 8000~K -- the wings of the Balmer lines are rather pressure insensitive \citep[see][and references therein]{Amarsi18} and have previously been fit to accurately constrain $T_\mathrm{eff}$ \citep[e.g.,][]{Barklem02}. Indeed, only in the wings of H$\beta$ and H$\alpha$, $\left|r_{\mathrm{FR},T_\mathrm{eff}}\right|$ reaches values $>0.97$. Because these wings are potentially very wide, the general method was altered to allow for a maximum dispersion spacing of 2000 pixels instead of 500 (corresponding to 34~{\AA} instead of 8.5~{\AA}). Hence, the FRs are in principle more prone to continuum variations, which seem to play a subordinate role, because we could not identify significant differences in the FRs among training spectra of the same targets from different spectrographs and therefore blaze functions.
  
  In order to describe the tightest relations analytically, we first deduced the individual linear trends. For this task we fitted the function
  \begin{equation}\label{Eq:Teff relation}
   T_\mathrm{eff}(\mathrm{FR}) = a \mathrm{FR} + b
  \end{equation}
  by performing an orthogonal distance regression (ODR), that is minimizing the sum of the squared normalized orthogonal distances
  \begin{equation}
   F = \frac{1}{2}\sum_{i=1}^{N} \frac{(a^2 \sigma_{T_\mathrm{eff},i}^2 + \sigma_{\mathrm{FR}_i}^2)(T_\mathrm{eff}(\mathrm{FR}_{i}) - T_{\mathrm{eff,ref},i})^2}{((a^2 + 1) \sigma_{\mathrm{FR}_i} \sigma_{T_\mathrm{eff,ref},i})^2}
  \end{equation}
  of the points $(\mathrm{FR}_{i},T_{\mathrm{eff},i})$ to the function. Here $a$ and $b$ denote the slope and intercept to be fit, $N$ is the number of FRs measured for the particular relation, whereas $\sigma_{T_\mathrm{eff,ref},i}$ and $\sigma_{\mathrm{FR}_i}$ represent the standard errors of the grid temperatures $T_{\mathrm{eff,ref},i}$ and measured flux ratios $\mathrm{FR}_{i}$, respectively. In a subsequent step the $T_\mathrm{eff}$ fit residuals were checked in a visual inspection for trends with metallicity and/or surface gravity. Fits showing (anti-) correlations in either of the two were rejected. In addition, we omitted sets of FRs where the minimum FR does not deviate by more than 15\% from the maximum FR. In doing so we reduced the influence of S/N, because larger spreads of possibly measurable FRs across the temperature range for one relation imply less sensitivity to S/N for that particular relation. 
  
  The aforementioned cleaning procedures left us with \numberofTeffrel FR-$T_\mathrm{eff}$ relations. The strongest for either of the two profiles H$\beta$ and H$\alpha$ is presented in Fig. \ref{Fig:ldr_teff_demo}. There, we show how $T_\mathrm{eff}$ behaves with the FRs measured using a bin width $w = 0.357$ {\AA} at $(\lambda_1,\lambda_2) = ($4860.279~\AA, 4854.482~\AA) and $(\lambda_1,\lambda_2) = ($6564.478~\AA, 6585.898~\AA), respectively. These wavelengths correspond to the blue wing of H$\beta$ and the red wing of H$\alpha$. We explored various realizations of $w$ and found a bin width of 0.357 {\AA}, that is 21 pixels in the training grid's dispersion direction, to be the best trade-off value between noise and resolution dependency on the one hand and information content on the other hand. A visual inspection of Fig. \ref{Fig:ldr_teff_demo} confirms the strong temperature trend with FR as already indicated by $r_{\mathrm{FR},T_\mathrm{eff}}$. The fit results along with their respective temperature residuals are indicated in the same figure. For the two extreme values in terms of $T_\mathrm{eff}$, for each of the two Balmer line regions, we show how the profile shapes and consequently the flux ratios of the wavelength regions of interest differ. We point out that $N$ is not the same in both panels and does not resemble the total number of stars in the training set but the number of spectra free of telluric absorption in the regions of interest. If a star has more than one available spectrum satisfying this condition, we averaged the deduced FRs to a single FR for that star. This ensures that intrinsically identical spectra are not over-represented in the fit. 
  
  \begin{figure*}
    \centering
    \resizebox{\hsize}{!}{\includegraphics{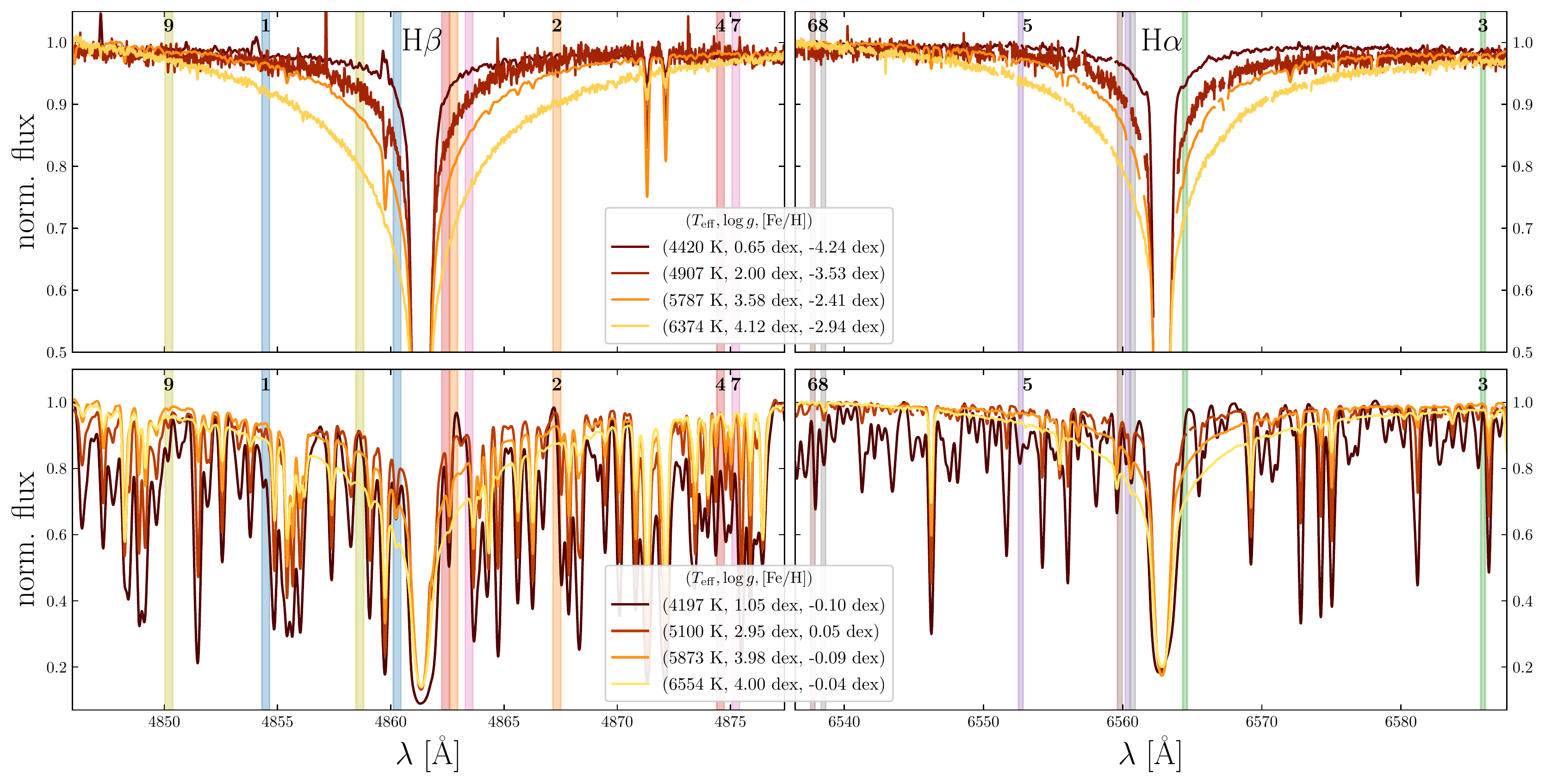}}
      \caption{Comparison of the shape changes with $T_\mathrm{eff}$ of the H$\beta$ (left) and H$\alpha$ (right) profiles at low (top) and high (bottom) metallicities. As in Fig. \ref{Fig:ldr_teff_demo}, the spectra were normalized by the 99th percentile of the flux in the shown wavelength intervals. Gaps in the spectra indicate telluric lines. Blue and green colored ranges denote the same FRs as in the inlays of Fig. \ref{Fig:ldr_teff_demo}, while the additional colored regions show the remaining seven relations ranked by the number at the top of each panel.
              }
      \label{Fig:best_spectral_range_demo}
  \end{figure*}
  We found that optimal solutions converge toward sets of two ranges obeying the following necessary conditions: At low metallicities and/or high temperatures, the ratio should incorporate one region with low and one region with high flux variance with temperature. In this case the first acts as pseudo-continuum while the latter carries the temperature  information. At high metallicities and/or low temperatures, the line depths at both wavelengths, $\lambda_1$ and $\lambda_2$, should be equally sensitive to metallicity (and, less importantly, gravity), hence assuring a constant FR at a given $T_\mathrm{eff}$. This behavior is similar to the one employed by the classical LDR approach of \citet{Kovtyukh03}.
  
  The above statements are bolstered by Fig. \ref{Fig:best_spectral_range_demo}, where we demonstrate how the profile shape changes with $T_\mathrm{eff}$ at low ($<-2.0$~dex) and high ($>-0.1$~dex) metallicities. We picked four representatives for each of the two [Fe/H] bins from the training set. While being clearly identifiable in the low-metallicity regime, the FR-$T_\mathrm{eff}$ trend is less obvious at high metallicities in combination with low temperatures. This is due to the features of species other than H dominating the wings of the Balmer lines, meaning that both components of the FR vary and have to be taken into account simultaneously. The latter observation is more pronounced in the H$\beta$ feature, because it lies in the blue part of the optical spectrum, where metal absorption is much more frequent. 
  
  Typically, the rms scatter of $T_\mathrm{eff}$ around an individual relation is of order $110$~K to $130$~K. This is close to the median uncertainty of the training values of 100~K. 
  \begin{figure}
    \centering
    \resizebox{\hsize}{!}{\includegraphics{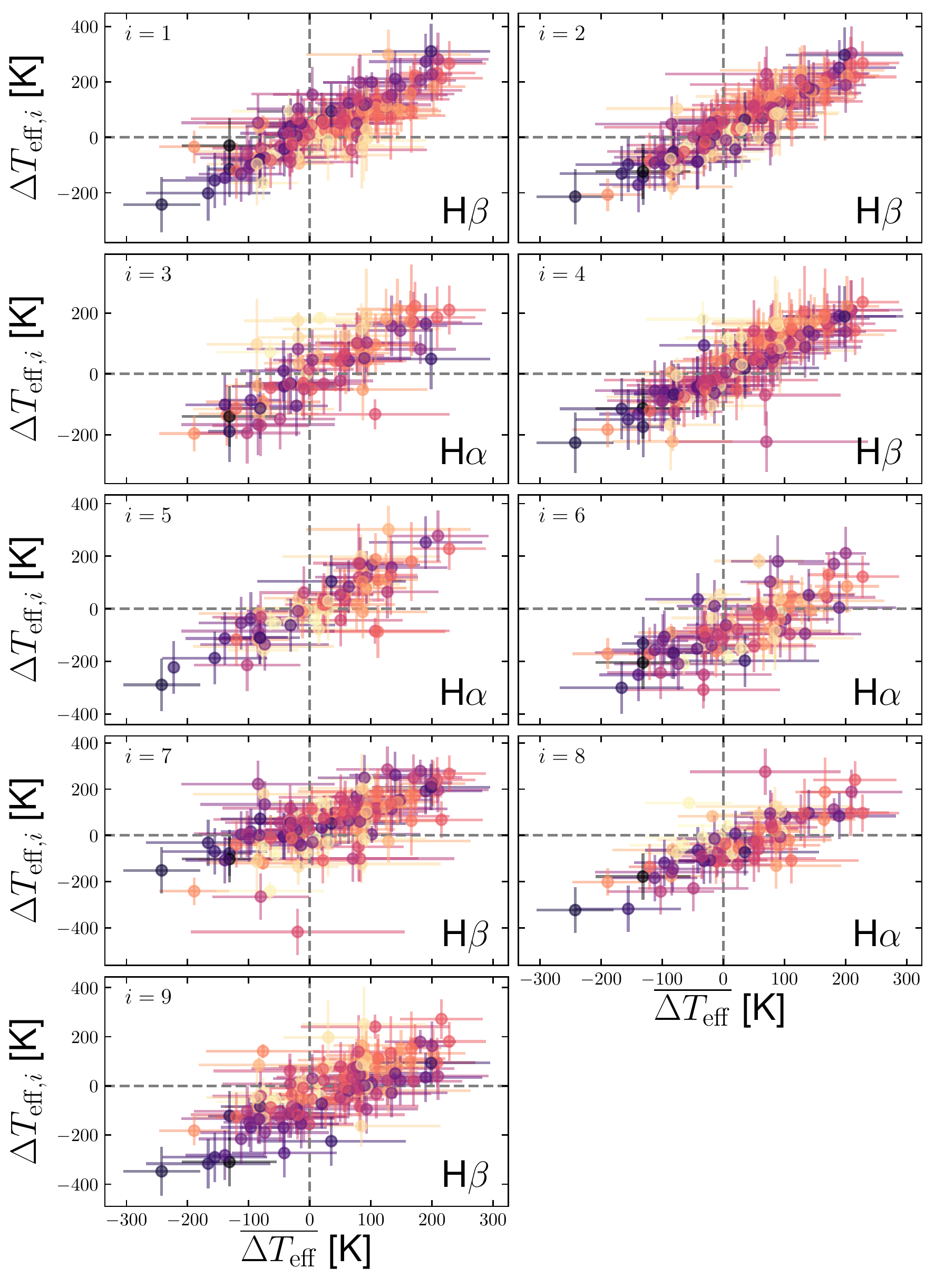}}
      \caption{Residuals of the inferred and the literature temperatures for the \numberofTeffrel strongest FR-$T_\mathrm{eff}$ relations versus the mean residuals. The color coding is the same as in Fig. \ref{Fig:grid_parameters}. Spectra with telluric contamination in one of the regions were omitted. Errors on the abscissa were computed via the standard deviation of the $\Delta T_{\mathrm{eff},i}$ (see Eq. \ref{Eq: statistical Teff error}), while the ordinate errors only denote the claimed uncertainty in the literature data.
              }
      \label{Fig:Teff_residuals}
  \end{figure}
  
  It is not straight forward to linearly combine results from the \numberofTeffrel individual measurements for each star to reduce $T_\mathrm{eff}$ uncertainties. As it turned out, the fit residuals are correlated as we show in Fig. \ref{Fig:Teff_residuals}. Ideally, this plot should consist of ellipses that are aligned with the coordinate axes, implying uncorrelated errors. Yet, the diagonal orientation of the ellipses indicates that the residuals are not purely noise-induced, but of systematic origin. 
  
  We discuss two possible reasons for this behavior, the first being the existence of hidden parameters. One or more additional parameters could affect the profile shape of Balmer lines and consequently lead to a correlation of the individual FRs and thus of the residuals of the linear FR-$T_\mathrm{eff}$ trends. Such parameters can either arise in the observations themselves or be of stellar origin. An observational bias could be introduced by the continuum shape (i.e. the blaze function of the spectrograph) in the spectral order where the profiles appear. We tested this possibility on a normalized version of the training grid and found no significant improvement as compared to the unnormalized case. Moreover, due to their large wavelength spacing, H$\alpha$ and H$\beta$ are commonly not located in the same spectral order and hence not subject to the same part of the blaze function. Yet, there is a non-negligible correlation of fit results from H$\alpha$ with the ones from H$\beta$. 
   
  As far as stellar parameters are concerned, [Fe/H] and $\log{g}$ can be ruled out to be responsible since correlations of the temperature residuals with these two were explicitly omitted. Moreover, we could not find any trends with the microturbulent velocity. For the GBS sample, there are $v\sin{i}$ measurements available (ranging from 0 to 13 km s$^{-1}$), which enable us to eliminate rotation as driving mechanism for the deviations, too. The last parameter is chemical peculiarity. Depending on the chemical enrichment history of the star, elements other than iron do not necessarily have to scale with metallicity (here [Fe/H]). Most importantly, $\alpha$-elements such as Mg are strong electron donors, so that their over- or under abundance can have effects on the electron pressure in the stellar atmosphere and accordingly the line formation. Using tabulated abundances for the GBS sample \citep{Jofre15} we could not find any trends of the $T_\mathrm{eff}$ residuals with abundances of any of the available chemical species. 
   
  The second plausible origin for the described systematics is the influence of inaccurate training values. So far, we have assumed that the training $T_\mathrm{eff}$ are of utmost accuracy, in other words the true individual temperatures should not deviate significantly from the training values when taking into account their uncertainties. If we dropped this hypothesis, correlations of the fit residuals to assumed uncorrelated FRs would indicate that either the error estimates in the sample temperatures are underestimated or that the procedures adopted to derive temperatures produce inaccurate temperatures. Considering the established accuracy of bolometric flux calibrations, which were used to determine $T_\mathrm{eff}$ for the GBS sample, this option seems unlikely. The spectroscopic determinations of $T_\mathrm{eff}$ for the remainder of the training set, however, might be subject to, for example, NLTE-, 3D-effects, or inaccurate atomic data. These can cause the true temperature of a star to expose non-zero slopes with excitation potential in an LTE treatment, which results in offset temperatures when enforcing excitation balance \citep[e.g.,][]{Hanke17}.
  
  For the above reasons we split the error budget on mean temperatures derived from our relations into a statistical and a systematic component. The latter is estimated by
  \begin{equation}\label{Eq: Teff systematic error}
   \sigma_{T_\mathrm{eff}, sys} = \sqrt{\sigma_\mathrm{tot}^2 - \sigma_\mathrm{stat}^2},
  \end{equation}
  with
  \begin{equation}
   \sigma_\mathrm{tot}^2 = \left\langle\left(T_\mathrm{eff}(\mathrm{FR}_i)_j - T_{\mathrm{eff,ref},j}\right)^2\right\rangle
  \end{equation}
  and
  \begin{equation}\label{Eq: statistical Teff error}
   \sigma_{\mathrm{stat}}^2 = \left\langle\left(T_\mathrm{eff}(\mathrm{FR}_i)_j - \left\langle T_\mathrm{eff}(\mathrm{FR}_i)\right\rangle_j\right)^2\right\rangle.
  \end{equation}
  Here, for the $j$th star in the training sample, $T_\mathrm{eff}(\mathrm{FR}_i)_j$ is the measured temperature for the $i$th relation and $T_{\mathrm{eff,ref},j}$ is the corresponding literature value. We find a value of $\sigma_{T_\mathrm{eff},sys} \approx \Teffsyserr$~K. Our fit results are summarized in Table \ref{Table:Teff fit results}.

 \subsection{Metallicity}\label{Subsec:Metallicity}
 \begin{figure*}
    \centering
    \resizebox{\hsize}{!}{\includegraphics{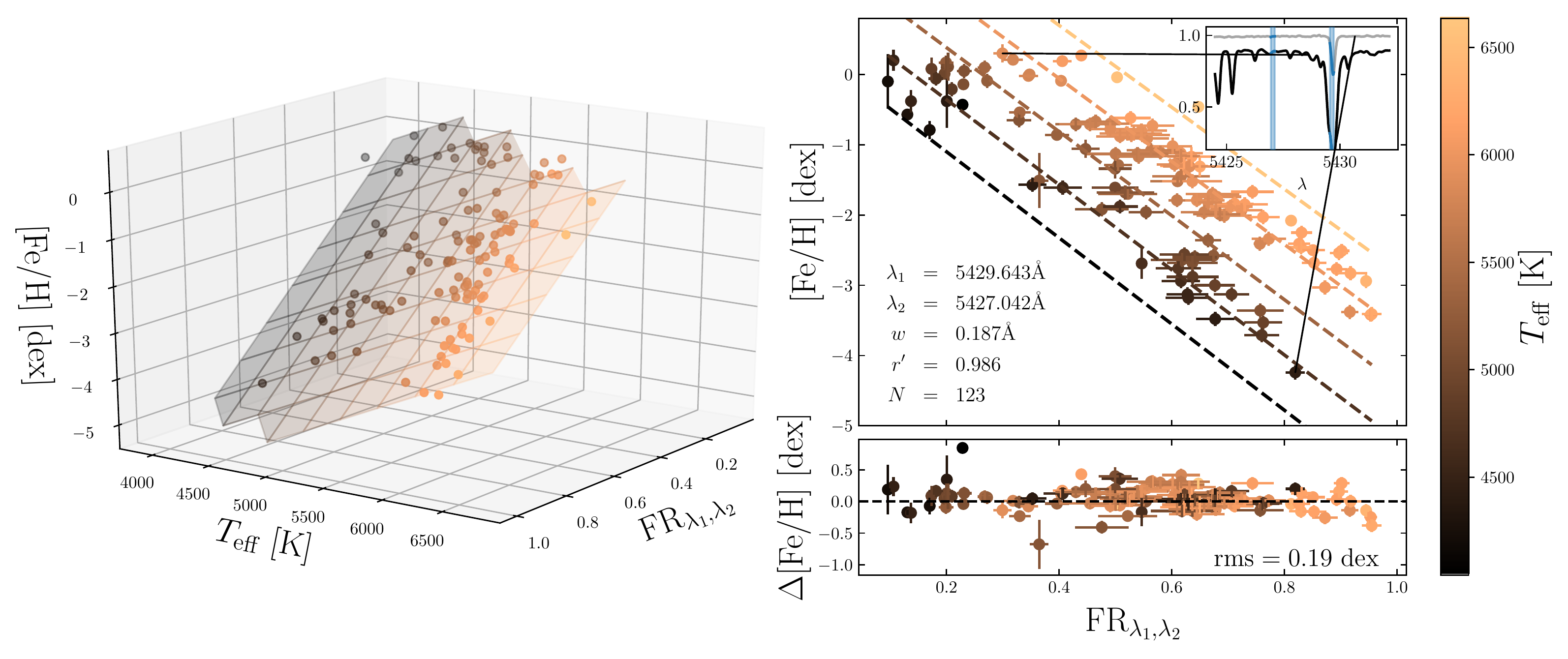}}
      \caption{[Fe/H] dependency on FR and $T_\mathrm{eff}$. The left panel illustrates the distribution of points of the tightest relation in FR--$T_\mathrm{eff}$--[Fe/H] space and the surface they span therein. To guide the eye, the best-fit surface according to Eq. \ref{Eq: [Fe/H] surface} is overlaid as a light-colored, distorted grid, where the grid lines indicate the isothermal and iso-FR lines. The same distribution of points is shown in the top right panel but in a coordinate frame that is rotated such that it is aligned with the FR and [Fe/H] axes. Dashed lines indicate the track of the isothermal lines on the surface. In analogy to Fig. \ref{Fig:ldr_teff_demo}, the inlay shows the most extreme spectra in terms of metallicity, as well as the two wavelength regions the FR was computed from (blue). The fit residuals  with respect to FR are presented in the lower right frame. All three scatter-plots follow the same $T_\mathrm{eff}$ color coding as indicated by the bar on the right.
              }
      \label{Fig:ldr_met_demo}
  \end{figure*}
 Our method revealed that, using FRs, a star's metallicity can be estimated best by investigating transitions of \ion{Fe}{i} with low-energy, lower level. The line strength (or depth) of these features is governed by the temperature and the abundance of the respective atoms in the atmospheric layers where the lines form, while $\log{g}$ and $v_\mathrm{t}$, in comparison, play a rather subordinate role. Assuming LTE, higher temperatures shift the excitation-de-excitation equilibrium toward favoring higher occupation numbers at high-energy levels and consequently lead to lower occupation in the lower levels. This results in a weakening of lines that are excited from these levels with respect to the ones at lower temperatures. The same effect would be observed at constant $T_\mathrm{eff}$ but at a lower [Fe/H], that is a lower number of atoms in the atmosphere column and thus less strong spectral features. We note that the former is a vastly simplified picture, which gets especially complicated by NLTE considerations, such as interactions between energy levels and over-ionization due to an enhanced UV-background \citep[see, e.g.,][for a detailed discussion on NLTE effects on Fe lines]{Lind12}. However, since this study concentrates on observables of real spectra and not the theoretical modeling thereof, these effects enter only to the extent that they affected the original determinations of the training values. Given that we can infer the effective temperature independently from other stellar quantities (see previous Sect.), we can break the degeneracy between line strength (by means of FRs), [Fe/H], and $T_\mathrm{eff}$ and hence determine [Fe/H].
 
 To this end, we first identified transitions that are readily described by FRs and detectable in all training sample spectra. We chose to pursue an empirically driven approach and only later consult predictions from literature atomic data in order to be unbiased and not to miss features that are susceptible to stellar parameters in our FR approach. Due to the expected degeneracy between [Fe/H] and $T_\mathrm{eff}$, the test statistic had to be modified to the multiple correlation coefficient 
 \begin{equation}\label{Eq: multiple correlation coefficient}
  r' = \sqrt{\frac{r_{\mathrm{FR,[Fe/H]}}^2 + r_{T_\mathrm{eff},\mathrm{[Fe/H]}}^2 - 2r_{\mathrm{FR,[Fe/H]}}r_{\mathrm{FR},T_\mathrm{eff}}r_{T_\mathrm{eff},\mathrm{[Fe/H]}}}{1 - r_{\mathrm{FR},T_\mathrm{eff}}^2}},
 \end{equation}
 where the various $r$ denote the correlation coefficients among the respective quantities as defined in Eq. \ref{Eq:corr coeff}. Here, once again, values of $r'$ close to unity indicate that [Fe/H] is strongly correlated with the two independent variables  $T_\mathrm{eff}$ and FR. If the latter is satisfied, points in FR-$T_\mathrm{eff}$-[Fe/H] space generate a two-dimensional hypersurface. As opposed to the FR-$T_\mathrm{eff}$ relations, it is not sufficient to describe these using only first-order terms of the independent variables. A better description can be achieved by allowing a second-order interaction term. We use the modified algebraic hypersurface
 \begin{equation}\label{Eq: [Fe/H] surface}
 \mathrm{[Fe/H]}\left(\mathrm{FR}, T_\mathrm{eff}\right) = \left(a \mathrm{FR} + b T_\mathrm{eff} + c \mathrm{FR}T_\mathrm{eff} + d\right) \left(1 + e^{\beta (\mathrm{FR} - \gamma)}\right)
 \end{equation}
 to describe the behavior. The exponential cut-off term with coefficients $\beta$ and $\gamma$ was introduced since for some relations [Fe/H] asymptotically drops for FRs above $\sim0.9$. This can be intuitively understood, because the line depth of a profile will inevitably go to zero as the metallicity decreases. As a consequence, any FR will approach unity. Therefore, depending on $T_\mathrm{eff}$, for some relations the metallicity sensitivity sharply decreases for [Fe/H$]\leq -2.5$~dex. Since those cases are still very sensitive indicators at higher metallicities, we decided to keep them and introduce the cut-off term. Neglecting the latter for those relations would lead to systematic overestimates of [Fe/H] by up 0.5~dex in the regime of very low metallicities ([Fe/H$]\leq-2.5$~dex). Fig. \ref{Fig:ldr_met_demo} presents the closest resemblance ($r'=0.986$) to a surface described by Eq. \ref{Eq: [Fe/H] surface} we could find. In contrast to the derivation of effective temperatures, here we are dealing with small-scale flux variations. Hence, the windows from which the mean flux levels were computed had to be decreased to 11 pixels, or 0.187~{\AA}, at the expense of stability against S/N.
 The information carrier in this best case is a \ion{Fe}{i} line at 5429.643~{\AA} with its pseudo-continuum slightly further in the blue in a region devoid of substantial absorption. The distribution in the FR-[Fe/H] plane is shown next to its three-dimensional counterpart.
 
 With in total 340 of these planes obeying $r'\geq0.95$ we found surprisingly many tight relations. Following the same approach as in the previous Sect., we lowered this number to 41 by demanding non-significant correlations of the fit residuals with $\log{g}$ and $v_\mathrm{t}$. Hence, while the total amount of the absorbed flux in the line surely depends on pressure and turbulence, we empirically deduced ratios of pairs of smaller ranges in stellar spectra that are not significantly influenced by these two quantities.
 
 Variations in the $\alpha$-element abundances are amongst the most common departures from the solar-scaled abundance distribution. Therefore, in order to not bias our method against $\alpha$-enhanced or $\alpha$-depleted stars, we had to ensure that no strong feature of the species O, Mg, Si, Ca, or Ti was present in the wavelength ranges considered for the computation of the FRs. VALD offers the extraction mode ``extract stellar'' to retrieve atomic data and estimates of the strength of transitions in a given wavelength interval with a particular set of stellar parameters. We employed this tool to cross-match the literature wavelengths from a line list for a sun-like atmosphere ($T_\mathrm{eff}=5750$~K, $\log{g}=4.50$~dex, [Fe/H$]=0.00$~dex) and a solar-metallicity giant atmosphere ($T_\mathrm{eff}=4500$~K, $\log{g}=1.00$~dex, [Fe/H$]=0.00$~dex) with our 41 values for $\lambda_1$ and $\lambda_2$. We excluded another ten pairs of FR regions with $\alpha$-element features in the vicinity ($\pm0.5$~{\AA}). To this end, the rejection threshold for unbroadened -- that is intrinsic, spectrograph-independent -- line depths in both hypothetical atmospheres was set to 0.2. Consequently, the final number of clean, metallicity-sensitive FRs is \numberofmetrel.
 
 The intra-relation rms metallicity scatter ranges from 0.16 up to 0.20 dex, while the inter-relation rms scatter for individual stars ranges from 0.01 to 0.31 dex. As fit residuals of individual hypersurfaces are correlated, in analogy to Eq. \ref{Eq: Teff systematic error}, we split the error into a statistical and a systematic part, that is $\sigma_{\mathrm{[Fe/H],sys}} = \metsyserr$~dex. The systematic error budget of \metsyserr~dex can be explained by the uncertainties in the training metallicities, only. We tabulate the wavelengths together with the ODR fit results for the \numberofmetrel most promising FRs in Table \ref{Table:met fit results}. 
 
 The species of the closest and anticipated strongest (in terms of line depth provided by VALD) theoretical transition in the Sun are provided in separate columns in Table \ref{Table:met fit results}. This information, however, has to be treated with caution, because first of all -- apart from lines contaminated by $\alpha$-elements -- we did not restrict our analysis to blend-free lines, but aimed for wavelength regions with a close to constant sensitivity to stellar parameters over a wide range of stellar parameters.  Secondly, and more importantly, even if a line was isolated in the solar spectrum, this does not necessarily mean that it will be isolated in a different star with significantly different parameters. Fortunately, the interplay between these factors is intrinsically accounted for in our approach. In fact, our strongest correlation (see inlay of Fig. \ref{Fig:ldr_met_demo}) incorporates not only the one \ion{Fe}{i} transition listed, but several other possible \ion{Fe}{i} blends on either side of the profile. We emphasize at this point that our method is prone to biases introduced by chemical peculiarity to the extent that any non-regular chemical behavior of a target star could affect the relative strength of a blend in our wavelength regions and therefore cause the corresponding FRs to deviate from the ones expected at the star's metallicity. This effect can, for example, be expected to be encountered in spectra of carbon-enhanced metal-poor (CEMP) stars, where absorption bands of carbonaceous molecules dominate the appearance of wide spectral bands.
 
 \subsection{Surface gravity}\label{Subsec:Gravity}
 \begin{figure*}
    \centering
    \resizebox{\hsize}{!}{\includegraphics{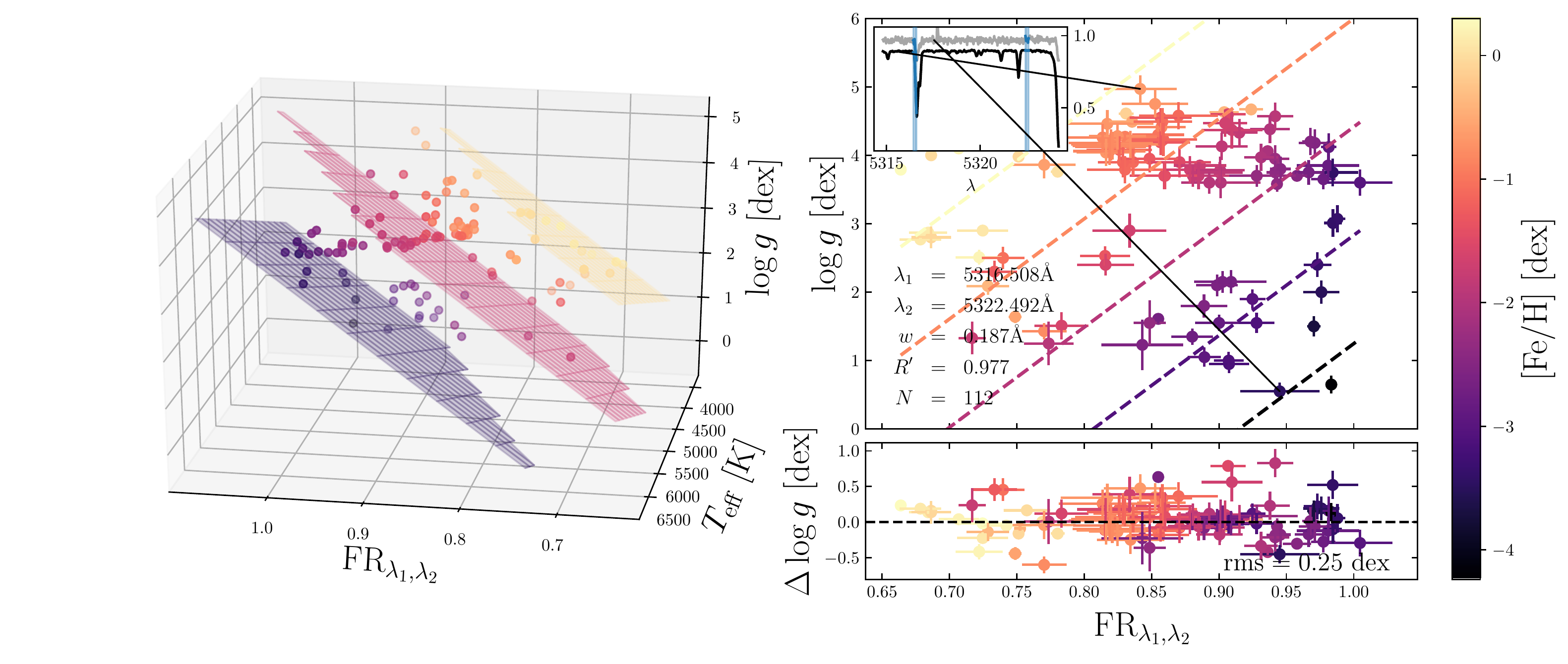}}
      \caption{$\log{g}$ dependency on FR, $T_\mathrm{eff}$, and [Fe/H]. The left panel illustrates the distribution of points of the tightest relation in FR--$T_\mathrm{eff}$--$\log{g}$ space, where the colors introduce metallicity as a fourth dimension. The fit solutions to Eq. \ref{Eq: gravity surface} at fixed [Fe/H] values of $-$3.5~dex, $-$1.75~dex, and 0.00~dex are shown as three parallel hyperplanes. The top right panel resembles the distribution in FR--$\log{g}$ space, only. There, dashed lines indicate the track of the iso-metallicity lines on the surface at a fixed temperature of $5000$~K. The inlay shows the most extreme spectra in terms of gravity, as well as the two wavelength regions the FR was computed from (blue). The fit residuals with respect to FR are presented in the lower right frame. All three scatter-plots follow the same [Fe/H] color coding as indicated by the bar on the right.
              }
      \label{Fig:ldr_logg_demo}
 \end{figure*}
 The derivation of stellar surface gravity using only dedicated FRs that are valid for the whole range of stellar parameters of our sample is less well defined. For dwarfs at moderately low to high metallicities, the wings of lines in the \ion{Mg}{i}~b triplet have proven to be good $\log{g}$ indicators \citep[e.g.,][]{Ramirez06}. However, for certain combinations of stellar parameters like low gravities and metallicities pronounced wings do not form and cannot be used as universal gravity indicators. In fact, the largest correlation for surface gravity with FRs in the grid spectra is found to be $r_{\mathrm{FR,}\log{g}}=0.89$. Unfortunately, strong $r_{\mathrm{FR,}\log{g}}$ seem to be mainly explained by the spurious correlation of $T_\mathrm{eff}$ and $\log{g}$ in the grid ($r_{T_\mathrm{eff},\log{g}} = 0.66$). Hence, it was necessary to once again increase the dimensionality of the product-moment correlation coefficient to include $T_\mathrm{eff}$ and [Fe/H] as independent variables to end up with
 \begin{equation}
  R'= \sqrt{\begin{pmatrix}r_{x_1, \log{g}}\\ r_{x_2, \log{g}}\\ r_{x_3, \log{g}}\end{pmatrix}^\mathrm{T} \begin{pmatrix} 1 & r_{x_1,x_2} & r_{x_1,x_3}\\
  r_{x_1,x_2} & 1 & r_{x_2,x_3}\\ r_{x_1,x_3} & r_{x_2,x_3} & 1\end{pmatrix}^{-1}\begin{pmatrix}r_{x_1, \log{g}}\\ r_{x_2, \log{g}}\\ r_{x_3, \log{g}}\end{pmatrix}},
 \end{equation}
 where
 \begin{equation}
  x_i \in \left\{ \mathrm{FR},T_\mathrm{eff},\mathrm{[Fe/H]} \right\}.
 \end{equation}
 33 combinations of wavelength ranges of width 0.187~{\AA} in the grid satisfy $R'\geq0.95$. Our analytical description for these 33 FRs is the hyperplane
 \begin{equation}\label{Eq: gravity surface}
  \log{g}(\mathrm{FR}) = a\mathrm{FR} + bT_\mathrm{eff} + c\mathrm{[Fe/H]} + d
 \end{equation}
 After carefully checking the residuals for small-scale structure and contamination by $\alpha$-element transitions (see previous Sect.), we ended up having 11 reliable relations. In Fig. \ref{Fig:ldr_logg_demo} we visualize the 4-dimensional plane in analogy to Fig. \ref{Fig:ldr_met_demo} for the strongest association at $R'=0.977$. In the particular case of the shown relation, the blue wavelength region includes a strong \ion{Fe}{ii} line at 5316.508~{\AA}, while the red range falls in a continuum window. Naturally, line strengths and therefore FRs involving \ion{Fe}{ii} lines have a strong [Fe/H] sensitivity. In fact, given gravity as prior, most FR combinations discussed here would offer a good metallicity indicator. Here, we use the fact that on top of their strong metallicity- and weak temperature-dependence profiles of ionized species expose a sensitivity to stellar surface gravity. Our 11 $\log{g}$ relations with the involved wavelength ranges and strongest contributing features in the Sun are listed in Table \ref{Table:logg fit results}.
 
 Fig. \ref{Fig:ldr_logg_demo} reveals the main weakness of our empirical approach to derive $\log{g}$. Ideally, an unbiased training sample would span a regularly spaced grid in parameter space. The sample used in this study, however, lacks coverage at the extreme values of almost all parameters (with the exception of high $\log{g}$ dwarfs, cf. Fig. \ref{Fig:grid_parameters}). Even in an idealized homogeneously distributed case, our sample size of $124\approx5^3$ stars would only allow limited explanatory power in three dimensions of independent variables (FR, $T_\mathrm{eff}$, and [Fe/H]). As a consequence, the low-number statistics prevents us from detecting subtle non-linearities causing systematic biases such as the ones seen in the high-FR regime in Sect. \ref{Subsec:Metallicity} at lower dimensionality (only two independent variables instead of three). A good example for potential non-linear behavior is the spectrum of the star $\beta$~Ara. Its gravity is underestimated by $\sim1$~dex in almost all relations, suggesting a breakdown of the linear relation towards $\beta$~Ara's corner in the parameter space ($T_\mathrm{eff}=4197$~K, [Fe/H$]=-0.05$~dex, $\log{g}=1.05$~dex). For the discussed reasons, one has to exercise caution when using the provided $\log{g}$ relations, especially for extreme cases corresponding to the edges of the parameter space studied here.
 
\section{ATHOS}\label{Sec: PAWN}
We implemented all the information provided in Tables \ref{Table:Teff fit results}, \ref{Table:met fit results}, and \ref{Table:logg fit results}, as well as the fit covariance matrices and systematic errors of Eqs. \ref{Eq:Teff relation}, \ref{Eq: [Fe/H] surface}, and \ref{Eq: gravity surface} in one stellar parameter tool, called ATHOS. ATHOS is a Python-based software capable of dealing with hundred thousands to millions of spectra within a short amount of time. In part, this is due to the incorporation of parallelization capabilities making use of modern multi-core CPU architectures. Using only FRs computed from input spectra and the dedicated analytical relations, the stellar parameters can be estimated in well under 30~ms. ATHOS' workflow for each spectrum is divided into four main phases:

The first step is the read phase. ATHOS operates on one-dimensional, RV-corrected input spectra of various file structures. Among these are the standard FITS spectrum formats and binary tables, as well as NumPy arrays, comma-, tab-, and whitespace-separated ASCII files. A minimal input consists of spectral fluxes and the corresponding dispersion information. At least one of the Balmer lines H$\beta$ and H$\alpha$ should fall in the covered wavelength. Otherwise, an external estimate of $T_\mathrm{eff}$ has to be given. For a proper treatment of noise, the error spectrum should be provided. Alternatively, a global S/N value for the entire spectrum will be set (not recommended). On a state-of-the-art machine with solid-state drive, the required time to read a spectrum with $2\cdot10^5$ dispersion points into memory ranges from 1~ms through 3~ms to 117~ms for NumPy arrays, FITS spectra, and ASCII files, respectively.

The second step masks out regions of potential telluric contamination. The correction is of utmost importance for the temperature determination from the heavily contaminated H$\alpha$ profile. For this purpose ATHOS internally stores the strongest topocentric wavelengths of our telluric model (see Sect. \ref{Subsec: Tellurics}). In order to exclude these from consideration, the velocity of the stellar rest frame with respect to the topocenter has to be provided. If H$\alpha$ is not included, or if the spectrum has been corrected for tellurics, this step can be omitted. The involved operations take about 0.5~ms per spectrum.

Next, the FRs are computed. To this end, the input spectrum is linearly interpolated between dispersion points in order to compute the mean fluxes in the interval $(\lambda - w/2, \lambda +  w/2)$. The typical execution time required here for a spectrum with $2\cdot10^5$ dispersion points is $\sim 13$~ms. In case the resolution of the input spectrum is less than 45000, the FRs are corrected using the relations discussed in Sect. \ref{Subsec:Resolution}.

The final phase is the parameter cascade. Starting from the FRs measured in the previous step, this routine computes the stellar parameters according to Eqs. \ref{Eq:Teff relation}, \ref{Eq: [Fe/H] surface}, and \ref{Eq: gravity surface} in the order $T_\mathrm{eff}$, [Fe/H], and finally $\log{g}$. As any subsequent relation depends on the result of the former, this structure is mandatory. By default, each final parameter and its (internal) error are estimated as the median and the median absolute deviation (mad) of the distribution of results from the different relations. These two measures were chosen, because they are more robust against outliers possibly arising from spectral artifacts compared to, for example, means or weighted means. In some situations, such as highly reddened stars, the S/N may strongly vary on larger scales from the blue to the red. To take this into account, we implemented the possibility of introducing wavelength-dependent weights for the individual FR-relations in the form of polynomials with user-defined degrees and coefficients. In these cases, the weighted median is used to compute the final set of parameters. Systematic errors are propagated throughout the whole cascade. The time demand of this computation step is $\sim0.6$~ms.

\section{Performance tests}\label{Sec: Discussion}
\subsection{Resolution dependencies}\label{Subsec:Resolution}
 All the FR-parameter relations implemented in ATHOS have been trained on a spectral grid at a fixed resolution of $R=45000$. In order to be applicable to a variety of spectroscopic surveys conducted with different instruments and resolutions, these relations have to be either insensitive to $R$, or exhibit predictable deviations with $R$. For this reason, we investigated $R$ dependencies by degrading our spectral grid to resolutions of $40000, 30000, 20000, 10000, 7500, 5000$, and $2500$ followed by a re-computation of all FRs which were established at $R=45000$. 
 
 \begin{figure*}
    \centering
    \resizebox{\hsize}{!}{\includegraphics{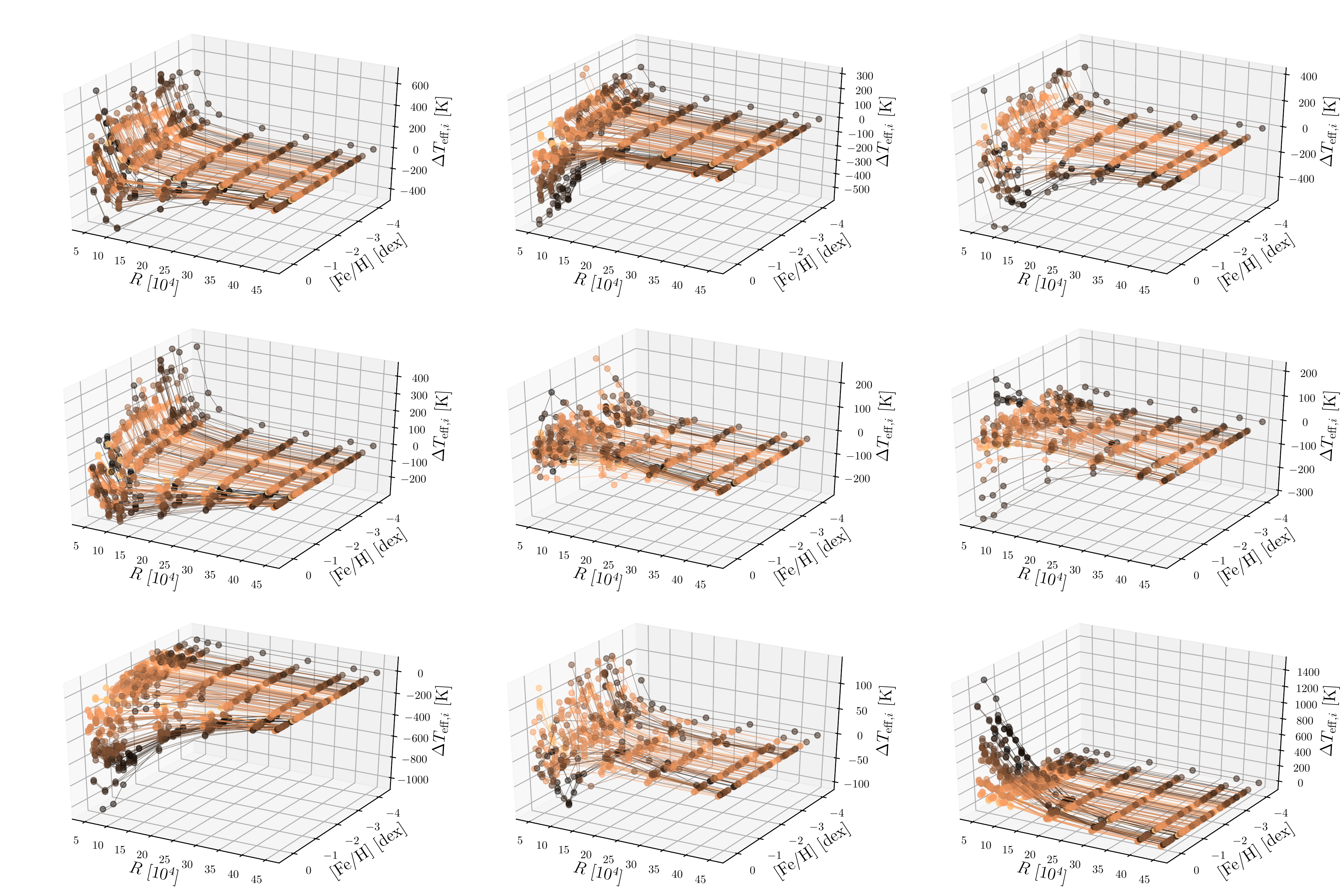}}
      \caption{$T_\mathrm{eff}$ deviations with respect to resolution and [Fe/H] ordered from the strongest to the weakest relation (see Table \ref{Table:Teff fit results} and Fig. \ref{Fig:best_spectral_range_demo}) from top left to bottom right. The deviations were computed by scaling $\Delta \mathrm{FR}$ with the slope of the respective relation. Points measured for the same star but at different $R$s are connected by solid lines. The color coding is the same $T_\mathrm{eff}$ scale as in Fig. \ref{Fig:ldr_met_demo}. 
              }
      \label{Fig:Teff_Rdependence}
 \end{figure*}
 Fig. \ref{Fig:Teff_Rdependence} shows how the newly computed FRs for our \numberofTeffrel $T_\mathrm{eff}$ indicators deviate from the originally derived ones. Since there is a simple linear relation between FR and $T_\mathrm{eff}$, the resulting temperature offset can be deduced by scaling the $\Delta \mathrm{FR}_i$ with the respective slope $a_i$. We found that down to $R\approx20000$ there is generally no significant deviation in FR, meaning the temperature deviations remain well below 150~K. For even lower values of $R$ the situation becomes non-trivial. Then, the absolute value and sign of the offset appears to not only depend on $R$, but on $T_\mathrm{eff}$ and [Fe/H], as well. For metallicities of $-1$~dex and below, we found significant deviations only as ``late'' as $R\lesssim10000$. At these low metallicities, the profile shapes of the Balmer lines are dominated by \ion{H}{i} itself and consequently expose larger scale lengths than just the line spread function of the spectrograph. Hence, convolving with Gaussians of FWHMs shorter than the range width of $w=0.357$~{\AA} -- the mathematical equivalent of employing a spectrograph of lower resolution -- does not disperse large amounts of flux out of the wavelength ranges from which the FRs are measured. Accordingly, the FRs remain largely unaffected. For higher metallicities than $-1$~dex, a non-linear offset trend of the FRs with decreasing resolution and temperature manifests itself already for $R\lesssim20000$. We suspect the driving mechanism for this behavior to be metal absorption from neighboring wavelengths being dispersed into the ranges of interest, or, vice versa, being dispersed out of the ranges. The strength of this additional (or lack of) absorption is mainly determined by $T_\mathrm{eff}$, $\log{g}$, and -- more importantly -- [Fe/H]. Consequently, the $T_\mathrm{eff}$ relations are not clean anymore, but get susceptible to pressure and metallicity. Depending on the degree of isolation of the FR ranges from neighboring metal lines, this effect is more or less severe, which explains the large variety of absolute temperature deviations between relations in Fig. \ref{Fig:Teff_Rdependence}.
 
 \begin{figure*}
    \centering
    \resizebox{\hsize}{!}{\includegraphics{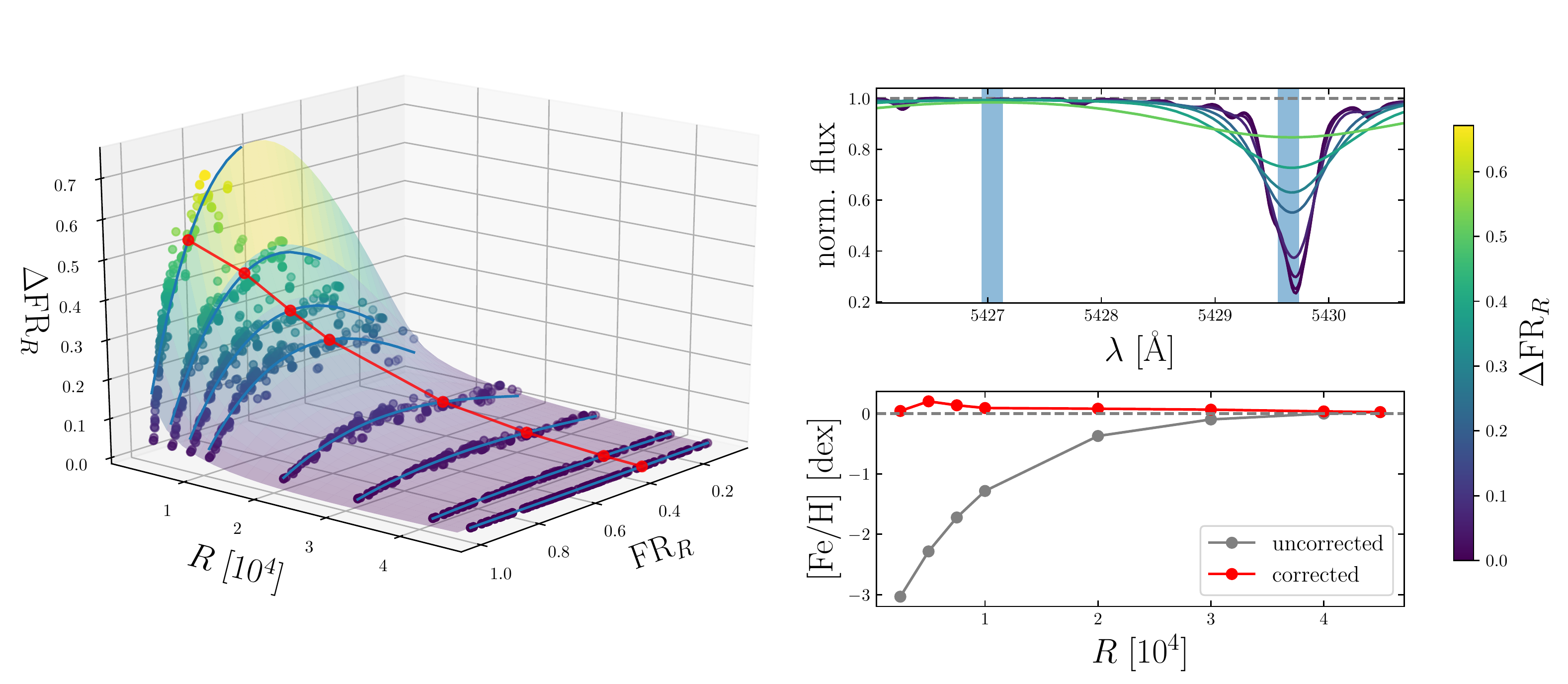}}
      \caption{Left panel: Deviations of the FRs measured in all training spectra for the strongest metallicity relation (see Table \ref{Table:met fit results}) at different resolutions with respect to the FRs computed at the original resolution of $45000$. In contrast to Fig. \ref{Fig:Teff_Rdependence}, we note that there is a non-linear relation between FR and [Fe/H], which is why we illustrate only the FR- and not the [Fe/H] deviations. The colored surface resembles the best-fit model to the data according to Eq. \ref{Eq: met_Rdependence}, where the color indicates the magnitude of the correction to be applied (see color bar on the right). Red points and their connecting line denote the track of the corrections to the FRs measured in the solar spectrum at different resolutions. Upper right panel: Portion of the solar spectrum at resolutions of 45000, 40000, 30000, 20000, 10000, 7500, 5000, and 2500 (dark- to light-colored). The color-coding refers to the same FR corrections as in the left panel. Blue vertical bars indicate the wavelength ranges from which the FR is computed. Lower right panel: Derived [Fe/H] for the Sun from uncorrected (gray) and corrected (red) FRs using Eq. \ref{Eq: [Fe/H] surface} and only the strongest metallicity relation shown in the upper right panel.  
              }
      \label{Fig:met_Rdependence}
 \end{figure*}
 The same reasoning holds for the $R$ dependence of the [Fe/H] relations, with the only difference being that optimal solutions usually consist of only one information carrier and a normalization component. In the vast majority of cases, resolution deviations from the training value only affect the information carrier, since the regions of the pseudo-continuum tend to be free of absorption, or at least have a negligible absorption component with respect to the information carrier. This behavior can be seen in the upper right panel of Fig. \ref{Fig:met_Rdependence}, where we show how resolutions between $45000$ and $2500$ affect the profiles in the solar spectrum in our strongest metallicity relation. Moreover, the FR deviations for all benchmark stars at their different metallicities and temperatures are presented in the left panel. It is obvious that higher metallicities, that is lower FRs, correspond to higher deviations in FR once the resolution is decreased. In general, low metallicities cause the line depths of the information carrier to be low and thus the FRs to reside close to unity. If the flux gets dispersed out of these profiles due to broader line spread functions, the FR is only marginally influenced. At higher metallicities, in turn, the FRs tend to be lower than unity and therefore the $R$-induced deviation can be much larger. Empirically, we found that all corrections can be well described by the relation
 \begin{equation}\label{Eq: met_Rdependence}
  \Delta \mathrm{FR}_R = \sum_{i=1}^3 \left(\sum_{j=1}^2 p_{ij} \ln{\left(\frac{R}{45000}\right)^j} \right) \mathrm{FR}_R ^{i}
 \end{equation}
 within an acceptable margin of error. Here, $\mathrm{FR}_R$ is the FR measured at the resolution $R$ and $p_{ij}$ is a coefficient matrix with nine entries. The best-fit surface that is spanned by Eq. \ref{Eq: met_Rdependence} for our strongest metallicity relation is illustrated in the left panel of Fig. \ref{Fig:met_Rdependence}. In the lower right panel of the same Fig. we show the growing importance of the FR corrections with decreasing resolution by comparing the metallicity from the uncorrected FR with the one derived using the corrected FR. Similar to [Fe/H], our $\log{g}$ surfaces are based on small-scale flux variations. Hence, we see comparable trends of the corresponding FR residuals with $R$ to the one shown in Fig. \ref{Fig:met_Rdependence}. The coefficient matrices $p_{ij}$ in Eq. \ref{Eq: met_Rdependence} for all analytical relations outlined in this work are tabulated in Table \ref{Table: R_correction_coefficients}. 
 
 \begin{figure*}
    \centering
    \resizebox{\hsize}{!}{\includegraphics{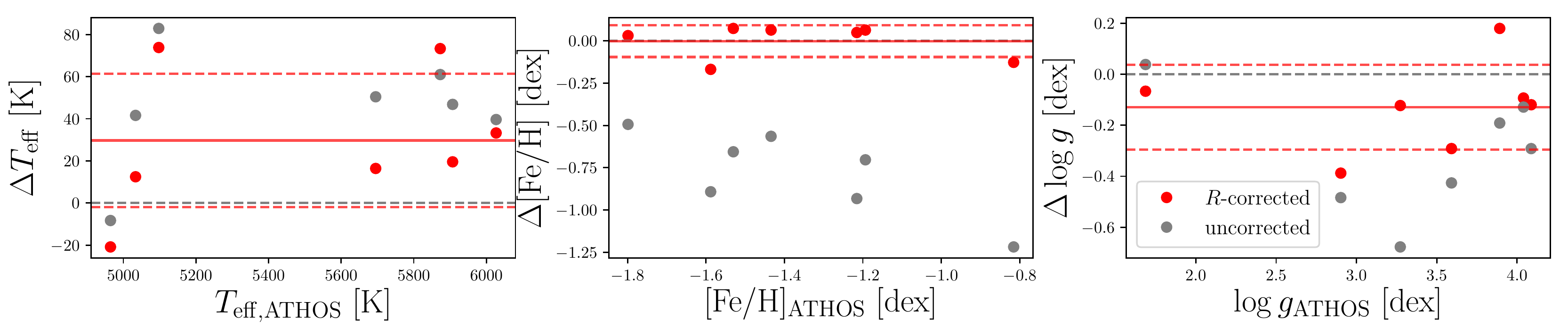}}
      \caption{Parameter deviations between ATHOS runs on X-shooter spectra and on high-resolution spectra of the same stars from our training sample. Here, gray circles indicate the values without any correction applied to the involved FRs, while red circles incorporate the corrections described by Eq. \ref{Eq: met_Rdependence}. The mean deviation and scatter for each stellar parameter are depicted by solid and dashed red lines, respectively.}
      \label{Fig:Xshooter_resolution_correction}
 \end{figure*}
 Despite the fact that Eq. \ref{Eq: met_Rdependence} seems to hold down to the lowest resolutions, there are additional circumstances to be considered. First of all, a decrease in resolution shrinks the range of possible manifestations of FRs, which increases the noise sensitivity of Eqs. \ref{Eq:Teff relation}, \ref{Eq: [Fe/H] surface}, and \ref{Eq: gravity surface}. Secondly, and more importantly, once the dispersion sampling of a real spectrum reaches the order of the width of the wavelength range we aim to measure -- e.g., $w=0.187$~{\AA} for the [Fe/H] and $\log{g}$ relations -- the mean of the fluxes within that width might correspond to only a fraction of the flux in one pixel. If we considered, for example, the $0.7\arcsec$ slit configuration of the X-shooter spectrograph at the VLT with its moderate resolution of $\approx10000$ and a sampling of $\sim5$~pixels~FWHM$^{-1}$, we would end up with only 1.87~pixels~$w^{-1}$ at a wavelength of $\lambda=5000$~{\AA}. 
 
 In testing ATHOS' sensitivity to spectral resolution under the aforementioned conditions, we retrieved and analyzed real spectra from the X-shooter spectral library \citep{Chen14} of the seven stars in common with our training sample. The results are presented in Fig. \ref{Fig:Xshooter_resolution_correction}. There, we show the departures of the parameter values at moderate resolution (10000) from the parameters obtained at high-resolution (45000) for corrected and uncorrected FRs, respectively. From the diagnostic plot for $T_\mathrm{eff}$ and the mean deviation and scatter of $29\pm32$~K in the $R$-corrected case we conclude that our temperature relations are very stable against $R$ -- at least in the parameter space covered by the seven stars discussed here. This holds true even for the computed temperatures without any applied correction, where we found a marginally significant deviation of $45\pm26$~K (rms). The importance of introducing resolution corrections becomes apparent when looking at the [Fe/H] findings. By not accounting for resolution effects, ATHOS would underestimate the metallicity from the X-shooter spectra by on average $-0.78\pm0.23$~dex, while the deviation vanishes for corrected FRs ($0.00\pm0.09$~dex). Similarly, an uncorrected mean deviation of $-0.31\pm0.22$~dex in the gravity results can be alleviated to $-0.13\pm0.17$~dex.
 
 \subsection{Spectrum noise and systematics}\label{Subsec:Noise}
 In order to test the stability of ATHOS against S/N we ran a series of Monte Carlo (MC) simulations by adding Poisson noise to high-S/N spectra of the Sun and $\alpha$~Boo as representatives of well studied and accurately parametrized dwarf and giant stars. The resulting spectra were fed to ATHOS to compute stellar parameters. This procedure was repeated $10^3$ times for S/N values of 100, 50, and 20. 
 \begin{figure}
    \centering
    \resizebox{\hsize}{!}{\includegraphics{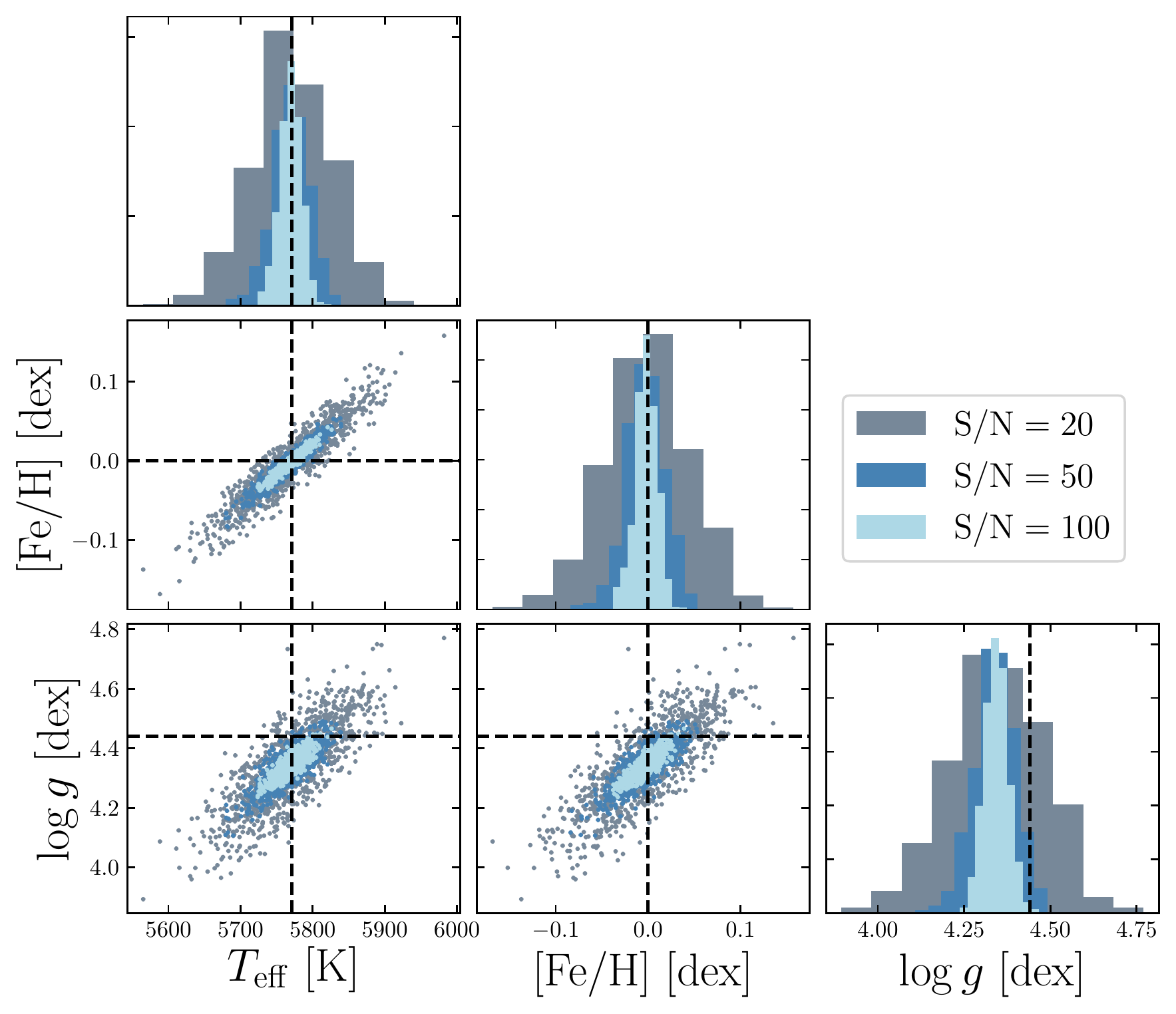}}
      \caption{ATHOS outputs from an MC-simulation on the solar spectrum. Noise was artificially induced such that S/N values of 100 (gray), 50 (dark blue), and 20 (light blue) with $10^3$ random initializations each were realized. The literature parameters \citep[5771~K, 0.00~dex, 4.44~dex; ][]{Heiter15} are indicated by black dashed lines.
              }
      \label{Fig:Sun_SNR_dep}
 \end{figure}
 Figures \ref{Fig:Sun_SNR_dep} and \ref{Fig:Arcturus_SNR_dep} show the distributions of the output. It is evident, for $\mathrm{S/N}=100$ and above, that ATHOS errors are not governed by random-noise effects in the input spectra, which can easily reach down to $\sigma_{T_\mathrm{eff}}=50$~K, $\sigma_{\mathrm{[Fe/H]}}=0.05$~dex, and $\sigma_{\log{g}}=0.05$~dex. This has been discussed already in Sects. \ref{Subsec:Effective temperature}, \ref{Subsec:Metallicity}, and \ref{Subsec:Gravity}. The systematic errors provided there ($\sigma_{T_\mathrm{eff},\mathrm{sys}}=\Teffsyserr$~K, $\sigma_{\mathrm{[Fe/H], sys}}=\metsyserr$~dex, $\sigma_{\log{g},\mathrm{sys}}=\loggsyserr$~dex), readily account for the deviations of non-stochastic origin found in the MC simulations. 
 
 Another implication of Figs. \ref{Fig:Sun_SNR_dep} and \ref{Fig:Arcturus_SNR_dep} is the strong inter-dependency among the stellar parameters visualized by the inclination of the distribution ellipses. This does not come by surprise as the parameter cascade (phase 4) within ATHOS requires the output of all previous steps as input. The only truly independent quantity is $T_\mathrm{eff}$, because we made sure that the respective relations are free of significant [Fe/H] or $\log{g}$ trends. Nevertheless, any deviation from the true value of $T_\mathrm{eff}$ enters the [Fe/H] relation Eq. \ref{Eq: [Fe/H] surface} as a constant and a (mild) additional slope with FR. Likewise, $T_\mathrm{eff}$ and [Fe/H] offsets propagate linearly onto the $\log{g}$ surfaces described by Eq. \ref{Eq: gravity surface}. We emphasize that this kind of behavior is also immanent to the established methods of using EWs or spectrum fitting to determine stellar parameters \citep[see, e.g., the detailed discussion in the appendix of][]{McWilliam95}. The inter-dependencies of the stellar parameters, however, tend to be neglected in most published studies; a practice that only started to change more recently \citep[see, e.g.,][]{GarciaPerez16}. As our approach is capable of predicting temperature irrespective of the other two quantities, we can at least break the degeneracy with $T_\mathrm{eff}$ similar to studies employing spectrum fitting of Balmer lines \citep[see, e.g.,][]{Barklem02}. 
 \begin{figure}
    \centering
    \resizebox{\hsize}{!}{\includegraphics{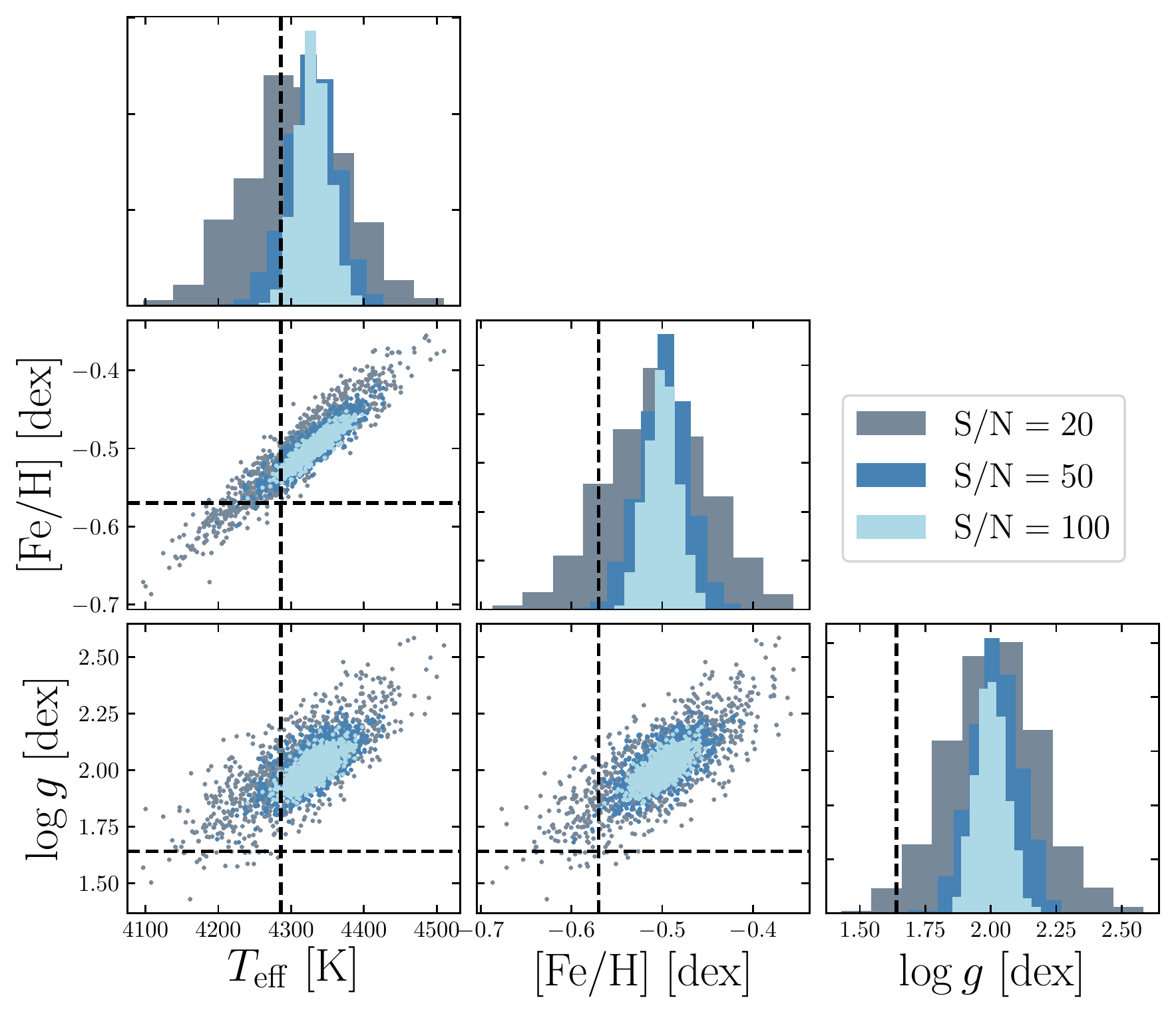}}
      \caption{Same as Fig. \ref{Fig:Sun_SNR_dep}, but for $\alpha$~Boo (4286~K, $-0.57$~dex, 1.64~dex).
              }
      \label{Fig:Arcturus_SNR_dep}
 \end{figure}

\subsection{Comparison to spectroscopic surveys: ELODIE 3.1}\label{Subsec: comparison surveys}
 \begin{figure*}
    \centering
    \resizebox{\hsize}{!}{\includegraphics{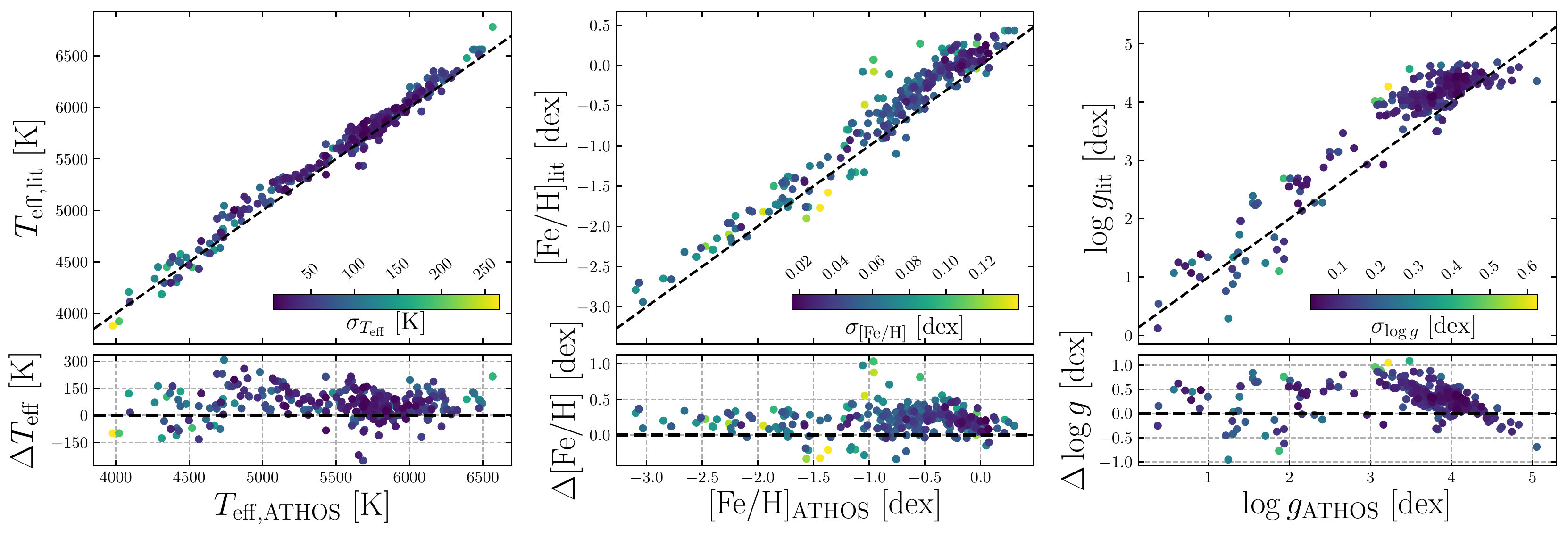}}
      \caption{Upper panels: Comparison of the ATHOS output for $T_\mathrm{eff}$, [Fe/H], and $\log{g}$ with literature results for the ELODIE spectral library \citep[version 3.1, ][see text for quality cuts made here]{Prugniel07}. In case a star occurs with multiple spectra in the library, it is reflected by more than one point here. Dashed lines represent the one-to-one relation. The colors reflect the internal statistical uncertainties computed with ATHOS. Lower panels: Residual distribution. All residuals are determined via $\Delta x = x_\mathrm{lit} - x_\mathrm{ATHOS}$.
      }
      \label{Fig:performance_ELODIE}
 \end{figure*}
 The ELODIE library \citep{Prugniel01} in its current version 3.1 \citep{Prugniel07} consists of 1959 spectra of 1389 stars obtained with the ELODIE spectrograph at the Observatoire de Haute-Provence. The spectra have a continuous coverage from $3900$ to $6800$~{\AA} at a nominal resolution of $42000$ and were released together with a catalog of stellar parameters compiled from literature data and quality flags. A restriction to the two best out of the four quality flags -- i.e. maximum uncertainties in $T_\mathrm{eff}$ and [Fe/H] of 115~K and 0.09~dex -- left us with 288 spectra of 201 stars. The median S/N of these spectra is 123 at 5550~{\AA}, while the minimum and maximum is 38 and 411, respectively. The library is provided such that the spectra are already shifted to the stellar rest-frame and tellurics were already masked, enabling an immediate analysis with ATHOS. The results of an ATHOS run on the 288 spectra and a comparison to the literature values are shown in Fig. \ref{Fig:performance_ELODIE}. 
 
 We find an excellent agreement between ATHOS values for $T_\mathrm{eff}$ and the Elodie compilation. The mean deviation and scatter over the entire temperature range is $64\pm76$~K (rms). For $4750\mathrm{~K}<T_\mathrm{eff}<5500$~K, there seems to be a systematic offset of 105~K from unity. Cross-matching our training sample with the ELODIE library resulted in five overlaps in the temperature range in question (Table \ref{Table:Teff training vs. ELODIE}). A possible source for the deviation could be the fact that our training $T_\mathrm{eff}$ are on average 71~K cooler. Two points in Fig. \ref{Fig:performance_ELODIE} around $5750$~K are clearly off from the overall trend by more than 200~K. Both correspond to the star HD~245, which ATHOS consistently finds to be $\sim240$~K warmer than the literature value of 5433~K. Comparing ELODIE spectra of the Sun with the two spectra of HD~245 shows a remarkable similarity of the Balmer profiles H$\alpha$ and H$\beta$ between the two stars. We conclude that both should have an almost identical temperature. Hence, the ATHOS result for HD~245 should be more reliable, because it is closer to the solar value ($5771$~K). A visual inspection of the Balmer profiles in some of the ELODIE spectra revealed another possible explanation for $T_\mathrm{eff}$ differences to be bad pixel artifacts. We found several unphysical spikes with heights of a few 10\% of the continuum level neighboring H$\alpha$ and H$\beta$. These are neither masked nor flagged in the ELODIE library and can possibly falsify parameter measurements.
 \begin{table}
  \caption{$T_\mathrm{eff}$ for overlapping stars between our training sample and the ELODIE library in the range $4750$~K to $5500$~K.}
  \label{Table:Teff training vs. ELODIE}
  \centering
  \begin{tabular}{l c c c}
  \hline\hline
  Name & $T_\mathrm{eff,training}$ & $T_\mathrm{eff,ELODIE}$ & $\Delta T_\mathrm{eff}$\\
   & [K] & [K] & [K]\\
  \hline                      
  \input{Teff_training_vs_ELODIE.dat}
    \hline
    \end{tabular}
 \end{table}
 
\begin{figure}
    \centering
    \resizebox{\hsize}{!}{\includegraphics{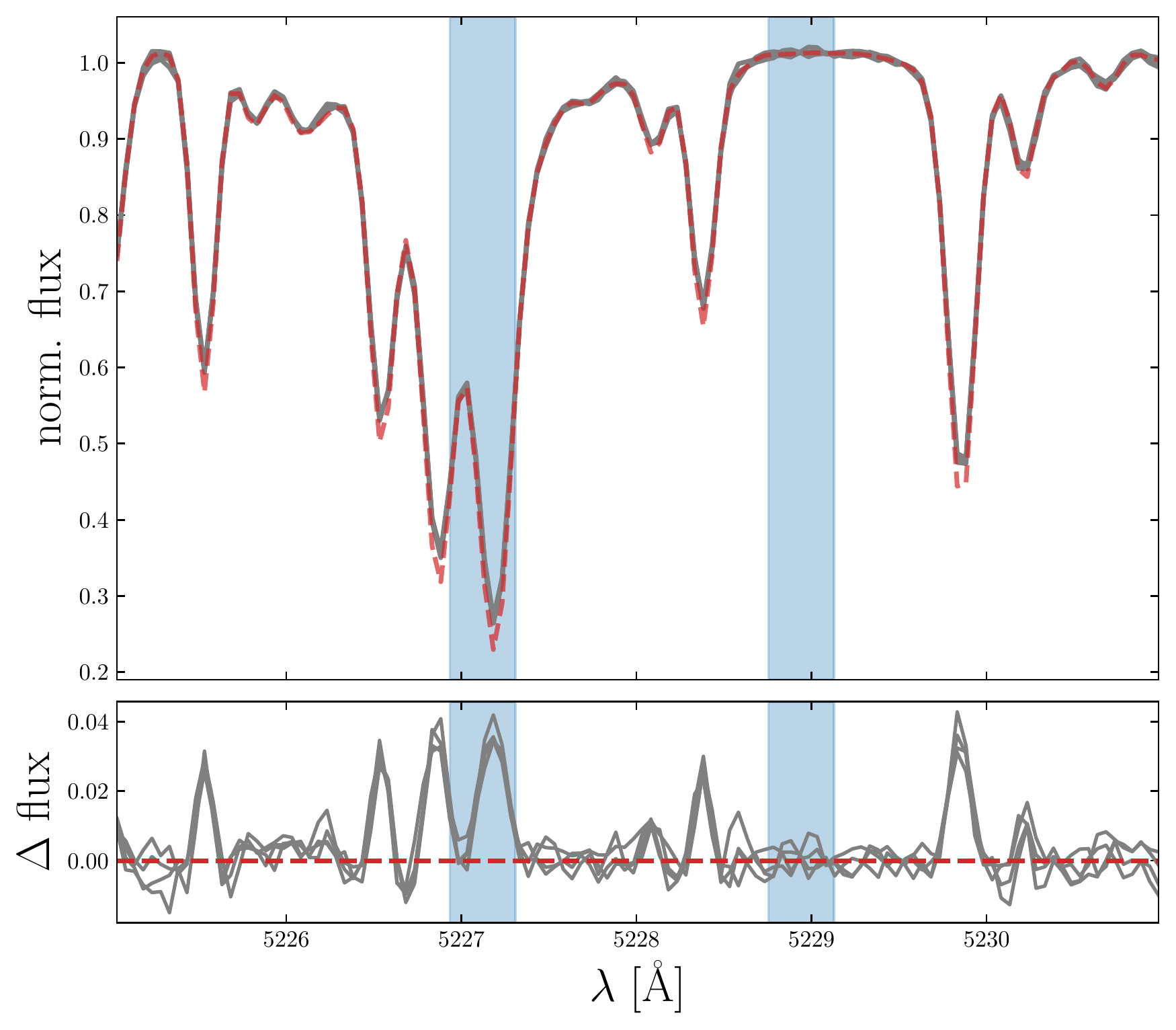}}
      \caption{Upper panel: Comparison of the four ELODIE spectra for the Sun with $\mathrm{S/N} > 100$ (gray) to the atlas spectrum by \citet{Hinkle00} (red dashed). The latter was degraded to match the resolution of our training grid and re-sampled to the ELODIE dispersion scale. The blue regions show the fluxes used in one of our [Fe/H] relations. All spectra were normalized to the mean flux in the rightmost blue band. Lower panel: Flux residuals.
              }
      \label{Fig:performance_ELODIE_additive_offset}
 \end{figure}

 The mean deviation in [Fe/H] is $0.21\pm0.18$~dex (rms). We investigated this behavior in the Sun. There are in total six solar spectra available through the ELODIE library, four of which satisfy $\mathrm{S/N}>100$. From these we derive [Fe/H$] = -0.19$~dex with a negligibly small scatter. In Fig. \ref{Fig:performance_ELODIE_additive_offset}, we compare a small wavelength portion to one of the solar spectra from our training grid. Here, the line cores in all ELODIE spectra do not reach as low as the ones in the reference spectrum. As a consequence, the overall line strength inferred from ELODIE is weaker, which translates into higher FRs and hence lower metallicities from ATHOS. Despite the fact that, compared to our training grid, ELODIE's resolution is slightly lower ($42000$ vs. $45000$), this finding cannot be attributed to resolution differences. If resolution were the sole reason, the missing line depth in the profile cores would be fully recovered in the wings, such that the total area of the profile stays constant and the integrated residuals evaluate to zero. This is not observed. We conclude that there must be an additive flux offset to ELODIE library spectra that leads to an unphysical weakening of absorption lines.
 
 Looking at $\log{g}$, we computed a mean difference and scatter of $0.29\pm0.30$~dex (rms). The residual distribution for giants ($\log{g}\leq2$~dex) is well described by random scatter, while for higher gravities there is a positive offset, which correlates strongly with $\Delta[$Fe/H] and $\Delta T_\mathrm{eff}$. This is an expected behavior, because, as discussed in Sect. \ref{Subsec:Noise}, our $\log{g}$ relations are very sensitive to prior metallicity estimates.

\subsection{Comparison with the S$^4$N library}\label{Subsec: comparison S4N}
 \begin{figure*}
    \centering
    \resizebox{\hsize}{!}{\includegraphics{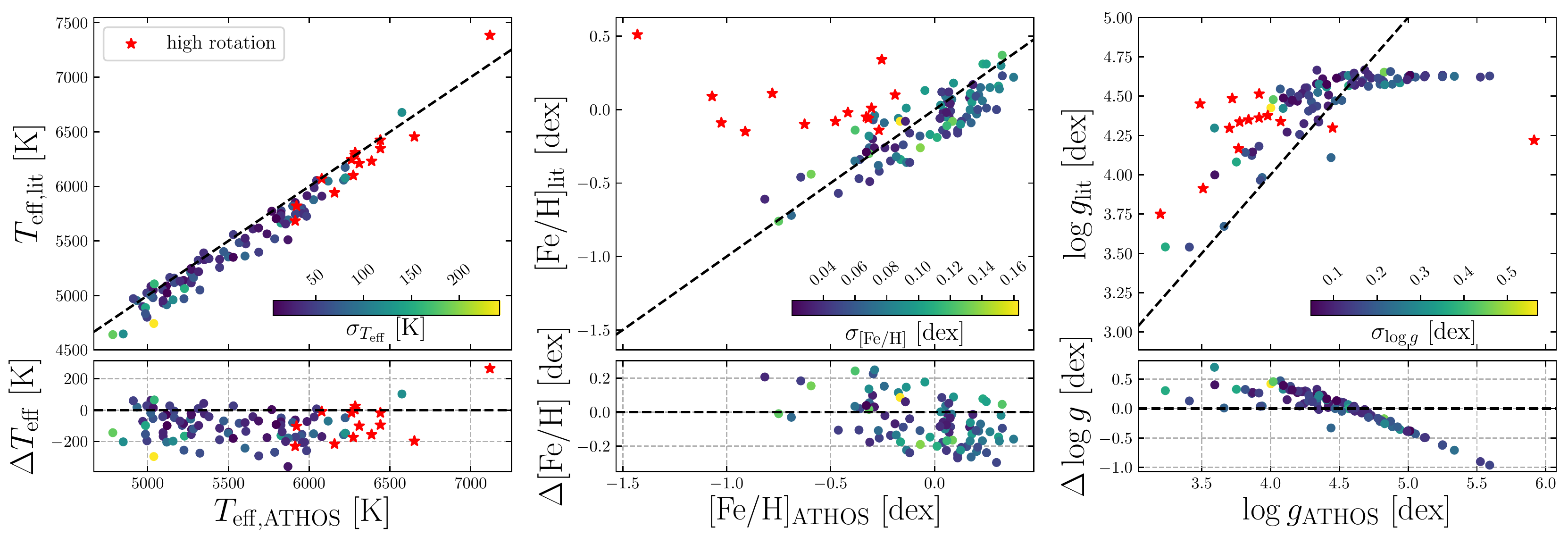}}
      \caption{Same as Fig. \ref{Fig:performance_ELODIE}, but for the S$^4$N library \citep{AllendePrieto04}. Red star symbols resemble stars with rotational velocities $v\sin{i}\geq5$~km~s$^{-1}$. For better visibility, we did not show these stars in the residual distributions of the middle and right panel, because they are far off (see discussion in the main text).
              }
      \label{Fig:performance_S4N}
 \end{figure*}
 S$^4$N is a high-resolution spectral library of 119 stars by \citet{AllendePrieto04}. The library encompasses a complete census of stars in the local volume ($R_\sun \leq 15$~pc) down to an absolute magnitude of $M_\mathrm{V}=6.5$~mag. The spectra were obtained with either the 2dcoudé spectrograph ($R\approx52000$) at McDonald Observatory, or the FEROS spectrograph ($R\approx45000$) at the ESO La Silla observatory. The S/N of the sample is very high (several 100s pixel$^{-1}$), so that noise-induced effects are uncritical. The spectra cover all wavelengths of all of our parameter relations ($3620$~{\AA} to at least $9210$~{\AA}). The provided $T_\mathrm{eff}$ was based on photometric colors assuming negligible reddening, while the metallicities were determined via full spectrum fitting of a 150~{\AA} range around H$\beta$ \citep[see][]{AllendePrieto03}. S$^4$N gravities were inferred from fitting theoretical isochrones using Hipparcos parallaxes. We note that the sample may not be as homogeneous as wished, for example, HD~82328 is a spectroscopic binary and HD~10780 is a BY Dra variable. 
 
 We computed stellar parameters for the S$^4$N library using ATHOS. To this end, we masked the H$\alpha$ profiles of the FEROS spectra, because \citet{AllendePrieto04} caution that those features have unreliable shapes due to fiber reflections within the spectrograph. Moreover, since the spectra are velocity-shifted but not corrected for tellurics, we had to cross-correlate our line list of tellurics with the library spectra in order to compute the respective velocities of the topocenter. This step was necessary to be able to include the H$\alpha$ relations of the 2dcoudé spectra without being biased by telluric contamination. Three binaries (HD~110379, HD~188088, and HD~223778) were excluded from consideration.
 
 \begin{figure}
    \centering
    \resizebox{\hsize}{!}{\includegraphics{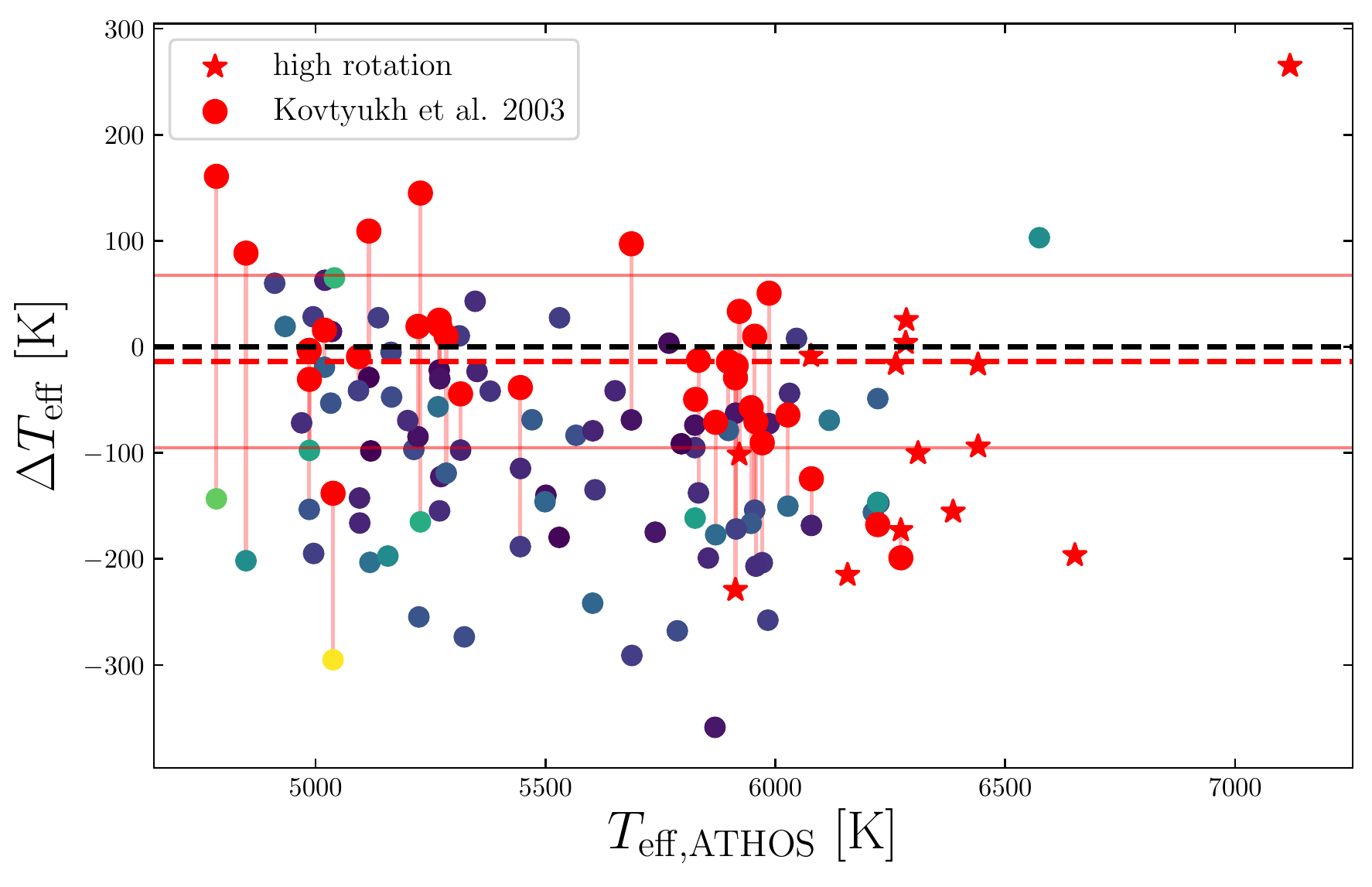}}
      \caption{Comparison of the residual temperature distribution of the S$^4$N literature temperatures (see bottom left panel of Fig. \ref{Fig:performance_S4N}) with the ones in common with \citet{Kovtyukh03} (red points). Vertical red lines connect points sharing the same spectrum and therefore ATHOS output. The red dashed line corresponds to the mean difference of 14~K when using the \citet{Kovtyukh03} reference and the solid red lines indicate the $\pm1\sigma$ scatter of 81~K.
              }
      \label{Fig:performance_S4N_Kov}
 \end{figure}
 For $T_\mathrm{eff}$, there seems to be a constant offset between the library parameters and ATHOS. Our temperatures are on average $105\pm92$~K (rms) warmer. We suspect that this is due to the literature values originating from photometric calibrations. The reader is referred, for example, to \citet{Lind08} for a comparison of the \citet{Alonso96, Alonso99a} temperature scale with the one obtained from fitting H$\alpha$. \citet{AllendePrieto04} cross-validated their photometric $T_\mathrm{eff}$ using the method of fitting Balmer lines by \citet{Barklem02}. Even though they have found almost negligibly warmer temperatures by on average $35\pm84$~K (rms), it seems that the small-scale trends in their Fig. 5 would -- taken as correction for the literature values -- improve our systematic discrepancies. Table \ref{Table:Teff training vs. S4n} shows a comparison of the temperatures of the stars in common between our sample and S$^4$N. All of them are also part of the GBS and therefore have highly reliable training temperatures, which are on average 88~K warmer. Another striking evidence showing that the photometric calibrations are systematically cooler than spectroscopic results can be seen when we compare our findings to those for the 32 stars in common with \citet{Kovtyukh03}. Their method for determining $T_\mathrm{eff}$ is based on spectroscopic line-depth ratios and has proven to have vanishingly small internal errors. In fact, as can be seen in Fig. \ref{Fig:performance_S4N_Kov}, employing their published temperatures reduces the mean deviation and scatter of ATHOS results to $14\pm81$~K (rms), in other words the scatter approaches the order of the systematic uncertainty expected for our relations, $\sigma_{T_\mathrm{eff},\mathrm{sys}}=\Teffsyserr$~K.
 \begin{table}
  \caption{$T_\mathrm{eff}$ for overlapping stars between our training sample and the S$^4$N library.}
  \label{Table:Teff training vs. S4n}
  \centering
  \begin{tabular}{l c c c}
  \hline\hline
  Name & $T_\mathrm{eff,training}$ & $T_\mathrm{eff,S4N}$ & $\Delta T_\mathrm{eff}$\\
   & [K] & [K] & [K]\\
  \hline                      
  \input{Teff_training_vs_S4N.dat}
    \hline
    \end{tabular}
 \end{table}
 
 \begin{figure}
    \centering
    \resizebox{\hsize}{!}{\includegraphics{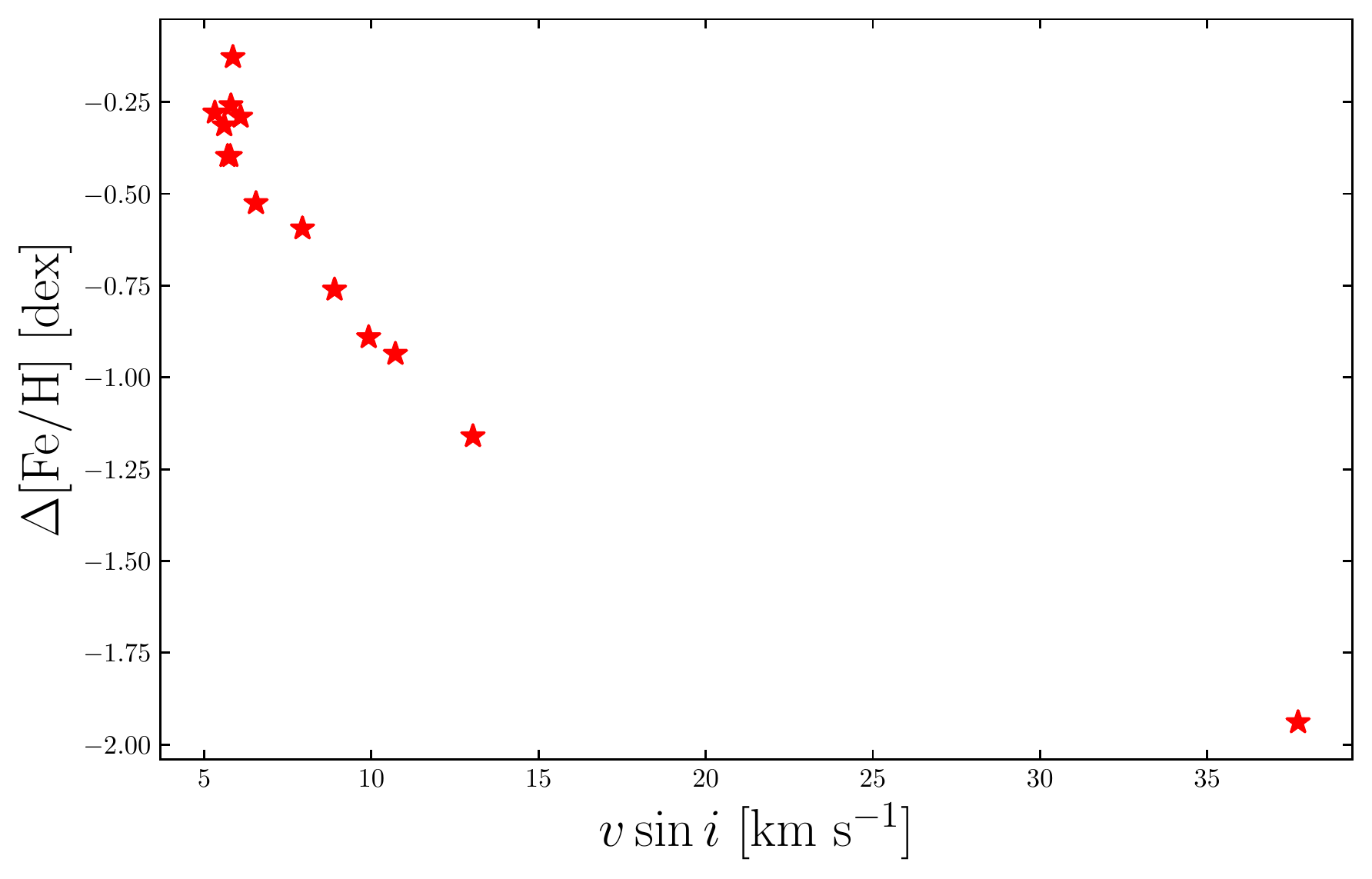}}
      \caption{Trend of $\Delta\mathrm{[Fe/H]}$ with $v\sin{i}$ for the red stars in Fig. \ref{Fig:performance_S4N}.
              }
      \label{Fig:performance_S4N_vsini}
 \end{figure}
 In the middle panel of Fig. \ref{Fig:performance_S4N}, there are several clear outliers (marked by red star symbols), which are predicted by ATHOS to have much lower metallicities than their tabulated reference values. Upon closer investigation, we found that the spectra expose significant rotational broadening. In fact, using $v\sin{i}$ values measured by \citet{AllendePrieto04}, there is a well defined behavior between $\Delta\mathrm{[Fe/H]}$ and $v\sin{i}$ (see Fig. \ref{Fig:performance_S4N_vsini}). To first order, rotational broadening looks similar to broadening by Gaussian line-spread functions. Thus the same reasoning as for the resolution dependencies in Sect. \ref{Subsec:Resolution} holds: Rotation disperses flux out of the small-scale profiles used to determine [Fe/H]. Despite its implemented correction for $R$, ATHOS is currently not capable of characterizing stars at rotational speeds $\gtrsim5$~km~s$^{-1}$.
 
 Overall, the [Fe/H] differences between S$^4$N literature values and ATHOS results are small, at an average of $-0.06\pm0.13$~dex (rms). In the high metallicity regime, we find higher metallicities by $0.14\pm0.10$~dex (rms). Assuming that the model temperatures for the full spectrum fitting of \citet{AllendePrieto04} were systematically too cool, the majority of the synthesized (neutral) profiles would have been stronger at the true metallicity. As a consequence, [Fe/H] would have been underestimated to match the observed spectrum. This probably explains part of the ATHOS metallicities being slightly higher. Given that the quoted errors on S$^4$N gravities are very small, we can conclude from the linear $\Delta \log{g}$ trend with $\log{g}_\mathrm{ATHOS}$ (see right panels of Fig. \ref{Fig:performance_S4N}) that our method's internal gravity error is relatively large. On average, however, the deviations do not exceed the $\sigma_{\log{g},\mathrm{sys}}=\loggsyserr$~dex provided for ATHOS (Sect. \ref{Subsec:Gravity}).
 
 \subsection{Comparison with the Gaia-ESO survey}
 \begin{figure*}
    \centering
    \resizebox{\hsize}{!}{\includegraphics{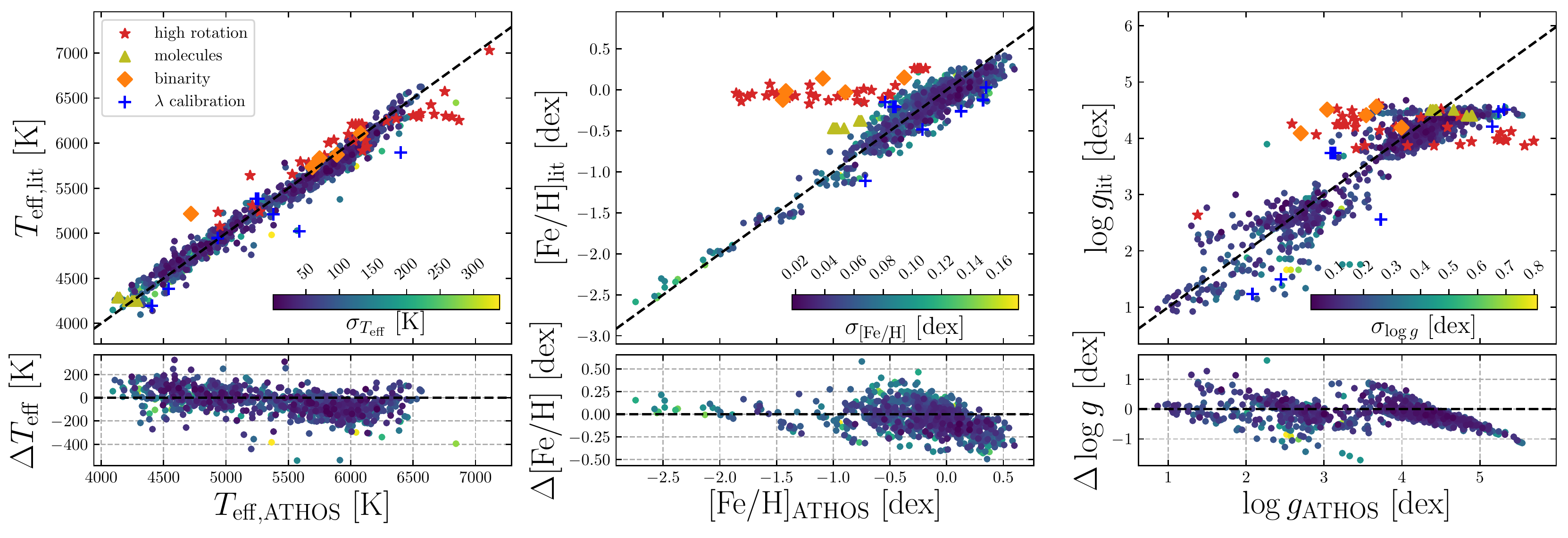}}
      \caption{Same as Fig. \ref{Fig:performance_ELODIE}, but for the high-resolution spectra and recommended parameters from the Gaia-ESO survey DR3. Red stars, golden triangles, orange diamonds, and blue pluses indicate the peculiar spectra we found to show high rotational broadening, strong molecular blending, binary components, or erroneous wavelength calibrations, respectively. In the residual distributions we excluded those, because they resemble extreme outliers (see text and Fig. \ref{Fig:performance_GES_deviation_origins}).
              }
      \label{Fig:performance_GES_UVES}
 \end{figure*}
 The Gaia-ESO survey \citep[henceforth GES,][]{Gilmore2012} is a large-scale spectroscopic survey of $\sim 10^5$ stars in the Milky Way. Here, we studied how ATHOS performs on the high-resolution ($\sim47000$) UVES spectra (4800 - 5800~{\AA}) of field and cluster stars that were released in data release 3.1 (DR3.1). GES published a catalog of recommended astrophysical parameters \citep{Smiljanic2014}. The spectra were analyzed by 13 different automated pipelines (``nodes''). Seven nodes measure EWs of \ion{Fe}{i} and \ion{Fe}{ii} (see Sect. \ref{Subsec: stellar parameters}) via fully automated codes. The remaining six nodes fit either observed or synthetic spectra. All subgroups employed the same GES line list. The recommended, publicly available set of parameters are the weighted median results from all 13 nodes. In order to construct the respective weights, the nodes were applied to the GBS and the results split in three groups: metal-rich dwarfs, metal-rich giants, and metal-poor stars (here grouping dwarfs and giants). The average difference for each group was then used to estimate the weights for the respective nodes. We point out that, while vaguely accounting for changing systematic errors with gravity (at least for high-metallicity stars) and metallicity, this treatment does not consider varying systematics of individual nodes over the considerably large temperature range of the survey. If any such differential systematic deviation existed, it would bias the recommended temperature value. In order to be consistent with our reference solar Fe-abundance $\log{\epsilon(\mathrm{Fe})}_\sun=7.50$~dex by \citet{Asplund09}, we subtracted 0.05~dex from GES metallicities, which are based on \citet{Grevesse07}.
 
 The spectra were shifted to their rest-wavelengths using RVs provided in the GES parameter catalog. Spectra with $\mathrm{S/N} < 40$ were omitted, as well as stars/spectra with GES peculiarity or technical flags including possible binarity, Balmer emission, strong molecular bands, strong rotation, radial velocity problems, or oversubtraction issues. The lower temperature limit was set to be $4000$~K\footnote{Thereby, we lost six stars.}. Moreover, in terms of parameter quality cuts, we set upper thresholds for the GES internal errors at 150~K and 0.2~dex for $T_\mathrm{eff}$ and [Fe/H], respectively. The final test sample comprised 1009 spectra of 912 stars. The median S/N is 67, which is considerably lower compared to the ELODIE and S$^4$N libraries. The results are illustrated in Fig. \ref{Fig:performance_GES_UVES}.
 
\begin{figure*}
    \centering
    \resizebox{\hsize}{!}{\includegraphics{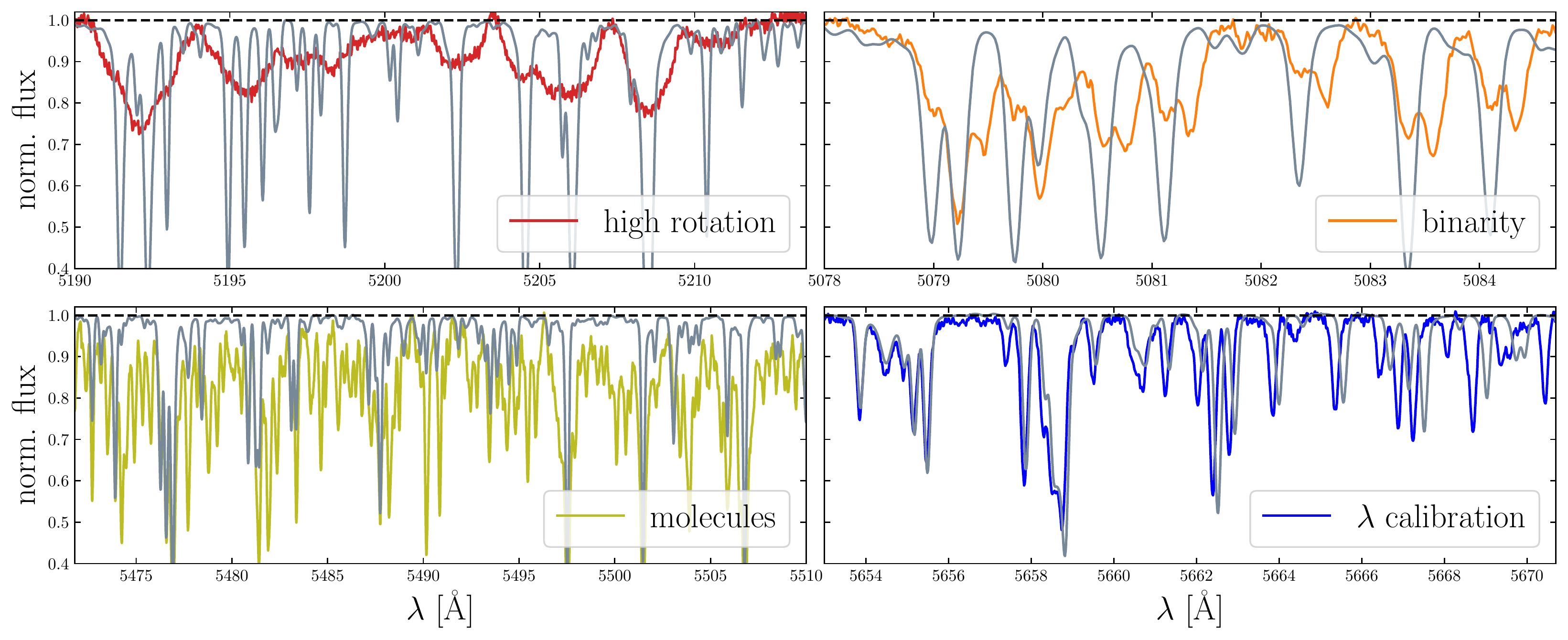}}
      \caption{Comparison of representative peculiar spectra (colored in the same way as in Fig. \ref{Fig:performance_GES_UVES}) to the solar spectrum (gray). An approximate normalization was performed by manually setting the continuum. The wavelength ranges were chosen such that the respective reason for peculiarity is illustrated best. The GES identifiers of the stars are (from upper left to lower right) 08100380-4901071, 18184436-4500066, 23001517-2231268, and 08090542-4740261.
              }
      \label{Fig:performance_GES_deviation_origins}
 \end{figure*}

 The most striking feature is an apparent bifurcation in the [Fe/H] comparison (upper plot in the middle panel) very similar to the one in Fig. \ref{Fig:performance_S4N} which is induced by strong rotational broadening of the underlying spectra. We visually inspected the spectra of the stars with the strongest disagreement between GES and ATHOS metallicities and identified four distinct reasons (see Fig. \ref{Fig:performance_GES_deviation_origins}). The most frequent reason for deviations is rotational broadening. This is surprising, because the GES spectra that were flagged as exposing fast rotation were explicitly omitted. The same applies to stars flagged as showing binary signatures. Yet, we identified several spectra showing either pronounced double-lined or asymmetrically distorted single-lined profiles over the entire spectral range. The third intrinsic effect is strong blending by molecular features and was identified in eight spectra of two stars. These molecular features prevail in cool, metal-rich dwarfs and affect most of the otherwise blend-free components of the FRs used in the [Fe/H] relations (Eq. \ref{Eq: [Fe/H] surface}). ATHOS wrongly identifies the affected stars as being more metal-poor, because the portions of metal lines and their normalizing component are closer in flux than they would usually be at a given metallicity. The effect is $~0.5$~dex for the two stars studied here. For the remaining outliers, we found that there are severe inconsistencies in the wavelength calibration of the spectra. This can be seen in the lower right panel of Fig. \ref{Fig:performance_GES_deviation_origins}: with increasing wavelength the blue spectrum appears compressed in comparison. We suspect that this is due to the wavelength solution diverging toward the edges of individual echelle orders. For some spectra the effect is so extreme that in the regions of overlapping orders, the order merging resulted in the spectrum appearing twice, spaced by up to 1.5~{\AA} apart. We conclude that the observed disagreements are not caused by the method implemented in ATHOS, but by issues in the reduction pipelines and quality control of GES.
 
 Over the entire $T_\mathrm{eff}$ range, ATHOS temperatures are marginally warmer by $39\pm101$~K (rms) compared to the recommended GES parameters. A closer look at the lower left panel of Fig. \ref{Fig:performance_GES_UVES} showing the residual distribution reveals two subgroups. The dividing $T_\mathrm{eff}$ is at around $5500$~K. Below this temperature, there is an excellent agreement at a mean difference and scatter of $14\pm97$~K. Above $5500$~K, on the other hand, the mean deviation is $-83\pm82$~K (rms), showing that GES is slightly cooler than our findings. Possible reasons are the potential systematics in the averaging process within GES on the one hand, or subtle non-linearities of $T_\mathrm{eff}$ with the FRs measured in the Balmer profiles, which were not detected. Nonetheless, given the rather low quality of the GES spectra in terms of S/N, it is remarkable that the scatter after removing the discussed biases of the two temperature groups is of order $\sim90$~K.
 
 ATHOS recovers the GES [Fe/H] with a mean deviation and scatter of $-0.04\pm0.15$~dex. Again, considering the low S/N of the GES products and the fact that metallicities are computed from only a few pixels in the spectra, this is an extraordinary result. At super-solar metallicities there is a slight disagreement in that ATHOS metallicities are higher by $\sim0.15$~dex. We argue that the trend arises due to the comparison sample (GES), which in \citet{Smiljanic2014}, Fig. 18, shows the same trend. We point out that our method returns an rms of 0.15~dex for the entire sample, while they only show the GES stars for which the dispersion is less than 0.20~dex (so the trend may be even more pronounced). ATHOS accurately reproduces GES gravities at mean residuals of $-0.18\pm0.35$~dex (rms). 
 
 \subsection{Comparison with globular cluster studies}\label{Subsec: GC surveys}
 While the previous sections focused on an assessment of ATHOS' performance on surveys targeting heterogeneous stellar populations, we also aimed at testing results from the very homogeneous populations of RGB stars in globular clusters (GCs).
 \begin{figure*}[ht!]
    \centering
    \resizebox{\hsize}{!}{\includegraphics{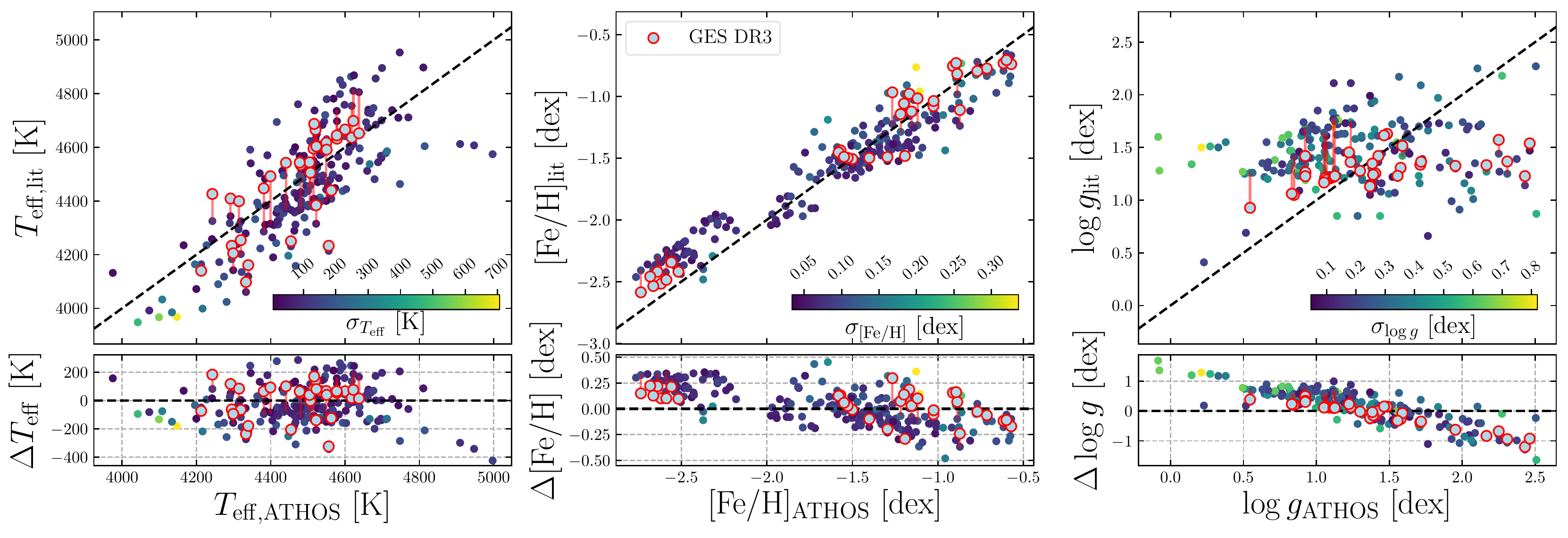}}
      \caption{Same as Fig. \ref{Fig:performance_ELODIE}, but for the high-resolution UVES sample of GC stars from the study of C09. Points connected by a red line represent the same spectrum but have two different literature values. The additional red open circles indicate the recommended parameters for DR3 of GES.
              }
      \label{Fig:performance_Carr09_GES}
 \end{figure*}
 \subsubsection{The Caretta et al. (2009) sample}\label{Subsubsec: The Caretta et al. (2009) sample}
 We compare our method to the UVES study by \citet{Carretta09} (henceforth C09) of 202 stars of 17 GCs spanning a wide range of metallicities and masses. Out of the 17 GCs, four -- NGC~104 (47~Tuc), NGC~2808, NGC~6752, and NGC~7078 (M~15) -- were re-reduced from the C09 archival data by GES and made publicly available through DR3\footnote{We note that compared to C09, two stars attributed to NGC~2808 are missing in DR3.}. Reduced spectra for the remaining 13 clusters were kindly provided by E. Carretta (private communication). For $T_\mathrm{eff}$, C09 relied on photometric calibrations and the assumption that the RGB of a GC is intrinsically narrow, which was used to infer $T_\mathrm{eff}$ from $V$ magnitudes alone. Despite the lower star-to-star errors, we emphasize that the approach might still be subject to biases introduced by the photometric calibrations, $E(B-V)$, or differential reddening. C09 determined surface gravities from isochrone fitting and Fe abundances using EWs of both ionization stages\footnote{We note that in the following we will refer to \ion{Fe}{ii} abundances whenever [Fe/H] is quoted in connection to C09.}. 
 
 We deduced a median S/N of 56 for the bluer spectral regions, where most of our FRs reside. Lacking information about the topocentric velocities by the time of the observations, H$\alpha$ profiles were masked out entirely, so that tellurics would not hamper the analysis. Fig. \ref{Fig:performance_Carr09_GES} presents our results. On top of the C09 literature values, we show the recommended parameters by GES, which were based on a re-reduction and analysis of C09 raw data and make up a considerable fraction of the metal-poor stars discussed in the previous Sect.
 
 \begin{table*}
  \caption{Mean $T_\mathrm{eff}$ and [Fe/H] residuals of C09 and GES with respect to ATHOS for individual GCs, as well as the mean [Fe/H] and scatter determined in the three studies.}
  \label{Table:C09_phot_scale_vs_spec}
  \centering
  \resizebox{\textwidth}{!}{%
  \begin{tabular}{@{\extracolsep{6pt}}lrrrrrrrrrrrrr}
  \hline\hline
   & \multicolumn{5}{c}{C09} & \multicolumn{5}{c}{GES DR3} & \multicolumn{2}{c}{ATHOS}\\
  \cline{2-6} \cline{7-11} \cline{12-13}
  GC ID & $\langle \Delta T_\mathrm{eff}\rangle$ & $\sigma_{\Delta T_\mathrm{eff}}$ & $\langle \mathrm{[Fe/H]} \rangle$ & $\sigma_\mathrm{[Fe/H]}$ & $\langle \Delta \mathrm{[Fe/H]}\rangle$ & $\langle \Delta T_\mathrm{eff}\rangle$ & $\sigma_{\Delta T_\mathrm{eff}}$ & $\langle \mathrm{[Fe/H]} \rangle$ & $\sigma_\mathrm{[Fe/H]}$ & $\langle \Delta \mathrm{[Fe/H]}\rangle$ & $\langle \mathrm{[Fe/H]} \rangle$ & $\sigma_\mathrm{[Fe/H]}$ & $N_\mathrm{stars}$\\
   & [K] & [K] & [dex] & [dex] & [dex] & [K] & [K] & [dex] & [dex] & [dex] & [dex] & [dex] &\\
  \hline                      
  \input{C09_phot_scale_vs_spec.dat}
    \hline
    \end{tabular}}
 \end{table*}
 ATHOS temperatures deviate on average by $-30\pm132$~K (rms) from C09, that is there is an excellent agreement between the entirely photometry-based C09 $T_\mathrm{eff}$-scale and our purely spectroscopically constrained one. We note that the intra-cluster temperature deviations are less scattered than the global scatter of 132~K. There seems to be a systematic offset between individual clusters on the two scales. In Table \ref{Table:C09_phot_scale_vs_spec}, we present the average temperature and [Fe/H] deviations between C09/GES and ATHOS. For example, the five stars of NGC~6171 constitute five of the six strongest deviating temperatures (lower left panel of Fig. \ref{Fig:performance_Carr09_GES}). The mean deviation for this GC is $-300$~K with a scatter of only 80~K. Likewise, ATHOS predicts NGC~6397 stars to be cooler than the C09 findings by on average $204\pm42$~K (rms). Generally, the very low intra-cluster scatters confirm the high precision of the Balmer-relations implemented in ATHOS. Systematic inter-cluster biases could either be founded in inaccuracies of the FR-relations for Balmer profiles in ATHOS, that is linked to unresolved systematics connected to additional parameters, or in the already mentioned possible caveats of the photometric relations of C09. For NGC~6171, we suspect dereddening to be the main source causing the offset \citep[cf.][]{OConnell11} and for NGC~6397, assuming a too warm estimate of $T_\mathrm{eff}$ for all cluster members in C09 could explain the discrepant metallicities of C09 and \citet{Koch11}.
 
 \begin{figure}
    \centering
    \resizebox{\hsize}{!}{\includegraphics{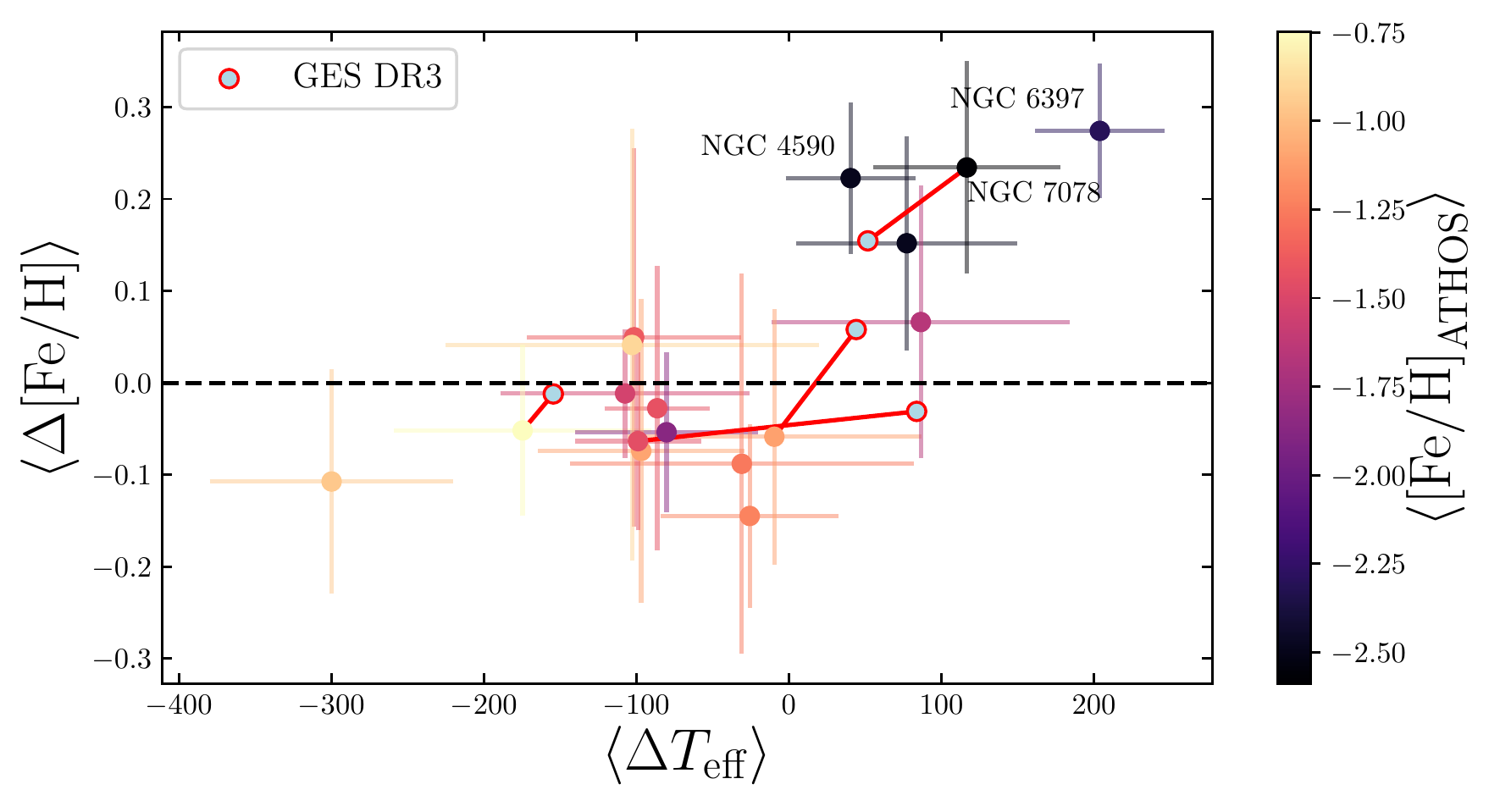}}
      \caption{[Fe/H] residuals averaged per GC versus mean temperature offset. The color coding resembles the cluster metallicity (see color bar on the right). Light colored error bars indicate the respective $\sigma_{\Delta T_\mathrm{eff}}$ and $\sigma_\mathrm{[Fe/H]}$ from Table \ref{Table:C09_phot_scale_vs_spec}. Red lines connect the same clusters corresponding to the same stars and spectra, but with stellar parameters either from C09, or GES DR3 (red circles).
              }
      \label{Fig:performance_Carr09_Teff_met_corr}
 \end{figure}
 Globally, our metallicity scale agrees within $\langle \Delta \mathrm{[Fe/H]} \rangle = 0.03\pm0.19$~dex (rms) with the one by C09. For the fairly large scatter, random uncertainties are likely to play a rather minor role. Fig. \ref{Fig:performance_Carr09_GES} implies a trend of $\Delta \mathrm{[Fe/H]}$ with metallicity, which is also reflected on the individual cluster scale (see Table \ref{Table:C09_phot_scale_vs_spec}). Deviations of clusters are most likely caused by a drift in the $T_\mathrm{eff}$ scale. This is confirmed by Fig. \ref{Fig:performance_Carr09_Teff_met_corr}, where we present a strong correlation between the mean temperature deviations and the mean metallicity residuals between C09 and ATHOS for individual GCs. In case of the strongest outlier in positive direction, NGC~6397, \citet{Koch11} found a metallicity of $-2.10$~dex compared to $-2.03$~dex. Adopting these values would reduce ATHOS' offset to $0.20$~dex. At this point it is noteworthy that \citet{Lind08} found much lower metallicities for NGC~6397 in better agreement with our findings for a stellar sample between the TO and the blue RGB. However, they showed that there is a significant trend of [Fe/H] with evolutionary stage ($-2.41$~dex to $-2.28$~dex).
 
 We found a marginal overall deviation of $0.17\pm0.55$~dex (rms) in the residual distribution of $\log{g}$ between C09 and ATHOS. Given that all stars considered here are giants clustering around $\log{g}\approx1.3$~dex, the apparent trend in the residual distribution can be attributed to larger errors on individual ATHOS gravities leading to an increased star-to-star scatter with respect to C09. The difference between the two distributions produces the observed anticorrelation of $\log{g}_\mathrm{\nameoftool}$ and $\Delta \log{g}$. Given that most of the information carried by the FRs is related to [Fe/H] (see Sect. \ref{Subsec:Gravity}), here the S/N of the giant spectra probably plays an important role for the large scatter in $\log{g}$.
 
 \subsubsection{The MIKE sample}\label{Subsubsec: The MIKE sample}
 \begin{figure*}
    \centering
    \resizebox{\hsize}{!}{\includegraphics{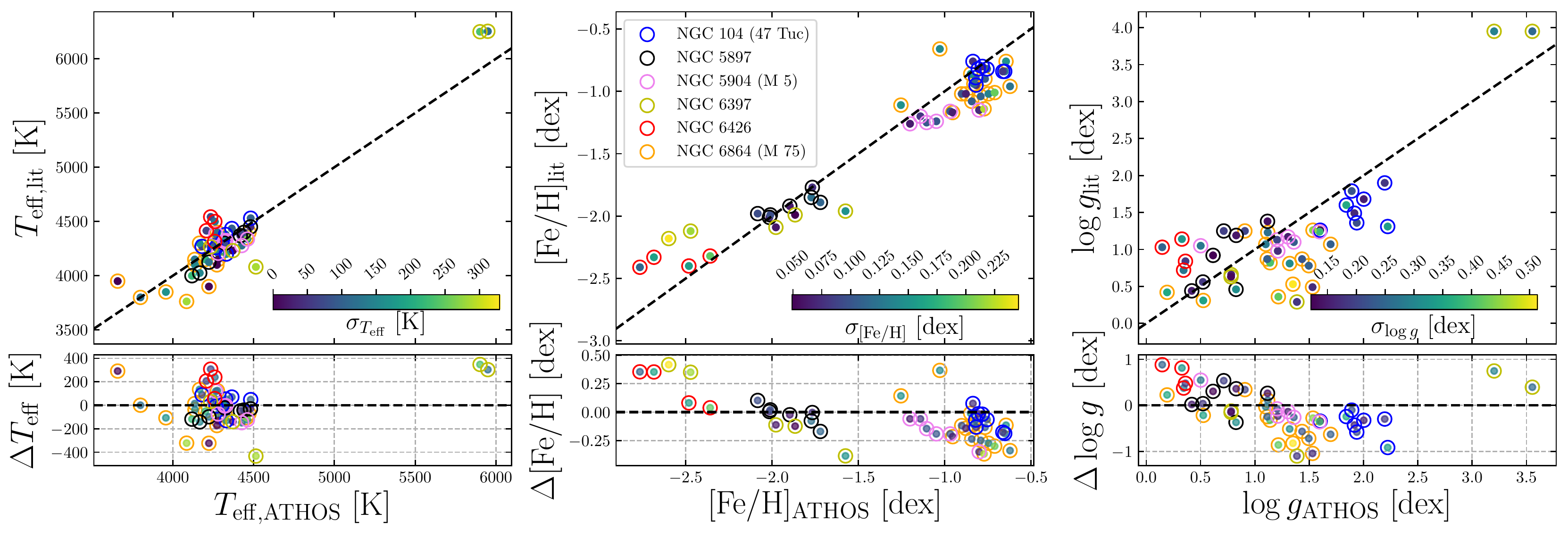}}
      \caption{Same as Fig. \ref{Fig:performance_ELODIE}, but for a sample of MIKE spectra of GC stars. The colored open circles indicate the membership in either of the GCs 47~Tuc \citep[blue, ][]{Koch08}, M~5 \citep[pink,][]{Koch10}, NGC~6397 \citep[yellow,][]{Koch11}, M~75 \citep[orange, ][]{Kacharov13}, NGC~5897 \citep[black, ][]{Koch14}, or NGC~6426 \citep[red,][]{Hanke17}.
              }
      \label{Fig:performance_MIKE_GCs}
 \end{figure*}
 \begin{table*}
  \caption{Comparison of the literature stellar parameters of the MIKE GC sample to ATHOS.}
  \label{Table:MIKE_resids}
  \centering
  \resizebox{\textwidth}{!}{%
  \begin{tabular}{@{\extracolsep{6pt}}lrrrrrlrrr}
  \hline\hline
   & & & & & & & \multicolumn{2}{c}{ATHOS} &\\
  \cline{8-9}
  GC ID & $\langle \Delta T_\mathrm{eff}\rangle$ & $\sigma_{\Delta T_\mathrm{eff}}$ & $\langle \mathrm{[Fe/H]} \rangle$ & $\sigma_\mathrm{[Fe/H]}$ & $\langle \Delta \mathrm{[Fe/H]}\rangle$ & reference & $\langle \mathrm{[Fe/H]} \rangle$ & $\sigma_\mathrm{[Fe/H]}$ & $N_\mathrm{stars}$\\
   & [K] & [K] & [dex] & [dex] & [dex] &  & [dex] & [dex] &\\
  \hline                      
  \input{MIKE_resids.dat}
    \hline
    \end{tabular}}
 \end{table*}
 During the last ten years, we have published a series of detailed spectroscopic studies of GC member stars. In total, 46 targets of six GCs (see Table \ref{Table:MIKE_resids} for details and references) have been observed using the high-resolution ($R\approx40000$) MIKE spectrograph mounted at the 6.5~m Clay telescope at Las Campanas Observatory. The stellar sample of NGC~6397 contains two MSTO stars, for which \citet{Koch11} used profile fits to H$\alpha$ to deduce $T_\mathrm{eff}$. With the exception of stars belonging to NGC~5897 and NGC~6864, all temperatures of RGB stars were derived in a differential approach, that is excitation balance was enforced differentially on abundances of \ion{Fe}{i} lines with respect to the same lines in well understood standard stars with reliable reference parameters \citep[see][]{Koch08}. These should be as close as possible to the stellar parameters of the stars to be analyzed. Then, the differential treatment cannot only forgo relying on uncertain oscillator strengths, but reduce the influence from NLTE-induced excitation imbalances as well \citep[][and references therein]{Hanke17}. In all studies, $\log{g}$ was computed from basic stellar structure equations. Here, we refer to abundances of \ion{Fe}{ii}, which were computed from EWs as [Fe/H]. 
 
 Fig. \ref{Fig:performance_MIKE_GCs} together with Table \ref{Table:MIKE_resids} show the comparison between the original works on the GCs and an ATHOS run. The mean $T_\mathrm{eff}$ difference is $-12\pm161$~K (rms), with a scatter dominated by differences between individual clusters. In terms of temperature, we found one cluster to systematically deviate from the one-to-one trend. 
 
 While exposing a rather small star-to-star dispersion in the residuals, all four stars of NGC~6426 are predicted to be cooler by $\sim203$~K compared to their respective reference values. One possible explanation for this observation is that none of the stars in \citet{Hanke17} satisfy excitation equilibrium of \ion{Fe}{i} in an absolute abundance treatment. In the differential approach pursued in that work the imbalances were compensated by the strong excitation imbalance (on the absolute scale) of the very similar, metal-poor benchmark star HD~122563. The imbalance in this star is likely to originate from NLTE effects \citep{Mashonkina11}. We showed that in order to achieve excitation equilibrium $T_\mathrm{eff}$ had to be lowered by 200~K compared to the differential case. As the majority of the temperatures of the metal-poor training stars in this study were determined using absolute abundances, the offset for NGC~6426 can be explained by ATHOS' temperature scale being tied to absolute rather than to differential methods. We point out that the other two metal-poor clusters, NGC~5897 and NGC~6397, do not show the same deviation. NGC~5897 was studied with an absolute treatment, hence the findings are accurately reproduced by ATHOS. For NGC~6397, the employed benchmark star was Arcturus, which does not show any excitation imbalances in an absolute LTE analysis \citep{Koch08}. Consequently, the differential temperature scale is essentially the same as the absolute one.
 
 Two GCs, namely NGC~6397 and NGC~6864, exhibit a large scatter (285~K and 179~K, respectively) with respect to what we would expect given the high S/N of the spectra. This was partly due to H$\beta$ not being covered by most of the MIKE spectra and additional loss of H$\alpha$ relations due to telluric contamination. An important source for the strongest offset ($\sim400$~K), which is shown by the RGB star \#13414 in NGC~6397, is its strong mass-loss that manifests in wind-induced emission spikes in the profile wings of H$\alpha$. For the two MSTO stars in NGC~6397, we identify the treatment of different broadening mechanisms in the LTE Balmer synthesis of the original work as possible origin of deviations in the temperature scales. The large intra-cluster scatter of $\Delta T_\mathrm{eff}$ for NGC~6864 (156~K) remains inexplicable.
 
 In terms of metallicity, the mean residual deviation and scatter is $-0.06\pm0.20$~dex. Similar to before, neither mean nor scatter are representative for all the subsamples and differences can be linked to deviating temperatures of individual stars. 
 
 The comparison of the literature gravities with our results for $\log{g}$ confirms that especially for gravities well below 1~dex ATHOS can only provide a rough estimate. Partly, this is also related to rather low S/N values of the spectra, which increase the internal $\log{g}$ errors of ATHOS (see right panels of Fig. \ref{Fig:performance_MIKE_GCs}).
 
 \subsection{Comparison with SDSS}\label{Subsec: SDSS}
 \begin{figure*}
    \centering
    \resizebox{\hsize}{!}{\includegraphics{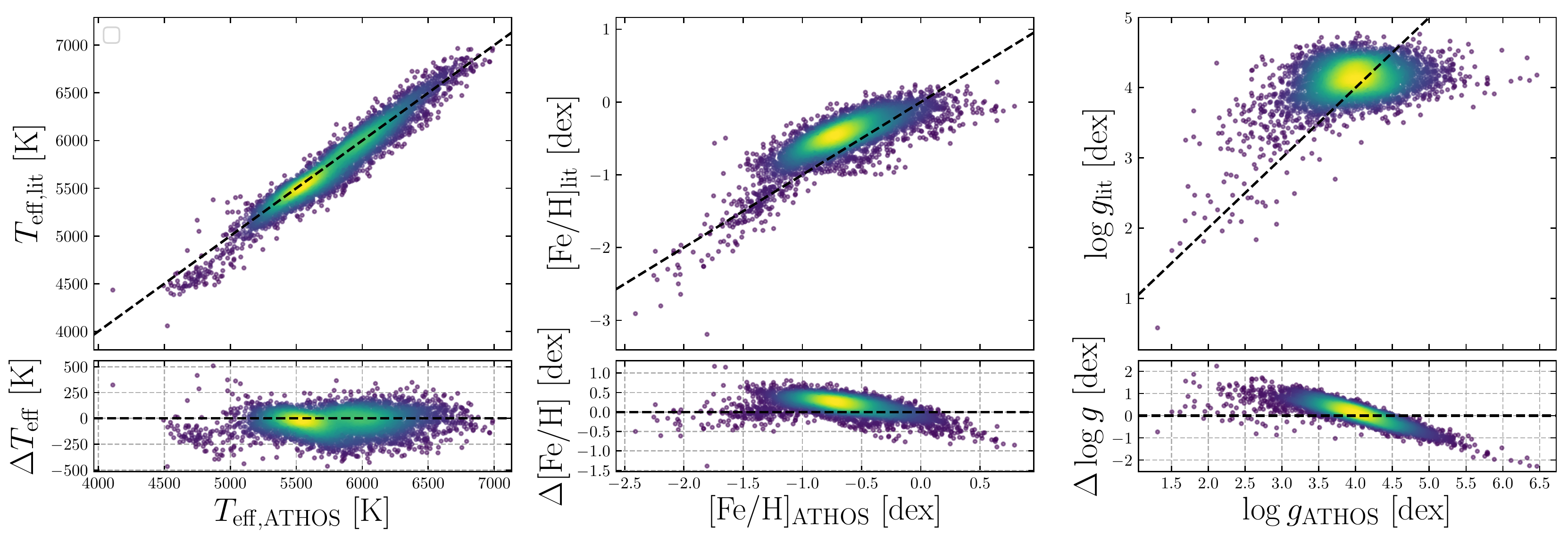}}
      \caption{Same as Fig. \ref{Fig:performance_ELODIE}, but for 3966 SDSS low-resolution ($R\approx1500$ to 2500) spectra. As opposed to earlier Figs., here the color does not represent ATHOS errors but the local point density (arbitrary units) obtained via a Gaussian kernel-density estimate.
              }
      \label{Fig:performance_SDSS}
 \end{figure*}
 The Sloan Digital Sky Survey \citep[SDSS, ][]{Abolfathi18} released several hundred thousand optical spectra at low resolution ($R\approx1500$ to 2500) with an accompanying catalog of stellar parameters for a subset thereof. These parameters were determined using the SEGUE Stellar Parameter Pipeline \citep[SSPP, ][]{Lee08}, which obtains stellar parameters from several photometric and spectroscopic estimators. The final adopted SSPP parameters are averaged from the individual measurements using a decision tree that excludes certain results based on photometric quality cuts and S/N values of the analyzed spectra \citep[see][for a detailed discussion]{Lee08}. 
 
 We used the DR14 SSPP catalog -- which has not changed since DR12 -- to retrieve and analyze spectra obeying our own quality cuts. These were set such that only spectra without peculiar flags, maximum SSPP uncertainties on $T_\mathrm{eff}$, [Fe/H], and $\log{g}$ of 50~K, 0.05~dex, and 0.1~dex, as well as with $\mathrm{S/N}>80$ were considered. The fact that we analyzed the resulting 3966 spectra within 16~s, that is 4~ms per spectrum, impressively shows that ATHOS' execution time is merely limited by the size of the input spectra (in this case $\sim3800$ dispersion points). We illustrate the comparison between ATHOS and SSPP results in Fig. \ref{Fig:performance_SDSS}.
 
 The mean offset and rms scatter in $T_\mathrm{eff}$ is $-37\pm115$~K, which proves our new temperature indicator to be accurate (on the scale of SSPP) and precise even at resolutions as low as $\sim2000$. Further, despite not being covered by our benchmark sample, the temperature range between 6500~K and 7000~K appears to be described well by extrapolating our linear FR-trends. 
 
 In terms of metallicity, we found a bias of $0.20\pm0.24$~dex (rms) and additional substructure in the residual distribution. This is not further surprising, because the pixel spacing in SDSS spectra resides mostly well above 1~{\AA}. Hence, ATHOS determines metallicities (and gravities) for the individual relations from wavelength ranges ($w=0.187$~{\AA}) that span less than a fifth of a pixel's width. Due to this, any inaccuracy in the wavelength solution or in the radial velocity determination of the input spectra adds a systematic contribution that potentially has a large influence on the deduced parameters. Our $\log{g}$ estimate is accurate on the SSPP scale on a 0.01~dex level with a scatter of 0.50~dex. While its $T_\mathrm{eff}$ method is highly accurate, ATHOS' metallicity and gravity estimates at these low resolutions are rather vague. Nevertheless, they can still be used, for example, as means to distinguish dwarfs from giants and metal-poor from metal-rich stars.
 
 \subsection{Comparison across surveys}
 \begin{figure*}
    \centering
    \resizebox{\hsize}{!}{\includegraphics{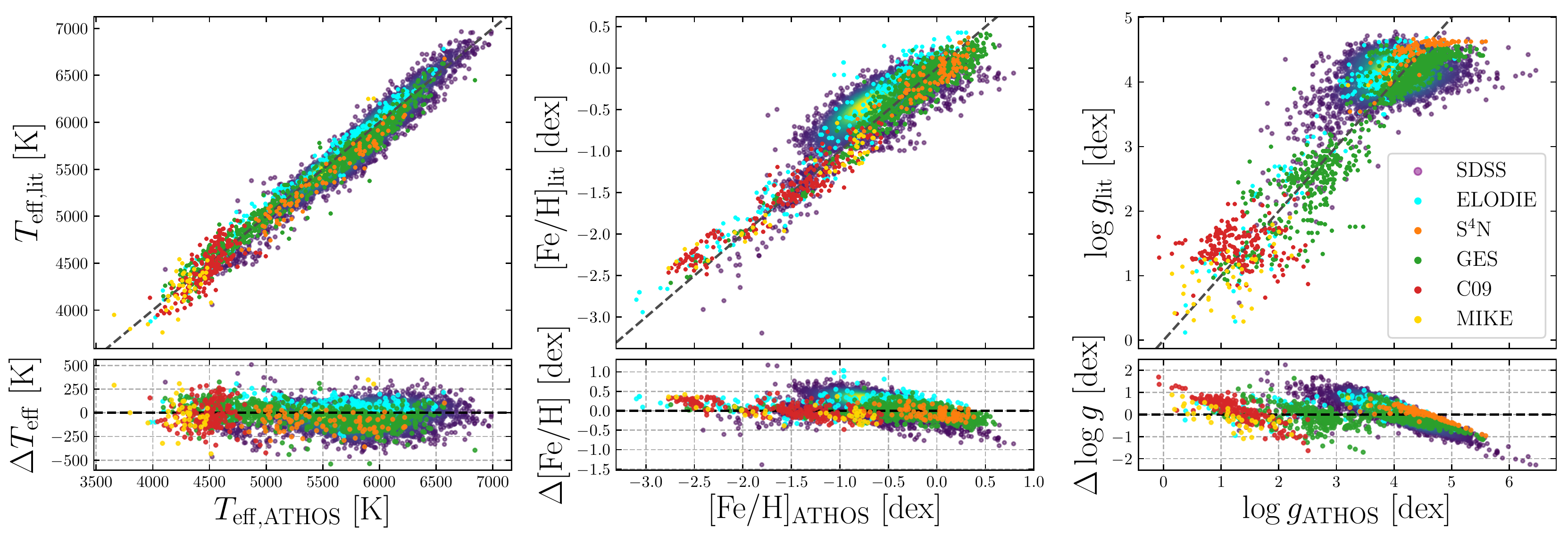}}
      \caption{Comparison of the results from all surveys discussed in Sections \ref{Subsec: comparison surveys} to \ref{Subsec: SDSS}. The color coding is indicated by the legend in the right panel.
              }
      \label{Fig:performance_survey_summary}
 \end{figure*}
 A key goal of this project was to enable cross-validation and parameter homogenization across various survey pipelines. This main strength of ATHOS is illustrated in Fig. \ref{Fig:performance_survey_summary}, where we show the combined results of all six data sets (ELODIE, S$^4$N, GES, C09, our MIKE GC studies, and SDSS/SEGUE) that have been described in detail before. It is noteworthy that all 1579 high-resolution spectra of different resolutions originating from six different instruments were analyzed within 43~s, corresponding to an average execution time of 27~ms per spectrum. For the 3966 low-resolution spectra the average time demand was considerably lower at 4~ms. We emphasize that this was achieved using only one CPU core on a regular laptop. 
 
 From Fig. \ref{Fig:performance_survey_summary} we conclude that using ATHOS' homogeneous parameter scale together with its wide-ranging applicability enables the detection of discrepancies between the methods employed by the different surveys. Our in-depth investigations of the individual samples (see earlier Sects.) revealed that departures from the ATHOS scale -- especially for $T_\mathrm{eff}$ -- are likely to be founded either in erroneous spectroscopic data, or possible shortcomings of the various survey pipelines.
 
\section{Conclusion}\label{Sec: Summary}
In this paper we presented a novel approach of deriving the fundamental stellar parameters $T_\mathrm{eff}$, $\log{g}$, and [Fe/H] by employing parameter-sensitive FRs of close-by wavelength windows in optical stellar spectra. Using a set of simple analytical relations, the FRs described here were tied to the well-calibrated parameters of a training set of 124 stars. The training sample comprised stars spanning a wide domain of stellar parameters, that is dwarfs and giants of spectral type F to K that range from very metal-poor ([Fe/H]$\approx-4.5$~dex) to super-solar metallicities. The introduced technique is non-iterative -- that is computationally inexpensive -- and depends on models only to the degree that the training parameters have been determined via model-dependent approaches. 

Our method was implemented into a Python-based code we call ATHOS, which is made publicly available\footnote{\url{https://github.com/mihanke/athos}}. Being fairly simple, it is straightforward to implement the relations established here in any other piece of spectrum analysis code. We have shown that our method accurately recovers the atmospheric parameters of the large samples of the ELODIE and S$^4$N libraries, the high-resolution part of the Gaia-ESO survey, the GC studies by C09 and our own group comprising in total 20 clusters, and -- at least in terms of $T_\mathrm{eff}$ -- the SDSS/SEGUE survey. Thus, ATHOS temperatures have proven to be  insensitive to spectral resolution over more than a magnitude of resolutions ($R\approx2000$ to 52000), while [Fe/H] and $\log{g}$ are reproduced accurately down to $R\approx10000$.

Our machine-learned, optimal FRs converged toward determining a star's effective temperature from wavelength regions around the Balmer lines H$\beta$ and H$\alpha$. As opposed to many other parameter-estimation methods, this can be done truly independent from any other stellar parameter, such as metallicity. We could show that our new temperature estimator agrees very well with the LDR-based scale of \citet{Kovtyukh03} ($\Delta T_\mathrm{eff} = 14\pm 81$~K, based on spectra of 32 stars). Our method, however, has a much more extended range of applicability, because it is also capable of analyzing giants and -- due to relying on Balmer- instead of metal lines -- metal-poor stars down to metallicities of $\sim-4.5$~dex. 

Based on the determined temperature, we can use FRs involving spectral features of atomic and ionized species to deduce [Fe/H] and $\log{g}$, respectively. We have been more lenient with the gravity estimates, since we assume that Gaia parallaxes \citep{Gaia18} will -- at least for the brighter and/or near-by stellar populations -- oust any spectroscopic method in the future.

Provided the optical spectra include parameter sensitive wavelength regions like H$\alpha$, ATHOS can efficiently and fast (a few milliseconds per spectrum) determine the three key stellar parameters, $T_\mathrm{eff}$, [Fe/H], and $\log{g}$. Unlike previously established automated methods of determining stellar parameters such as EW-matching algorithms \citep[e.g.,][and references therein]{Smiljanic2014} or full spectrum fitting techniques \citep[e.g., The Cannon,][]{Ness15}, the approach introduced here is largely insensitive to continuum normalization and resolution of the spectra. Hence the strength of ATHOS is that it is not confined to a specific survey or spectrograph and that it does not require a separate training for each individual survey. Furthermore, it can deal with any spectral type (within the training sample's limitations) irrespective of a priori assumptions or priors. These aspects constitute ATHOS' main strength, which is the determination of homogeneous stellar parameters across surveys.

Our new tool is capable of parameterizing one million stars in just about eight hours on a single CPU core. Hence, in its current version, ATHOS can homogeneously and robustly analyze all high-resolution spectra ever to be observed with 4MOST and WEAVE in less than one day, without any manual interference or having to use extensive computational resources. In addition, it is a very useful tool for flagging outliers (binaries, fast rotators, chemically peculiar stars) which previous surveys like GES may have overlooked.

Future add-ons in ATHOS will focus on the investigation of peculiar stars like CEMP stars, fast rotators, and low-gravity ($\log{g}<1$~dex) stars. Another class of peculiar stars we desire to explore is the collection of variable stars, such as RR-Lyrae or Cepheid variables, which can exhibit atmospheric shocks that might distort spectral profiles like H$\alpha$.  Finally, we will examine the possibility to extend our FR-based method to near-infrared spectra in order to enlarge ATHOS' wavelength applicability.

\begin{acknowledgements}
A.K. and E.K.G. gratefully acknowledge support from the Collaborative Research Cluster ``The Milky Way System'' (SFB~881) of the German Research Foundation (DFG), particularly through subprojects A5, A8, and A9. We thank the referee for helpful and constructive inputs. The authors are grateful to E. Carretta et al. for providing ready-to-use spectra of their large-scale GC program. Based in part on data products from observations made with ESO Telescopes at the Paranal Observatory under program ID 188.B-3002 (GES). Based in part on data obtained from the ESO Science Archive Facility. Furthermore, this paper includes data gathered with the 6.5 m Magellan telescopes located at Las Campanas Observatory, Chile. This work has made use of the VALD database, operated at Uppsala University, the Institute of Astronomy RAS in Moscow, and the University of Vienna.
\end{acknowledgements}

\bibliographystyle{aa.bst}
\bibliography{sources.bib}

\begin{appendix}
\section{Additional tables}
\begin{table*}
\caption{Fit information on FR-$T_\mathrm{eff}$ relations (Eq. \ref{Eq:Teff relation}) sorted by decreasing $\left|r_{\mathrm{FR,}T_\mathrm{eff}}\right|$. The width of the bands around the wavelengths $\lambda_1$ and $\lambda_2$ is 0.357~{\AA}.}
\label{Table:Teff fit results}
\centering
\begin{tabular}{c c c c c c}
\hline\hline
$r_{\mathrm{FR,}T_\mathrm{eff}}$ & $\sigma_{T_\mathrm{eff}}$ & $\lambda_1$ & $\lambda_2$ & $a$ & $b$\\
 & [K] & [{\AA}] & [{\AA}] & [K] & [K]\\
\hline                      
\input{Teff_fit_results.dat}
  \hline
  \end{tabular}
  \end{table*}

\begin{table*}
\caption{Information on the strongest FR-$T_\mathrm{eff}$-[Fe/H] relations (Eq. \ref{Eq: [Fe/H] surface}) sorted by decreasing $r'$. The width of the bands around the wavelengths $\lambda_1$ and $\lambda_2$ is 0.187~{\AA}.}
\label{Table:met fit results}
\centering
\resizebox{\textwidth}{!}{%
\begin{tabular}{ccccccccrrrrrr}
\hline\hline
$r'$ & $r_{\mathrm{FR,[Fe/H]}}$ & $\sigma_\mathrm{[Fe/H]}$ & $\lambda_1$ & $\lambda_2$ & $\lambda_\mathrm{ref}^{(a)}$ & Species$^{(a)}$ & depth$_\odot^{(a)}$ & $a$ & $b$ & $c$ & $d$ & $\beta$ & $\gamma$\\
 & & [dex] & [{\AA}] & [{\AA}] &  [{\AA}] &  &  & [dex] & [$10^{-4}$~dex~K$^{-1}$] & [$10^{-5}$~dex~K$^{-1}$] & [dex] & &\\
\hline                      
\input{met_fit_results.dat}
  \hline
  \end{tabular}}
  \tablefoot{
  \tablefoottext{a}{Strongest line in the vicinity of $\lambda_1$ in a sun-like star according to VALD.}
  }
  \end{table*}
  
\begin{table*}
\caption{Information on the strongest FR-$T_\mathrm{eff}$-[Fe/H]-$\log{g}$ relations (Eq. \ref{Eq: gravity surface}) sorted by decreasing $R'$. The width of the bands around the wavelengths $\lambda_1$ and $\lambda_2$ is 0.187~{\AA}.}
\label{Table:logg fit results}
\centering
\resizebox{\textwidth}{!}{%
\begin{tabular}{cccccccrrrr}
\hline\hline
$R'$ & $\sigma_{\log{g}}$ & $\lambda_1$ & $\lambda_2$ & $\lambda_\mathrm{ref}^{(a)}$ & Species$^{(a)}$ & depth$_\odot^{(a)}$ & $a$ & $b$ & $c$ & $d$\\
 & [dex] & [{\AA}] & [{\AA}] &  [{\AA}] &  &  & [dex] & [dex~K$^{-1}$] &  & [dex]\\
\hline
\input{logg_fit_results.dat}
  \hline
  \end{tabular}}
  \tablefoot{
  \tablefoottext{a}{Strongest participating ionized line in a sun-like star according to VALD.}
  }
  \end{table*} 
  
 \begin{table*}
\caption{Coefficients for the resolution-dependent correction to measured FRs (Eq. \ref{Eq: met_Rdependence}). The order of appearance is the same as in Tables \ref{Table:Teff fit results}, \ref{Table:met fit results}, and \ref{Table:logg fit results}.}
\label{Table: R_correction_coefficients}
\centering
\small
%\resizebox{\textwidth}{!}{%
\begin{tabular}{c@{\hskip 1cm}c@{\hskip 1cm}c@{\hskip 1cm}c@{\hskip 1cm}c@{\hskip 1cm}c}
\hline\hline
$p_{11}$ & $p_{12}$ & $p_{21}$ & $p_{22}$ & $p_{31}$ & $p_{32}$\\
\hline
\multicolumn{6}{c}{$T_\mathrm{eff}$}\\
\hline
\input{R_correction_coefficients.tab}
  \hline
  \end{tabular}%}
  \end{table*} 
 
\end{appendix}

\end{document}

%% file: Teff_training_vs_ELODIE.dat
HD~103095  & 5140\phantom{$^{(a)}$} & 5064 &   76\\
HD~175305  & 5059\phantom{$^{(a)}$} & 5040 &   19\\
HD~188510  & 5400\phantom{$^{(a)}$} & 5510 & -110\\
HD~45282   & 5148\phantom{$^{(a)}$} & 5273 & -125\\
HD~108317  & 5027\phantom{$^{(a)}$} & 5244 & -217\\

%% file: Teff_training_vs_S4N.dat
 $\alpha$~Boo        & 4286\phantom{$^{(a)}$} & 4158 &  128\\
 $\beta$~Gem         & 4858\phantom{$^{(a)}$} & 4666 &  192\\
 $\delta$~Eri        & 4954\phantom{$^{(a)}$} & 5023 &  -69\\
 $\epsilon$~Eri      & 5076\phantom{$^{(a)}$} & 5052 &   24\\
 $\alpha$~Cen~B      & 5231\phantom{$^{(a)}$} & 4970 &  261\\
 $\mu$~Cas           & 5308\phantom{$^{(a)}$} & 5323 &  -15\\
 $\tau$~Cet          & 5414\phantom{$^{(a)}$} & 5328 &   86\\
 $\alpha$~Cen~A      & 5792\phantom{$^{(a)}$} & 5519 &  273\\
 18~Sco              & 5810\phantom{$^{(a)}$} & 5693 &  117\\
 $\beta$~Hyi         & 5873\phantom{$^{(a)}$} & 5772 &  101\\
 $\beta$~Vir         & 6083\phantom{$^{(a)}$} & 6076 &    7\\
 $\eta$~Boo          & 6099\phantom{$^{(a)}$} & 5942 &  157\\
 $\alpha$~CMi        & 6554\phantom{$^{(a)}$} & 6677 & -123\\

%% file: C09_phot_scale_vs_spec.dat
NGC~104  (47~Tuc) & $-175$ &  84 & $-0.80$ & 0.09 & $-0.05$ & $-154$ &  90 & $-0.76$ & 0.03 & $-0.01$ & $-0.75$ & 0.12 & 11\\
NGC~288           & $ -31$ & 113 & $-1.35$ & 0.08 & $-0.09$ & $ ...$ & ... & $  ...$ &  ... & $  ...$ & $-1.26$ & 0.14 & 10\\
NGC~1904 (M~79)   & $-107$ &  82 & $-1.55$ & 0.04 & $-0.01$ & $ ...$ & ... & $  ...$ &  ... & $  ...$ & $-1.54$ & 0.06 & 10\\
NGC~2808          & $ -10$ &  96 & $-1.17$ & 0.09 & $-0.06$ & $  44$ & 106 & $-1.06$ & 0.06 & $ 0.06$ & $-1.11$ & 0.11 & 10\\
NGC~3201          & $ -86$ &  35 & $-1.46$ & 0.07 & $-0.03$ & $ ...$ & ... & $  ...$ &  ... & $  ...$ & $-1.44$ & 0.13 & 13\\
NGC~4590 (M~68)   & $  41$ &  43 & $-2.27$ & 0.09 & $ 0.22$ & $ ...$ & ... & $  ...$ &  ... & $  ...$ & $-2.49$ & 0.07 & 13\\
NGC~5904 (M~5)    & $-101$ &  70 & $-1.34$ & 0.06 & $ 0.05$ & $ ...$ & ... & $  ...$ &  ... & $  ...$ & $-1.39$ & 0.17 & 14\\
NGC~6121 (M~4)    & $ -97$ &  68 & $-1.17$ & 0.05 & $-0.07$ & $ ...$ & ... & $  ...$ &  ... & $  ...$ & $-1.10$ & 0.16 & 14\\
NGC~6171 (M~107)  & $-300$ &  80 & $-1.07$ & 0.05 & $-0.11$ & $ ...$ & ... & $  ...$ &  ... & $  ...$ & $-0.96$ & 0.12 &  5\\
NGC~6218 (M~12)   & $ -26$ &  58 & $-1.37$ & 0.05 & $-0.14$ & $ ...$ & ... & $  ...$ &  ... & $  ...$ & $-1.22$ & 0.09 & 11\\
NGC~6254 (M~10)   & $  87$ &  98 & $-1.60$ & 0.07 & $ 0.07$ & $ ...$ & ... & $  ...$ &  ... & $  ...$ & $-1.66$ & 0.13 & 14\\
NGC~6397          & $ 204$ &  42 & $-2.03$ & 0.04 & $ 0.27$ & $ ...$ & ... & $  ...$ &  ... & $  ...$ & $-2.30$ & 0.07 & 13\\
NGC~6752          & $ -99$ &  42 & $-1.52$ & 0.05 & $-0.06$ & $  84$ &  38 & $-1.48$ & 0.02 & $-0.03$ & $-1.46$ & 0.11 & 14\\
NGC~6809 (M~55)   & $ -80$ &  60 & $-1.92$ & 0.07 & $-0.05$ & $ ...$ & ... & $  ...$ &  ... & $  ...$ & $-1.87$ & 0.08 & 14\\
NGC~6838 (M~71)   & $-103$ & 122 & $-0.86$ & 0.08 & $ 0.04$ & $ ...$ & ... & $  ...$ &  ... & $  ...$ & $-0.90$ & 0.23 & 11\\
NGC~7078 (M~15)   & $ 117$ &  62 & $-2.35$ & 0.07 & $ 0.23$ & $  52$ &  24 & $-2.47$ & 0.07 & $ 0.15$ & $-2.59$ & 0.13 & 13\\
NGC~7099 (M~30)   & $  77$ &  72 & $-2.34$ & 0.06 & $ 0.15$ & $ ...$ & ... & $  ...$ &  ... & $  ...$ & $-2.49$ & 0.09 & 10\\

%% file: MIKE_resids.dat
NGC~104  (47~Tuc) & $  14$ &  65 & $-0.84$ & 0.05 & $-0.07$ & \citet{Koch08}     & $-0.77$ & 0.07 &  8\\
NGC~5897          & $ -71$ &  44 & $-1.92$ & 0.08 & $-0.02$ & \citet{Koch14}     & $-1.89$ & 0.13 &  7\\
NGC~5904 (M~5)    & $ -39$ &  89 & $-1.21$ & 0.04 & $-0.17$ & \citet{Koch10}     & $-1.04$ & 0.13 &  6\\
NGC~6397          & $  22$ & 289 & $-2.07$ & 0.08 & $ 0.03$ & \citet{Koch11}     & $-2.10$ & 0.38 &  5\\
NGC~6426          & $ 203$ &  92 & $-2.37$ & 0.04 & $ 0.21$ & \citet{Hanke17}    & $-2.57$ & 0.16 &  4\\
NGC~6864 (M~75)   & $ -55$ & 156 & $-0.98$ & 0.13 & $-0.14$ & \citet{Kacharov13} & $-0.84$ & 0.15 & 15\\

%% file: Teff_fit_results.dat
$-0.986$ & 117 & 4860.279 & 4854.482 & $  -7095$ & 11059\\
$-0.986$ & 129 & 4862.778 & 4867.334 & $  -8586$ & 12755\\
$-0.985$ & 118 & 6564.478 & 6585.898 & $  -8131$ & 12234\\
$-0.985$ & 116 & 4862.421 & 4874.559 & $  -6405$ & 10497\\
$-0.983$ & 123 & 6560.347 & 6552.680 & $ -11576$ & 15705\\
$-0.982$ & 131 & 6559.803 & 6537.737 & $ -10724$ & 15004\\
$-0.981$ & 132 & 4863.458 & 4875.239 & $  -8602$ & 12942\\
$-0.981$ & 122 & 6560.738 & 6538.502 & $  -8741$ & 12896\\
$-0.979$ & 124 & 4858.630 & 4850.198 & $ -10689$ & 15128\\

%% file: met_fit_results.dat
0.986 & $-0.819$ & 0.19 & 5429.643 & 5427.042 & 5429.696 & \ion{Fe}{i}  & 0.88 & $-6.610$ & 11.198 & $ 11.829$ & $-4.402$ & $\infty$ & $\infty$\\
0.985 & $-0.852$ & 0.19 & 5227.122 & 5228.941 & 5227.189 & \ion{Fe}{i}  & 0.88 & $-6.379$ & 10.285 & $  5.500$ & $-3.719$ & $\infty$ & $\infty$\\
0.984 & $-0.803$ & 0.20 & 5446.847 & 5443.974 & 5446.916 & \ion{Fe}{i}  & 0.87 & $-4.924$ & 14.706 & $-31.008$ & $-5.604$ & $\infty$ & $\infty$\\
0.983 & $-0.904$ & 0.16 & 5455.466 & 5459.801 & 5455.453 & \ion{Fe}{i}  & 0.70 & $-6.981$ &  5.615 & $ 39.246$ & $-1.442$ &  37.220 & 1.0199\\
0.983 & $-0.902$ & 0.19 & 4872.094 & 4879.234 & 4872.137 & \ion{Fe}{i}  & 0.83 & $-4.787$ &  7.924 & $ -4.738$ & $-2.575$ &  28.044 & 1.0428\\
0.982 & $-0.837$ & 0.17 & 5328.476 & 5334.579 & 5328.531 & \ion{Fe}{i}  & 0.83 & $-4.723$ & 11.695 & $-11.561$ & $-4.524$ &  82.200 & 0.9965\\
0.982 & $-0.837$ & 0.20 & 5340.971 & 5347.703 & 5341.023 & \ion{Fe}{i}  & 0.81 & $-4.828$ & 12.867 & $-21.451$ & $-4.458$ &  78.494 & 1.0090\\
0.980 & $-0.891$ & 0.20 & 5266.494 & 5259.218 & 5266.554 & \ion{Fe}{i}  & 0.81 & $-4.553$ &  9.122 & $-12.306$ & $-3.004$ &  23.424 & 1.0597\\
0.980 & $-0.931$ & 0.19 & 5404.075 & 5408.087 & 5404.151 & \ion{Fe}{i}  & 0.71 & $-5.907$ &  5.086 & $ 10.134$ & $-0.351$ &  32.793 & 1.0361\\
0.980 & $-0.900$ & 0.19 & 5006.054 & 5011.409 & 5006.118 & \ion{Fe}{i}  & 0.82 & $-6.822$ &  6.248 & $ 23.255$ & $-1.100$ &  58.534 & 1.0147\\
0.979 & $-0.934$ & 0.19 & 5424.016 & 5415.839 & 5424.067 & \ion{Fe}{i}  & 0.72 & $-6.299$ &  3.973 & $ 17.797$ & $ 0.213$ &  22.887 & 1.0530\\
0.979 & $-0.925$ & 0.18 & 5133.639 & 5140.099 & 5133.688 & \ion{Fe}{i}  & 0.72 & $-5.537$ &  6.529 & $ -3.726$ & $-0.763$ &  53.246 & 1.0165\\
0.978 & $-0.858$ & 0.19 & 5049.880 & 5046.854 & 5049.819 & \ion{Fe}{i}  & 0.81 & $-4.362$ & 12.375 & $-31.623$ & $-3.964$ & 137.924 & 1.0021\\
0.977 & $-0.889$ & 0.19 & 4903.255 & 4903.833 & 4903.310 & \ion{Fe}{i}  & 0.80 & $-5.014$ &  8.576 & $ -4.300$ & $-2.399$ &  50.381 & 1.0174\\
0.977 & $-0.929$ & 0.18 & 5369.888 & 5367.763 & 5369.961 & \ion{Fe}{i}  & 0.72 & $-8.099$ &  3.443 & $ 26.448$ & $ 1.966$ &  38.024 & 1.0275\\
0.977 & $-0.931$ & 0.19 & 5369.922 & 5362.051 & 5369.961 & \ion{Fe}{i}  & 0.72 & $-6.822$ &  3.292 & $ 27.035$ & $ 0.739$ &  29.052 & 1.0395\\
0.977 & $-0.932$ & 0.20 & 5415.159 & 5414.462 & 5415.198 & \ion{Fe}{i}  & 0.73 & $-6.860$ &  3.586 & $ 25.306$ & $ 0.580$ &  42.760 & 1.0199\\
0.976 & $-0.928$ & 0.20 & 5415.227 & 5417.607 & 5415.198 & \ion{Fe}{i}  & 0.73 & $-6.287$ &  4.085 & $ 21.904$ & $-0.076$ &  43.529 & 1.0146\\
0.976 & $-0.896$ & 0.19 & 5281.726 & 5277.272 & 5281.789 & \ion{Fe}{i}  & 0.75 & $-5.972$ &  7.677 & $  1.553$ & $-1.171$ &  51.788 & 1.0255\\
0.976 & $-0.931$ & 0.19 & 5162.216 & 5160.652 & 5162.272 & \ion{Fe}{i}  & 0.71 & $-7.568$ &  3.086 & $ 31.213$ & $ 1.384$ &  36.542 & 1.0291\\
0.976 & $-0.802$ & 0.19 & 5110.332 & 5117.285 & 5110.413 & \ion{Fe}{i}  & 0.85 & $-5.952$ & 13.363 & $ -6.359$ & $-4.283$ &  53.898 & 1.0178\\
0.974 & $-0.931$ & 0.20 & 5162.335 & 5159.496 & 5162.272 & \ion{Fe}{i}  & 0.71 & $-8.036$ &  2.606 & $ 38.862$ & $ 1.703$ &  45.191 & 1.0202\\
0.974 & $-0.838$ & 0.20 & 6546.220 & 6550.045 & 6546.237 & \ion{Fe}{i}  & 0.61 & $-5.765$ & 11.342 & $ -8.249$ & $-2.602$ &  91.439 & 1.0184\\
0.972 & $-0.904$ & 0.20 & 6400.071 & 6406.735 & 6400.000 & \ion{Fe}{i}  & 0.66 & $-6.359$ &  6.320 & $ 11.849$ & $-0.569$ &  50.731 & 1.0214\\
0.972 & $-0.912$ & 0.20 & 5014.877 & 5008.145 & 5014.942 & \ion{Fe}{i}  & 0.71 & $-7.053$ &  6.110 & $  6.324$ & $ 0.638$ &  83.241 & 1.0078\\
0.970 & $-0.844$ & 0.19 & 5142.887 & 5143.363 & 5142.928 & \ion{Fe}{i}  & 0.80 & $-5.292$ &  9.575 & $ 12.631$ & $-3.506$ &  70.293 & 1.0072\\
0.969 & $-0.919$ & 0.19 & 5444.994 & 5440.931 & 5445.041 & \ion{Fe}{i}  & 0.64 & $-6.421$ &  4.837 & $  7.639$ & $ 0.802$ &  43.797 & 1.0184\\
0.968 & $-0.894$ & 0.19 & 5215.137 & 5222.022 & 5215.180 & \ion{Fe}{i}  & 0.72 & $-6.704$ &  6.178 & $ 20.460$ & $-0.459$ &  75.759 & 1.0123\\
0.966 & $-0.910$ & 0.19 & 4983.189 & 4979.432 & 4983.250 & \ion{Fe}{i}  & 0.71 & $-7.336$ &  5.430 & $ 11.336$ & $ 1.220$ &  88.469 & 1.0078\\
0.965 & $-0.912$ & 0.20 & 5463.235 & 5467.587 & 5463.275 & \ion{Fe}{i}  & 0.65 & $-8.174$ &  2.550 & $ 42.073$ & $ 2.020$ &  56.474 & 1.0105\\
0.962 & $-0.910$ & 0.19 & 5078.967 & 5078.559 & 5078.974 & \ion{Fe}{i}  & 0.65 & $-6.356$ &  5.470 & $  8.597$ & $ 0.422$ &  76.379 & 1.0071\\

%% file: logg_fit_results.dat
0.977 & 0.25 & 5316.508 & 5322.492 & 5316.609 & \ion{Fe}{ii} & 0.68 & $ 14.635$ &  8.522 & $  1.395$ & $-11.732$\\
0.974 & 0.26 & 5197.474 & 5189.348 & 5197.568 & \ion{Fe}{ii} & 0.64 & $ 14.344$ &  9.288 & $  1.318$ & $-12.549$\\
0.973 & 0.27 & 5275.895 & 5273.022 & 5275.997 & \ion{Fe}{ii} & 0.66 & $ 13.314$ &  9.859 & $  1.173$ & $-12.282$\\
0.971 & 0.29 & 5234.500 & 5229.553 & 5234.623 & \ion{Fe}{ii} & 0.61 & $ 15.691$ & 11.450 & $  1.032$ & $-15.961$\\
0.969 & 0.30 & 5169.135 & 5170.291 & 5169.028 & \ion{Fe}{ii} & 0.80 & $  8.678$ &  7.940 & $  1.372$ & $ -5.012$\\
0.968 & 0.31 & 5018.549 & 5018.923 & 5018.436 & \ion{Fe}{ii} & 0.81 & $  9.935$ &  7.606 & $  1.390$ & $ -6.053$\\
0.967 & 0.30 & 5234.738 & 5229.366 & 5234.623 & \ion{Fe}{ii} & 0.61 & $ 14.429$ &  9.649 & $  1.319$ & $-12.900$\\
0.966 & 0.32 & 4924.046 & 4917.654 & 4923.921 & \ion{Fe}{ii} & 0.81 & $ 10.001$ &  8.929 & $  1.256$ & $ -7.585$\\
0.966 & 0.32 & 5197.661 & 5199.191 & 5197.568 & \ion{Fe}{ii} & 0.64 & $ 15.213$ &  7.582 & $  1.557$ & $-11.717$\\
0.965 & 0.31 & 4923.791 & 4921.360 & 4923.921 & \ion{Fe}{ii} & 0.81 & $  9.285$ & 11.712 & $  0.977$ & $ -9.534$\\
0.961 & 0.33 & 5362.986 & 5369.310 & 5362.861 & \ion{Fe}{ii} & 0.53 & $ 18.580$ & 10.704 & $  1.151$ & $-18.198$\\

%% file: R_correction_coefficients.tab
$ -0.1145$ & $\phantom{-}0.1006$ & $\phantom{-}0.2395$ & $ -0.3236$ & $ -0.1422$ & $\phantom{-}0.2233$\\
$ -0.1932$ & $ -0.0055$ & $\phantom{-}0.4174$ & $ -0.0443$ & $ -0.2273$ & $\phantom{-}0.0564$\\
$\phantom{-}0.0997$ & $\phantom{-}0.0383$ & $ -0.1507$ & $ -0.0629$ & $\phantom{-}0.0213$ & $\phantom{-}0.0087$\\
$ -0.0367$ & $\phantom{-}0.0029$ & $\phantom{-}0.1005$ & $ -0.0362$ & $ -0.0957$ & $\phantom{-}0.0225$\\
$ -0.0147$ & $ -0.0048$ & $\phantom{-}0.0206$ & $\phantom{-}0.0051$ & $ -0.0036$ & $ -0.0010$\\
$ -0.0083$ & $ -0.0030$ & $\phantom{-}0.0146$ & $\phantom{-}0.0046$ & $ -0.0021$ & $ -0.0007$\\
$ -0.4459$ & $ -0.0867$ & $\phantom{-}1.1138$ & $\phantom{-}0.2134$ & $ -0.7034$ & $ -0.1342$\\
$ -0.0044$ & $ -0.0028$ & $\phantom{-}0.0030$ & $ -0.0008$ & $ -0.0011$ & $ -0.0001$\\
$\phantom{-}0.3741$ & $\phantom{-}0.6467$ & $ -0.8781$ & $ -1.4578$ & $\phantom{-}0.5119$ & $\phantom{-}0.8113$\\
\hline
\multicolumn{6}{c}{[Fe/H]}\\
\hline
$ -0.1628$ & $\phantom{-}0.0963$ & $\phantom{-}0.0211$ & $\phantom{-}0.1719$ & $\phantom{-}0.1624$ & $ -0.2435$\\
$ -0.0957$ & $\phantom{-}0.1268$ & $\phantom{-}0.0235$ & $ -0.0305$ & $\phantom{-}0.0900$ & $ -0.0769$\\
$ -0.1974$ & $\phantom{-}0.0114$ & $\phantom{-}0.2249$ & $\phantom{-}0.3543$ & $ -0.0019$ & $ -0.3430$\\
$ -0.5505$ & $\phantom{-}0.0105$ & $\phantom{-}1.5792$ & $\phantom{-}0.6570$ & $ -0.9952$ & $ -0.6611$\\
$ -0.1128$ & $\phantom{-}0.0785$ & $ -0.1788$ & $\phantom{-}0.0939$ & $\phantom{-}0.3126$ & $ -0.1597$\\
$ -0.0506$ & $\phantom{-}0.1028$ & $ -0.1990$ & $ -0.0592$ & $\phantom{-}0.2545$ & $ -0.0481$\\
$ -0.4901$ & $ -0.1390$ & $\phantom{-}0.8957$ & $\phantom{-}0.8738$ & $ -0.3928$ & $ -0.7218$\\
$ -0.2962$ & $\phantom{-}0.0304$ & $\phantom{-}0.3082$ & $\phantom{-}0.2296$ & $ -0.0013$ & $ -0.2497$\\
$ -0.3275$ & $ -0.2857$ & $\phantom{-}0.5860$ & $\phantom{-}1.0933$ & $ -0.2469$ & $ -0.7979$\\
$ -0.3663$ & $ -0.1501$ & $\phantom{-}0.6220$ & $\phantom{-}0.6880$ & $ -0.2349$ & $ -0.5207$\\
$ -0.4833$ & $ -0.3609$ & $\phantom{-}0.7898$ & $\phantom{-}1.3104$ & $ -0.2771$ & $ -0.9227$\\
$\phantom{-}0.0921$ & $\phantom{-}0.4613$ & $ -0.6880$ & $ -0.6347$ & $\phantom{-}0.6762$ & $\phantom{-}0.2447$\\
$ -0.5689$ & $ -0.2943$ & $\phantom{-}0.9994$ & $\phantom{-}1.1321$ & $ -0.4193$ & $ -0.8307$\\
$ -0.7387$ & $ -0.2816$ & $\phantom{-}1.5571$ & $\phantom{-}1.4245$ & $ -0.7521$ & $ -1.0862$\\
$ -0.4483$ & $ -1.0653$ & $\phantom{-}0.7227$ & $\phantom{-}2.8071$ & $ -0.2652$ & $ -1.7269$\\
$ -0.5011$ & $ -0.5237$ & $\phantom{-}0.6973$ & $\phantom{-}1.5339$ & $ -0.1853$ & $ -0.9987$\\
$ -0.5269$ & $ -0.4523$ & $\phantom{-}0.7844$ & $\phantom{-}1.5684$ & $ -0.2024$ & $ -1.0724$\\
$ -0.7816$ & $ -0.4215$ & $\phantom{-}1.4134$ & $\phantom{-}1.5670$ & $ -0.6101$ & $ -1.1310$\\
$ -0.4919$ & $ -0.0894$ & $\phantom{-}0.9176$ & $\phantom{-}0.6867$ & $ -0.4131$ & $ -0.5857$\\
$ -0.4279$ & $ -0.2774$ & $\phantom{-}0.6881$ & $\phantom{-}1.0599$ & $ -0.2449$ & $ -0.7715$\\
$ -0.4665$ & $\phantom{-}0.1212$ & $\phantom{-}1.0117$ & $\phantom{-}0.2511$ & $ -0.5373$ & $ -0.3664$\\
$ -0.8631$ & $ -0.6335$ & $\phantom{-}1.8016$ & $\phantom{-}2.0952$ & $ -0.9253$ & $ -1.4544$\\
$ -0.9674$ & $ -0.0418$ & $\phantom{-}1.7919$ & $\phantom{-}0.7950$ & $ -0.8198$ & $ -0.7393$\\
$ -0.1267$ & $ -0.3806$ & $ -0.0206$ & $\phantom{-}1.2404$ & $\phantom{-}0.1690$ & $ -0.8449$\\
$ -0.1583$ & $ -0.2914$ & $\phantom{-}0.0661$ & $\phantom{-}0.9410$ & $\phantom{-}0.1003$ & $ -0.6466$\\
$ -0.6306$ & $ -0.0556$ & $\phantom{-}1.4125$ & $\phantom{-}0.8338$ & $ -0.7524$ & $ -0.7490$\\
$ -0.6343$ & $ -0.5944$ & $\phantom{-}1.1681$ & $\phantom{-}1.8994$ & $ -0.5308$ & $ -1.3066$\\
$ -0.4738$ & $ -0.0811$ & $\phantom{-}0.8130$ & $\phantom{-}0.7534$ & $ -0.3386$ & $ -0.6707$\\
$ -0.5036$ & $ -0.1870$ & $\phantom{-}0.6535$ & $\phantom{-}0.5247$ & $ -0.1492$ & $ -0.3366$\\
$ -0.5608$ & $ -0.4396$ & $\phantom{-}1.0014$ & $\phantom{-}1.2750$ & $ -0.4357$ & $ -0.8335$\\
$ -0.7883$ & $ -0.7793$ & $\phantom{-}1.5488$ & $\phantom{-}2.1325$ & $ -0.7461$ & $ -1.3466$\\
\hline
\multicolumn{6}{c}{$\log{g}$}\\
\hline
$\phantom{-}0.0053$ & $\phantom{-}0.0607$ & $\phantom{-}0.2061$ & $\phantom{-}0.1452$ & $ -0.1970$ & $ -0.1910$\\
$ -0.1542$ & $\phantom{-}0.8918$ & $\phantom{-}0.4985$ & $ -1.5541$ & $ -0.3362$ & $\phantom{-}0.6851$\\
$\phantom{-}0.5844$ & $ -0.2724$ & $ -0.9297$ & $\phantom{-}1.0836$ & $\phantom{-}0.3515$ & $ -0.7988$\\
$ -1.1834$ & $\phantom{-}0.6540$ & $\phantom{-}2.9930$ & $ -0.8632$ & $ -1.8135$ & $\phantom{-}0.2200$\\
$ -0.2574$ & $\phantom{-}0.0419$ & $\phantom{-}0.9056$ & $\phantom{-}0.2766$ & $ -0.6439$ & $ -0.3034$\\
$ -0.5870$ & $ -0.2431$ & $\phantom{-}1.5675$ & $\phantom{-}1.1114$ & $ -0.9139$ & $ -0.8202$\\
$ -1.4071$ & $ -0.8273$ & $\phantom{-}3.3310$ & $\phantom{-}2.2169$ & $ -1.9190$ & $ -1.3806$\\
$ -0.4093$ & $\phantom{-}0.3062$ & $\phantom{-}1.2904$ & $ -0.1234$ & $ -0.8358$ & $ -0.1546$\\
$ -0.4633$ & $ -0.1567$ & $\phantom{-}1.1246$ & $\phantom{-}0.7837$ & $ -0.6525$ & $ -0.6126$\\
$ -0.4241$ & $\phantom{-}0.2582$ & $\phantom{-}1.3438$ & $ -0.2326$ & $ -0.8688$ & $\phantom{-}0.0201$\\
$ -1.0676$ & $\phantom{-}1.1142$ & $\phantom{-}2.7772$ & $ -1.9071$ & $ -1.7034$ & $\phantom{-}0.8097$\\